\documentclass[reqno]{amsart}
\usepackage{amssymb}
\usepackage{hyperref}
\usepackage{upref}

\newtheorem{theorem}{Theorem}[section]
\newtheorem{lemma}[theorem]{Lemma}

\newtheorem{hypothesis}[theorem]{Hypothesis}

\theoremstyle{definition}

\newtheorem{remark}[theorem]{Remark}

\allowdisplaybreaks

\newcommand{\bbN}{{\mathbb{N}}}
\newcommand{\bbR}{{\mathbb{R}}}
\newcommand{\bbZ}{{\mathbb{Z}}}
\newcommand{\bbC}{{\mathbb{C}}}
\newcommand{\bbT}{{\mathbb{T}}}

\newcommand{\calB}{{\mathcal B}}
\newcommand{\calC}{{\mathcal C}}
\newcommand{\calD}{{\mathcal D}}
\newcommand{\calE}{{\mathcal E}}
\newcommand{\calF}{{\mathcal F}}
\newcommand{\calI}{{\mathcal I}}
\newcommand{\calK}{{\mathcal K}}
\newcommand{\calL}{{\mathcal L}}
\newcommand{\calM}{{\mathcal M}}

\newcommand{\N}{p}
\newcommand{\f}{\frac}
\newcommand{\lb}{\label}
\newcommand{\no}{\nonumber}
\newcommand{\be}{\begin{equation}}
\newcommand{\ee}{\end{equation}}
\newcommand{\bea}{\begin{eqnarray}}
\newcommand{\eea}{\end{eqnarray}}
\newcommand{\ul}{\underline}
\newcommand{\ol}{\overline}
\newcommand{\ti}{\widetilde}

\renewcommand{\Im}{\text{\rm Im}}
\newcommand{\deven}{\delta_{\mathrm{even}}}
\newcommand{\dodd}{\delta_{\mathrm{odd}}}

\DeclareMathOperator{\sAL}{s-AL}
\DeclareMathOperator{\shAL}{s-\hatt {AL}}
\DeclareMathOperator{\AL}{AL}

\newcommand{\humu}{{ \hat{\underline{\mu} }}}
\newcommand{\hunu}{{\underline{\hat{\nu}}}}
\newcommand{\hmu}{{\hat{\mu} }}
\newcommand{\hnu}{{\hat{\nu}}}

\newcommand{\uz}{{\underline{z}}}

\newcommand{\ual}{{\underline{\alpha}}}
\newcommand{\ua}{{\underline{A}}}

\newcommand{\uxi}{{\underline{\Xi}}}

\newcommand{\Div}{\operatorname{Div}}
\newcommand{\Amap}{\ul{A}_{Q_0}}
\newcommand{\amap}{\ul{\alpha}_{Q_0}}

\newcommand{\Oh}{O}

\newcommand{\dott}{\,\cdot\,}
\newcommand{\hatt}{\widehat}  
\newcommand{\Pinfp}{{P_{\infty_+}}}
\newcommand{\Pinfm}{{P_{\infty_-}}}
\newcommand{\Pzp}{{P_{0,+}}}
\newcommand{\Pzm}{{P_{0,-}}}
\newcommand{\Pinfpm}{{P_{\infty_\pm}}}
\newcommand{\Pzpm}{{P_{0,\pm}}}

\DeclareMathOperator{\sym}{Sym}

\newcommand{\symq}{{\sym^p (\calK_p)}}

\newcommand{\pgam}{\Gamma}

\newcommand{\eps}{\varepsilon}

\numberwithin{equation}{section}

\begin{document}

\title[Algebro-Geometric Solutions of the Ablowitz--Ladik Hierarchy]{Algebro-Geometric Finite-Band Solutions of the Ablowitz--Ladik Hierarchy}
\author[F.\ Gesztesy]{Fritz Gesztesy}
\address{Department of Mathematics,
University of Missouri,
Columbia, MO 65211, USA}
\email{\href{mailto:fritz@math.missouri.edu}{fritz@math.missouri.edu}}
\urladdr{\href{http://www.math.missouri.edu/personnel/faculty/gesztesyf.html}{http://www.math.missouri.edu/personnel/faculty/gesztesyf.html}}

\author[H.\ Holden]{Helge Holden}
\address{Department of Mathematical Sciences,
Norwegian University of
Science and Technology, NO--7491 Trondheim, Norway}
\email{\href{mailto:holden@math.ntnu.no}{holden@math.ntnu.no}}
\urladdr{\href{http://www.math.ntnu.no/\~{}holden/}{http://www.math.ntnu.no/\~{}holden/}}

\author[J. Michor]{Johanna Michor}
\address{Faculty of Mathematics\\
University of Vienna\\
Nordbergstrasse 15\\ 1090 Wien\\ Austria\\ and International Erwin Schr\"odinger
Institute for Mathematical Physics, Boltzmanngasse 9\\ 1090 Wien\\ Austria}
\email{\href{mailto:Johanna.Michor@esi.ac.at}{Johanna.Michor@esi.ac.at}}
\urladdr{\href{http://www.mat.univie.ac.at/~jmichor/}{http://www.mat.univie.ac.at/\~{}jmichor/}}

\author[G. Teschl]{Gerald Teschl}
\address{Faculty of Mathematics\\
University of Vienna\\
Nordbergstrasse 15\\ 1090 Wien\\ Austria\\ and International Erwin Schr\"odinger
Institute for Mathematical Physics, Boltzmanngasse 9\\ 1090 Wien\\ Austria}
\email{\href{mailto:Gerald.Teschl@univie.ac.at}{Gerald.Teschl@univie.ac.at}}
\urladdr{\href{http://www.mat.univie.ac.at/~gerald/}{http://www.mat.univie.ac.at/\~{}gerald/}}

\dedicatory{To Walter Thirring, on the occasion of his 80th birthday, \\ theoretical and mathematical physicist extraordinaire.}

\thanks{Research supported in part by the Research Council of Norway,  
the US National Science Foundation under Grant No.\ DMS-0405526, and
the Austrian Science Fund (FWF) under Grant No.\ Y330.}

\thanks{{\it Int. Math. Res. Notices} 2007, rnm082, 1--55.}


\begin{abstract}
We provide a detailed derivation of all complex-valued
algebro-geometric finite-band solutions of the Ablowitz--Ladik hierarchy. In addition, we survey a recursive construction of the Ablowitz--Ladik hierarchy and its zero-curvature and Lax formalism.
\end{abstract}

\maketitle

\section{Introduction}\lb{sAL1}

In the mid-seventies, Ablowitz and Ladik, in a series of papers
\cite{AblowitzLadik:1975}--\cite{AblowitzLadik2:1976} (see also
\cite{Ablowitz:1977}, \cite[Sect.\ 3.2.2]{AblowitzClarkson:1991},
\cite[Ch.\ 3]{AblowitzPrinariTrubatch:2004},
\cite{ChiuLadik:1977}), used inverse scattering methods to analyze certain
integrable differential-difference systems. One of their integrable
variants of such systems included a discretization of the celebrated AKNS-ZS system, the pair of coupled nonlinear differential difference
equations,
\begin{align}
\begin{split}
-i\alpha_t - (1 - \alpha \beta) (\alpha^- +\alpha^+) + 2\alpha &=0, \lb{AL1.12} \\
-i\beta_t + (1 - \alpha\beta)(\beta^- + \beta^+) - 2\beta &=0   
\end{split}
\end{align}
with $\alpha=\alpha(n,t)$, $\beta=\beta(n,t)$, $(n,t)\in\bbZ\times\bbR$. 
Here we used the notation $f^{\pm}(n) = f(n\pm 1)$, $n\in\bbZ$, for complex-valued 
sequences $f=\{f(n)\}_{n\in\bbZ}$. In particular, Ablowitz and Ladik 
\cite{AblowitzLadik:1976} (see also \cite[Ch.\ 3]{AblowitzPrinariTrubatch:2004}) showed that in the focusing case, where $\beta = -\ol\alpha$, and in the defocusing case, where $\beta = \ol\alpha$, \eqref{AL1.12} yields the discrete analog of the nonlinear Schr\"odinger equation
\begin{equation}
-i\alpha_t - (1 \pm |\alpha|^2)(\alpha^- + \alpha^+) + 2\alpha =0. \lb{AL1.13}
\end{equation}
We will refer to \eqref{AL1.12} as the Ablowitz--Ladik system. The principal theme of this paper will be to derive the algebro-geometric finite-band solutions of the Ablowitz--Ladik (AL) hierarchy, a completely integrable sequence of systems of nonlinear evolution equations on the lattice $\bbZ$ whose first nonlinear member is the Ablowitz--Ladik system \eqref{AL1.12}.

Since the mid-seventies there has been an enormous amount of activity in the area of integrable differential-difference equations. Two principal directions of research are responsible for this development: Originally, the development  was driven by the theory of completely integrable systems and its applications to fields such as nonlinear optics, and more recently, it gained additional momentum due to its intimate connections with the theory of orthogonal polynomials. In the following we first briefly recall the development in connection with integrable systems and subsequently turn to the one influenced by research on orthogonal polynomials.

The first systematic discussion of the Ablowitz--Ladik hierarchy appears to be due to Schilling \cite{Schilling:1989} (cf.\ also \cite{TamizhmaniMa:2000}, 
\cite{Vekslerchik:2002}, \cite{ZengRauchWojciechowski:1995}); infinitely many conservation laws are derived, for instance, by Ding, Sun, and Xu \cite{DingSunXu:2006}; the bi-Hamiltonian structure of the AL hierarchy is considered by Ercolani and Lozano \cite{ErcolaniLozano:2006}; connections
between the AL hierarchy and the motion of a piecewise linear curve have
been established by Doliwa and Santini \cite{DoliwaSantini:1995}; B\"acklund and Darboux transformations were studied by Geng \cite{Geng:1989} and 
Vekslerchik \cite{Vekslerchik:2006}; the Hirota bilinear formalism, AL $\tau$-functions, etc., were considered by Vekslerchik \cite{Vekslerchik:2002}. The initial value problem for half-infinite AL systems was discussed by Common \cite{Common:1992}, for an application of the inverse scattering method to \eqref{AL1.13} we refer to Vekslerchik and Konotop \cite{VekslerchikKonotop:1992}. This just scratches the surface of these developments and the interested reader will find much more material in the references cited in these papers and the ones discussed below.

Algebro-geometric (and periodic) solutions of the AL system \eqref{AL1.12} 
have briefly been studied by Ahmad and Chowdhury \cite{AhmadChowdhury:1987}, 
\cite{AhmadChowdhury:1987a}, Bogolyubov, Prikarpatskii, and Samoilenko
\cite{BogolyubovPrikarpatskiiSamoilenko:1981},  Bogolyubov and
Prikarpatskii \cite{BogolyubovPrikarpatskii:1982}, 
Chow, Conte, and Xu \cite{ChowConteXu:2006}, Geng, Dai, and Cao 
\cite{GengDaiCao:2003}, and Vaninsky \cite{Vaninsky:2001}. In an effort to analyze models describing oscillations in nonlinear dispersive wave systems, Miller, Ercolani, Krichever, and Levermore \cite{MillerErcolaniKricheverLevermore:1995} (see also \cite{Miller:1995}) gave a detailed analysis of algebro-geometric solutions of the AL system 
\eqref{AL1.12}. Introducing 
\begin{equation}
U(z)=\begin{pmatrix} z & \alpha \\ \beta z & 1 \end{pmatrix},
\quad
V(z)=i\begin{pmatrix}
z-1-\alpha\beta^- & \alpha - \alpha^- z^{-1}\\ \beta^- z -\beta &
1+ \alpha^- \beta -z^{-1} \end{pmatrix}    \lb{AL1.20}
\end{equation}
with $z\in\bbC\setminus\{0\}$ a spectral parameter,  the authors in
\cite{MillerErcolaniKricheverLevermore:1995} relied on the fact that the 
Ablowitz--Ladik system \eqref{AL1.12} is equivalent to the zero-curvature equations
\begin{equation}
U_t+UV -V^+ U=0,  \lb{AL1.21}
\end{equation} 
the latter being the compatibility relation for the spatial and temporal linear problems 
\begin{equation}
\Phi= U \Phi^-,  \quad  \Phi_t^- = V \Phi^-.  \lb{AL1.22}
\end{equation}
Here we extended the notation $f^{\pm}(n) = f(n\pm 1)$, $n\in\bbZ$, to 
$\bbC^2$-valued and $2\times 2$-matrix-valued sequences with complex-valued entries.

Miller, Ercolani, Krichever, and Levermore
\cite{MillerErcolaniKricheverLevermore:1995} then performed a thorough
analysis of the solutions $\Phi=\Phi(z,n,t)$ associated with the pair
$(U,V)$ and derived the theta function representations of $\alpha,
\beta$ satisfying the AL system \eqref{AL1.12}. In the
particular focusing and defocusing cases they also discussed periodic
and quasi-periodic solutions $\alpha$ with respect to $n\in\bbZ$ and 
$t\in\bbR$. Vekslerchik \cite{Vekslerchik:1999} also studied finite-genus solutions for the AL hierarchy by establishing connections with Fay's identity for theta functions.

The connection between the Ablowitz--Ladik system \eqref{AL1.12} and orthogonal polynomials comes about as follows: Let $\{\alpha (n)\}_{n\in\bbN}\subset\bbC$ be a sequence of complex
numbers subject to the condition $|\alpha (n)|<1$, $n\in\bbN$, and define the transfer matrix
\begin{equation}
T(z)=\begin{pmatrix} z & \alpha \\ \overline{\alpha} z &1
\end{pmatrix}, \quad z\in\bbT,     \lb{AL1.3}
\end{equation}
with spectral parameter $z$ on the unit circle
$\bbT=\{z\in\bbC\,|\,|z|=1\}$. Consider the system of difference equations
\begin{equation}
\Phi(z,n) = T(z,n)\Phi(z,n-1), \quad (z,n)\in\bbT\times\bbN,    \lb{AL1.1}
\end{equation}
with initial condition $\Phi (z,0) = \begin{pmatrix} 1\\ 1 \end{pmatrix}$, where
\begin{equation}
\Phi(z,n)=\begin{pmatrix} \varphi(z,n)\\ z^n{\ol \varphi}(1/z,n) \end{pmatrix},
\quad (z,n)\in\bbT\times\bbN_0.
\lb{AL1.4}
\end{equation}
(Here $\bbN_0=\bbN\cup\{0\}$.) Then $\varphi (\dott,n)$ are monic
polynomials of degree $n$ first introduced by Szeg\H o in the
1920's in his seminal work on the asymptotic distribution of eigenvalues of
sections of Toeplitz forms \cite{Szego:1920}, \cite{Szego:1921}  (see
also \cite[Ch.\ XI]{Szego:1978}). Szeg\H o's point of departure was the
trigonometric moment problem and hence the theory of orthogonal
polynomials on the unit circle. Indeed, given a probability measure $d\sigma$
supported on an infinite set on the unit circle, one is interested in finding monic polynomials $\chi(\dott,n)$ of degree $n\in\bbN_0$ in $z=e^{i\theta}$, $\theta\in [0,2\pi]$, such that
\begin{equation}
\int_{0}^{2\pi}d\sigma(e^{i\theta})\,
\overline{\chi (e^{i\theta},m)}
\chi (e^{i\theta},n) =  w(n)^{-2} \delta_{m,n}, \quad m,n\in\bbN_0,
\lb{AL1.8}
\end{equation}
where $w(0)^2=1$, $w(n)^2=\prod_{j=1}^n \big(1-|\alpha(j)|^2\big)^{-1}$, $n\in\bbN$. Szeg\H o showed that the corresponding polynomials \eqref{AL1.4} with $\varphi$ replaced by $\chi$ satisfy the recurrence formula \eqref{AL1.1}.  Early work in this area includes important contributions by Akhiezer, Geronimus, Krein, Tom{\v c}uk, Verblunsky, Widom, and others, and is summarized in the books by Akhiezer \cite{Akhiezer:1965}, Geronimus \cite{Geronimus:1961}, Szeg{\H o} \cite{Szego:1978}, and especially in the recent two-volume treatise by Simon \cite{Simon:2005}.

Unaware of the paper \cite{MillerErcolaniKricheverLevermore:1995},
Geronimo and Johnson \cite{GeronimoJohnson:1998} studied \eqref{AL1.1} in
the case where the coefficients $\alpha$ are random variables. Under appropriate ergodicity assumptions on $\alpha$ and the
hypothesis of a vanishing Lyapunov exponent on prescribed spectral
arcs on the unit circle $\bbT$, Geronimo and Johnson 
\cite{GeronimoJohnson:1998} (cf.\ also 
\cite{GeronimoJohnson:1996}, \cite{GeronimoTeplyaev:1994}) developed the corresponding spectral theory associated with \eqref{AL1.1} and the unitary operator it generates in $\ell^2(\bbZ)$. In this sense the discussion in \cite{GeronimoJohnson:1998} is a purely stationary one and connections with a zero-curvature formalism, theta function representations, and integrable hierarchies are not made in 
\cite{GeronimoJohnson:1998} (but in this context we refer to the discussion concerning  
\cite{GeronimoGesztesyHolden:2004} in the next paragraph). More recently, the defocusing case with periodic and quasi-periodic coefficients was studied in great detail by Deift \cite{Deift:2007}, Golinskii and Nevai 
\cite{GolinskiiNevai:2001}, Killip and Nenciu \cite{KillipNenciu:2006}, 
Li \cite{Li:2005}, Nenciu  \cite{Nenciu:2005}, \cite{Nenciu:2006}, and 
Simon \cite[Ch.\ 11]{Simon:2005}, \cite{Simon:2005a}, \cite{Simon:2006}. 

An important extension of \eqref{AL1.1} was developed by Baxter in a
series of papers on Toeplitz forms \cite{Baxter:1960}--\cite{Baxter:1963} in 1960--63. In these papers the
transfer matrix $T$ in \eqref{AL1.3} is replaced by the more general (complexified) transfer matrix $U(z)=\left(\begin{smallmatrix} z & \alpha \\ 
\beta z & 1 \end{smallmatrix}\right)$ in \eqref{AL1.22}, that is, precisely the matrix $U$ responsible for the spatial part in the Ablowitz--Ladik system in its zero-curvature formulation \eqref{AL1.20}--\eqref{AL1.22}. Here $\alpha=\{\alpha(n)\}_{n\in\bbN}$, $\beta=\{\beta(n)\}_{n\in\bbN}$ are  subject to the condition $\alpha (n)\beta (n)\ne 1$, $n\in\bbN$. Studying the following extension of \eqref{AL1.1},
\begin{equation}
\Psi(z,n) = U(z,n)\Psi(z,n-1), \quad 
(z,n)\in\bbT\times\bbN,  \lb{AL1.10a}
\end{equation}
Baxter was led to biorthogonal polynomials on the unit circle with respect
to a complex-valued measure on $\bbT$. In this context of biorthogonal Laurent polynomials we refer to \cite{BertolaGekhtman:2005} and 
\cite{BultheelGonzalezVeraHendriksenNjastad:1999}. Reference 
\cite{BertolaGekhtman:2005}, in particular, deals with isomonodromic tau functions and is applicable to generalized integrable lattices of the Toda-type. Baxter's $U$ matrix in \eqref{AL1.22} led to a new hierarchy of nonlinear difference equations, called the  Szeg{\H o}--Baxter (SB) hierarchy in 
\cite{GeronimoGesztesyHolden:2004}, in honor of these two pioneers of orthogonal polynomials on the unit circle. The latter reference also contains an in depth study of algebro-geometric solutions of \eqref{AL1.1}.

In addition to these recent  developments on the AL system and the AL hierarchy, we offer a variety of results in this paper apparently not covered before. These include: 

$\bullet$ An elementary, yet effective recursive construction of the AL hierarchy using Laurent polynomials.

$\bullet$ Explicit formulas for Lax pairs for the AL hierarchy.

$\bullet$ The detailed connection between the AL hierarchy and a ``complexified'' version of transfer matrices first introduced by Baxter.

$\bullet$ A unified treatment of stationary algebro-geometric finite-band solutions  and their theta function representations of the entire AL hierarchy. 

$\bullet$ A unified treatment of algebro-geometric solutions  and their theta function representations of the time-dependent AL hierarchy by solving the $\ul r$th AL flow with initial data given by stationary algebro-geometric finite-band  solutions. 

The structure of this paper is as follows: In Section \ref{sAL2} we
describe our zero-curvature formalism for the Ablowitz--Ladik (AL) hierarchy.
Extending a recursive polynomial approach discussed in great detail in 
\cite{GesztesyHolden:2003} in the continuous case and in 
\cite{BullaGesztesyHoldenTeschl:1997}, \cite[Ch.\ 4]{GesztesyHolden:2005}, 
\cite[Chs.\ 6, 12]{Teschl:2000} in the discrete context to the case of Laurent polynomials with respect to the spectral parameter, we derive the AL hierarchy of systems of nonlinear evolution equations whose first nonlinear member is the Ablowitz--Ladik system \eqref{AL1.12}. Section \ref{sALstat} is devoted to a detailed study of the stationary AL hierarchy. We employ the recursive Laurent polynomial formalism of Section \ref{sAL2} to describe nonnegative divisors of degree $p$ on a hyperelliptic curve $\calK_p$ of genus $p$ associated with the $\ul p$th system in the stationary AL hierarchy. By means of a fundamental meromorphic function $\phi$ on $\calK_p$ (an analog of the Weyl--Titchmarsh function for the system \eqref{AL1.1}) we then proceed to derive the theta function representations of the associated Baker--Akhiezer vector and all stationary algebro-geometric finite-band solutions of the AL hierarchy. The corresponding time-dependent results for the AL hierarchy are presented in detail in Section \ref{sALtime}. Appendix \ref{sAL.A} collects relevant material on hyperelliptic curves and their theta functions and introduces the terminology freely used in Sections \ref{sALstat} and \ref{sALtime}. Appendix \ref{ALApp.high} is of a technical nature and summarizes expansions of various key quantities related to the Laurent polynomial recursion formalism as the spectral parameter tends to zero and to infinity.

\section{The Ablowitz--Ladik Hierarchy, Recursion Relations, 
Zero-Curvature Pairs, and Hyperelliptic Curves} \label{sAL2}

In this section we summarize the construction of the Ablowitz--Ladik hierarchy employing a Laurent polynomial recursion formalism and derive the associated sequence of Ablowitz--Ladik zero-curvature pairs (we also hint at Lax pairs). Moreover, we discuss the Burchnall--Chaundy Laurent polynomial in connection with the stationary Ablowitz--Ladik hierarchy and the underlying hyperelliptic curve. For a detailed treatment of this material we refer to 
\cite{GesztesyHolden:2005}, \cite{GesztesyHoldenMichorTeschl:2007}.

We denote by $\bbC^{\bbZ}$ the set of complex-valued sequences indexed by $\bbZ$. 

Throughout this section we suppose the following hypothesis. 

\begin{hypothesis} \lb{hAL2.1} 
In the stationary case we assume that $\alpha, \beta$ satisfy
\begin{equation}
\alpha, \beta\in \bbC^{\bbZ}, \quad  \alpha(n)\beta(n)\notin \{0,1\}, \; n\in\bbZ.   
\lb{AL2.01}
\end{equation}
In the time-dependent case we assume that $\alpha, \beta$ satisfy
\begin{align}
\begin{split}
& \alpha(\dott,t), \beta(\dott,t) \in \bbC^{\bbZ}, \; t\in\bbR, \quad 
\alpha(n,\dott), \beta(n,\dott)\in C^1(\bbR), \; n\in\bbZ,   \lb{AL2.01a}  \\
& \alpha(n,t)\beta(n,t)\notin \{0,1\}, \; (n,t)\in\bbZ\times \bbR.
\end{split}
\end{align}
\end{hypothesis}

We denote by $S^\pm$ the shift operators acting on complex-valued sequences 
$f=\{f(n)\}_{n\in\bbZ} \in\bbC^{\bbZ}$ according to
\begin{equation}
(S^\pm f)(n)=f(n\pm1), \quad n\in\bbZ. \lb{AL2.02}
\end{equation}
Moreover, we will frequently use the notation
\begin{equation}
f^\pm = S^{\pm} f, \quad f\in\bbC^{\bbZ}. 
\end{equation}

To construct the  Ablowitz--Ladik hierarchy one typically introduces appropriate zero-curvature pairs of  $2\times2$ matrices, denoted by $U(z)$ and $V_{\ul p}(z)$, 
$\ul p=(p_-,p_+)\in\bbN_0^2$  (with $z$ a certain spectral parameter to be discussed later), defined recursively in the following. We take the shortest route to the construction of $V_{\ul p}$ and hence to that of the Ablowitz--Ladik hierarchy by starting from the recursion relation \eqref{AL0+}--\eqref{ALh_l-} below.   

Define sequences $\{f_{\ell,\pm}\}_{\ell\in \bbN_0}$, $\{g_{\ell,\pm}\}_{\ell\in \bbN_0}$, and $\{h_{\ell,\pm}\}_{\ell\in \bbN_0}$ recursively by
\begin{align} \label{AL0+}
g_{0,+} &= \tfrac12 c_{0,+}, \quad f_{0,+} = - c_{0,+}\alpha^+, 
\quad h_{0,+} = c_{0,+}\beta, \\ \label{ALg_l+}
g_{\ell+1,+} - g_{\ell+1,+}^- &= \alpha h_{\ell,+}^- + \beta f_{\ell,+}, \quad \ell\in \bbN_0,\\ \label{ALf_l+}
f_{\ell+1,+}^- &= f_{\ell,+} - \alpha (g_{\ell+1,+} + g_{\ell+1,+}^-), \quad \ell\in \bbN_0, \\  \label{ALh_l+}
h_{\ell+1,+} &= h_{\ell,+}^- + \beta (g_{\ell+1,+} + g_{\ell+1,+}^-), \quad \ell\in \bbN_0,  
\end{align}
and
\begin{align} \label{AL0-}
g_{0,-} &= \tfrac12 c_{0,-}, \quad f_{0,-} = c_{0,-}\alpha, 
\quad h_{0,-} = - c_{0,-}\beta^+, \\ \label{ALg_l-}
g_{\ell+1,-} - g_{\ell+1,-}^- &= \alpha h_{\ell,-} + \beta f_{\ell,-}^-, \quad \ell\in \bbN_0,\\ \label{ALf_l-}
f_{\ell+1,-} &= f_{\ell,-}^- + \alpha (g_{\ell+1,-} + g_{\ell+1,-}^-), \quad \ell\in \bbN_0, \\ \label{ALh_l-}
h_{\ell+1,-}^- &= h_{\ell,-} - \beta (g_{\ell+1,-} + g_{\ell+1,-}^-), \quad \ell\in \bbN_0.
\end{align}
Here $c_{0,\pm}\in\bbC$ are given constants. For later use we also introduce
\begin{equation}\lb{ALminus}
f_{-1,\pm}= h_{-1,\pm}=0.
\end{equation}

\begin{remark}\lb{rAL2.2}
The sequences $\{f_{\ell,+}\}_{\ell\in \bbN_0}$, 
$\{g_{\ell,+}\}_{\ell\in \bbN_0}$, and
$\{h_{\ell,+}\}_{\ell\in \bbN_0}$ can be computed recursively as follows: 
Assume that $f_{\ell,+}$,
$g_{\ell,+}$, and $h_{\ell,+}$ are known.  Equation \eqref{ALg_l+} is a 
first-order difference equation in $g_{\ell+1,+}$ that can be solved directly
and yields a local lattice function that is determined up to a new constant denoted
by $c_{\ell+1,+}\in\bbC$. Relations \eqref{ALf_l+} and \eqref{ALh_l+}
then determine $f_{\ell+1,+}$ and $h_{\ell+1,+}$, etc.  The sequences 
$\{f_{\ell,-}\}_{\ell\in \bbN_0}$, $\{g_{\ell,-}\}_{\ell\in \bbN_0}$, and 
$\{h_{\ell,-}\}_{\ell\in \bbN_0}$ are determined similarly.
\end{remark}

Upon setting 
\begin{equation}
\gamma = 1 - \alpha \beta, \lb{ALgamma}
\end{equation}
one explicitly obtains 
\begin{align}
\begin{split}
f_{0,+} &= c_{0,+}(-\alpha^+), \quad 
f_{1,+} = c_{0,+}\big(- \gamma^+ \alpha^{++} + (\alpha^+)^2 \beta\big) 
+ c_{1,+} (-\alpha^+), \\
g_{0,+} &= \tfrac{1}{2}c_{0,+},  \quad 
 g_{1,+} = c_{0,+}(-\alpha^+ \beta) + \tfrac{1}{2}c_{1,+}, \\
h_{0,+} &= c_{0,+}\beta, \quad 
 h_{1,+} = c_{0,+}\big(\gamma \beta^- - \alpha^+ \beta^2\big) 
+ c_{1,+} \beta, \\
f_{0,-} &= c_{0,-}\alpha, \quad  
f_{1,-} = c_{0,-}\big(\gamma \alpha^- - \alpha^2 \beta^+\big) + c_{1,-} \alpha, \\
g_{0,-} &= \tfrac{1}{2}c_{0,-}, \quad  
g_{1,-} = c_{0,-}(-\alpha \beta^+) + \tfrac{1}{2}c_{1,-}, \\
h_{0,-} &= c_{0,-}(-\beta^+), \quad  
h_{1,-} = c_{0,-}\big(- \gamma^+ \beta^{++} 
+ \alpha (\beta^+)^2 \big) + c_{1,-} (- \beta^+), \, \text{ etc.}
\end{split}
\end{align}
Here $\{c_{\ell,\pm}\}_{\ell \in \bbN}$ denote summation constants
which naturally arise when solving the difference equations for 
$g_{\ell, \pm}$ in \eqref{ALg_l+}, \eqref{ALg_l-}.  
Subsequently, it will also be useful to work with the corresponding homogeneous coefficients $\hat f_{\ell, \pm}$,
$\hat g_{\ell, \pm}$, and $\hat h_{\ell, \pm}$, defined by the vanishing of all summation constants $c_{k,\pm}$ for $k=1,\dots,\ell$, and choosing $c_{0,\pm}=1$,
\begin{align}
& \hat f_{0,+}=-\alpha^+, \quad \hat f_{0,-}=\alpha, \quad 
 \hat f_{\ell,\pm}=f_{\ell,\pm}|_{c_{0,\pm}=1, \, c_{j,\pm}=0, j=1,\dots,\ell},  \quad \ell\in\bbN, 
 \lb{AL2.04a} \\
& \hat g_{0,\pm}=\tfrac12, \quad 
\hat g_{\ell,\pm}=g_{\ell,\pm}|_{c_{0,\pm}=1, \, c_{j,\pm}=0, j=1,\dots,\ell}, 
\quad \ell\in\bbN,  \lb{AL2.04b} \\
& \hat h_{0,+}=\beta, \quad \hat h_{0,-}=-\beta^+,  \quad 
\hat h_{\ell,\pm}=h_{\ell,\pm}|_{c_{0,\pm}=1, \, c_{j,\pm}=0, j=1,\dots,\ell}, 
\quad \ell\in\bbN.  \lb{AL2.04c}
\end{align}
By induction one infers that
\begin{equation} \label{ALhat f}
f_{\ell, \pm} = \sum_{k=0}^\ell c_{\ell-k, \pm} \hat f_{k, \pm}, \quad
g_{\ell, \pm} = \sum_{k=0}^\ell c_{\ell-k, \pm} \hat g_{k, \pm}, \quad 
h_{\ell, \pm} = \sum_{k=0}^\ell c_{\ell-k, \pm} \hat h_{k, \pm}.  
\end{equation} 
In a slight abuse of notation we will occasionally stress the dependence of $f_{\ell,\pm}$, 
$g_{\ell,\pm}$, and $h_{\ell,\pm}$ on $\alpha, \beta$ by writing  
$f_{\ell,\pm}(\alpha,\beta)$, $g_{\ell,\pm}(\alpha,\beta)$, and 
$h_{\ell,\pm}(\alpha,\beta)$. 

One can show (cf.\ \cite{GesztesyHoldenMichorTeschl:2007}) that all homogeneous elements 
$\hat f_{\ell,\pm}$, $\hat g_{\ell,\pm}$, and $\hat h_{\ell,\pm}$, $\ell\in\bbN_0$, are polynomials in $\alpha, \beta$, and some of their shifts. 

\begin{remark}\lb{rAL2.4}
As an efficient tool to later distinguish between nonhomogeneous and homogeneous 
quantities $f_{\ell,\pm}$, $g_{\ell,\pm}$, $h_{\ell,\pm}$, and $\hat f_{\ell,\pm}$, 
$\hat g_{\ell,\pm}$, $\hat h_{\ell,\pm}$, respectively, we now introduce the notion of degree as follows. Denote  
\begin{align}
f^{(r)}=S^{(r)}f, \quad f=\{f(n)\}_{n\in\bbZ}\in\bbC^{\bbZ}, \quad
   S^{(r)}&=\begin{cases}(S^+)^r, &\text{$r\ge 0$},\\
(S^-)^{-r}, &\text{$r< 0$},\end{cases} \quad
r\in \bbZ,    \lb{AL2.1AA}
\end{align}
and define
\begin{equation}
\deg \big(\alpha^{(r)}\big)=r, \quad \deg \big(\beta^{(r)}\big)=-r, \quad r\in\bbZ.  
\lb{3.2.15aa}
\end{equation}
This then results in 
\begin{align}
\begin{split}
\deg\big(\hat f_{\ell,+}^{(r)}\big)&= \ell+1+r, \quad \deg\big(\hat f_{\ell,-}^{(r)}\big)
= -\ell+r, \quad \deg\big(\hat g_{\ell, \pm}^{(r)}\big)= \pm\ell, \\
\deg\big(\hat h_{\ell,+}^{(r)}\big)&= \ell-r, \quad \deg\big(\hat h_{\ell,-}^{(r)}\big)
= -\ell-1-r, \quad \ell\in\bbN_0, \; r\in\bbZ. 
\end{split}
\end{align}
\end{remark}

We also note the following useful result (cf.\ \cite{GesztesyHoldenMichorTeschl:2007}):  Assume \eqref{AL2.01}, then,
\begin{align} \lb{ALadd}
\begin{split}
g_{\ell,+} - g_{\ell,+}^- &= \alpha h_{\ell,+} + \beta f_{\ell,+}^-, \quad \ell\in\bbN_0, \\
g_{\ell,-} - g_{\ell,-}^- &= \alpha h_{\ell,-}^- + \beta f_{\ell,-}, \quad \ell\in\bbN_0.   
\end{split}
\end{align}
Moreover, we note the following symmetries, 
\begin{equation}\lb{ALsym}
\hat f_{\ell,\pm}(c_{0,\pm},\alpha,\beta)=\hat h_{\ell,\mp}(c_{0,\mp},\beta, \alpha), \quad 
\hat g_{\ell,\pm}(c_{0,\pm},\alpha,\beta)=\hat g_{\ell,\mp}(c_{0,\mp},\beta, \alpha), \quad \ell\in \bbN_0.
\end{equation}

Next we relate the homogeneous coefficients $\hat f_{\ell,\pm}$, 
$\hat g_{\ell,\pm}$, and $\hat h_{\ell,\pm}$ to certain matrix elements of $L$, where $L$ will later be identified as the Lax difference expression associated with the 
Ablowitz--Ladik hierarchy. For this purpose it is useful to introduce the standard basis 
$\{\delta_m\}_{m\in\bbZ}$ in $\ell^2(\bbZ)$ by
\begin{equation}
\delta_m=\{\delta_{m,n}\}_{n\in\bbZ}, \; m\in\bbZ, \quad
\delta_{m,n}=\begin{cases} 1, &m=n, \\ 0, & m\neq n. \end{cases}
\lb{ALbasis}
\end{equation}
The scalar product in $\ell^2(\bbZ)$, denoted by $(\dott,\dott) $, is defined by
\begin{equation}
(f,g) =\sum_{n\in \bbZ} \ol{f(n)}g(n), \quad f,g \in \ell^2(\bbZ).
\lb{ALsp}
\end{equation}

In the standard basis just defined, we introduce the difference expression $L$ by
\begin{align}
L &= \left(\begin{smallmatrix} \ddots &&\hspace*{-8mm}\ddots
&\hspace*{-10mm}\ddots &\hspace*{-12mm}\ddots
&\hspace*{-14mm}\ddots &&&
\raisebox{-3mm}[0mm][0mm]{\hspace*{-6mm}{\Huge $0$}}
\\
&0& -\alpha(0) \rho(-1) & -\beta(-1)\alpha(0) &
-\alpha(1)\rho(0) & \rho(0) \rho(1)
\\
&& \rho(-1) \rho(0) & \beta(-1) \rho(0) &
-\beta(0) \alpha(1) & \beta(0) \rho(1) & 0
\\
&&&0& -\alpha(2) \rho(1) & -\beta(1) \alpha(2) &
-\alpha(3) \rho(2) & \rho(2) \rho(3)
\\
&&\raisebox{-4mm}[0mm][0mm]{\hspace*{-6mm}{\Huge $0$}} &&
\rho(1) \rho(2) & \beta(1) \rho(2) & -\beta(2) \alpha(3)
& \beta(2) \rho(3) & 0
\\
&&&&&\hspace*{-14mm}\ddots &\hspace*{-14mm}\ddots
&\hspace*{-14mm}\ddots &\hspace*{-8mm}\ddots &\ddots
\end{smallmatrix}\right)    \lb{ALLop} \\
&= \rho^- \rho \, \deven \, S^{--} + (\beta^-\rho \, \deven - \alpha^+\rho \, \dodd) S^- 
- \beta\alpha^+   \no \\
& \quad + (\beta \rho^+ \, \deven - \alpha^{++} \rho^+ \, \dodd) S^+ 
+ \rho^+ \rho^{++} \, \dodd \, S^{++},   \no 
\end{align}
where 
\begin{equation} 
\deven = \chi_{_{2\bbZ}}, \quad \dodd = 1 - \deven = \chi_{_{2\bbZ +1}}.
\end{equation}
In particular, terms of the form $-\beta(n) \alpha(n+1)$ 
represent the diagonal $(n,n)$-entries, $n\in\bbZ$, in the infinite matrix
\eqref{ALLop}. In addition, we used the abbreviation
\begin{equation}
\rho = \gamma^{1/2} = (1-\alpha \beta)^{1/2}.   \lb{ALLPrho}
\end{equation}

Next, we introduce the unitary operator $U_{\tilde\varepsilon}$ in $\ell^2(\bbZ)$ by 
\begin{equation}
U_{\tilde \varepsilon} = \big( \tilde \varepsilon(n) \delta_{m,n} \big)_{(m,n)\in\bbZ^2}, 
\quad \tilde\varepsilon(n)\in\{1,-1\},  \; n\in\bbZ,   
\lb{ALUeps}
\end{equation}
and the sequence $\varepsilon = \{\varepsilon(n)\}_{n\in\bbZ}\in\bbC^\bbZ$ by
\begin{equation}
\varepsilon(n) = \tilde\varepsilon(n-1)\tilde\varepsilon(n), \; n\in\bbZ.  
\end{equation}
A straightforward computation then shows that 
\begin{equation}
U_{\tilde \varepsilon} L U_{\tilde \varepsilon}^{-1} = \widetilde L_\varepsilon,   
\lb{ALUnEq}
\end{equation}
where $\widetilde L_{\varepsilon}$ is associated with the sequences $\tilde{\alpha}=\alpha$, $\tilde{\beta}=\beta$, and $\tilde{\rho}= \eps \rho$. Moreover, the recursion formalism in \eqref{AL0+}--\eqref{ALh_l-} yields coefficients which are polynomials in $\alpha$, $\beta$ and some of their shifts and hence depends only quadratically on $\rho$. As a result, the choice of square root of $\rho(n)$, $n\in\bbZ$, in \eqref{ALLPrho} is immaterial when introducing the AL hierarchy via the Lax equations \eqref{ALLaxEq}.  

The half-lattice (i.e., semi-infinite) version of
$L$ was rediscovered by Cantero, Moral, and Vel\'azquez
\cite{CanteroMoralVelazquez:2003} in 2003 in the special case where 
$\beta=\ol{\alpha}$ (see also Simon  
\cite{Simon:2005}--\cite{Simon:2006} who coined the term CMV matrix 
in this context). 

The next result details the connections between $L$ and the recursion coefficients $f_{\ell,\pm}$, $g_{\ell,\pm}$, and $h_{\ell,\pm}$:

\begin{lemma} \lb{lAL2.6}
Let $n\in\bbZ$. Then the homogeneous coefficients 
$\{\hat f_{\ell,\pm}\}_{\ell\in\bbN_0}$, $\{\hat g_{\ell,\pm}\}_{\ell\in\bbN_0}$, and 
$\{\hat h_{\ell,\pm}\}_{\ell\in\bbN_0}$ satisfy the following relations:
\begin{align}
\hat f_{\ell,+}(n) &=  \alpha(n) (\delta_{n},L^{ \ell+1} \delta_{n})  + 
\rho(n)\begin{cases}  
(\delta_{n-1},L^{ \ell+1} \delta_{n}) ,  & n \text{ even,}\\
(\delta_{n},L^{ \ell+1} \delta_{n-1}) ,  & n \text{ odd,} \end{cases}  
\quad \ell\in\bbN_0,  \no \\
 \hat f_{\ell,-}(n) &= \alpha(n) (\delta_{n},L^{- \ell} \delta_{n})  + 
 \rho(n)\begin{cases}  
(\delta_{n-1},L^{- \ell} \delta_{n}) ,  & n \text{ even,}\\
(\delta_{n},L^{- \ell} \delta_{n-1}) ,  & n \text{ odd,} \end{cases}  
\quad \ell\in\bbN_0, \no \\
\hat g_{0,\pm} &= 1/2, \quad 
\hat g_{\ell,\pm}(n) = (\delta_{n},L^{\pm \ell} \delta_{n}) ,  \quad 
\ell\in\bbN,  \\
\hat h_{\ell,+}(n) &= \beta(n) (\delta_{n},L^{ \ell} \delta_{n})  + 
\rho(n)\begin{cases}  
(\delta_{n},L^{ \ell} \delta_{n-1}) ,  & n \text{ even,}\\
(\delta_{n-1},L^{ \ell} \delta_{n}) ,  & n \text{ odd,} \end{cases}  
\quad \ell\in\bbN_0, \no  \\
 \hat h_{\ell,-}(n) &= \beta(n) (\delta_{n},L^{- \ell-1} \delta_{n})  + 
 \rho(n)\begin{cases}  
(\delta_{n},L^{- \ell-1} \delta_{n-1}) ,  & n \text{ even,}\\
(\delta_{n-1},L^{- \ell-1} \delta_{n}) ,  & n \text{ odd,} \end{cases} 
\;\; \ell\in\bbN_0.  \no
\end{align}
\end{lemma}

For the proof of Lemma \ref{lAL2.6} and some of its applications in connection with conservation laws and the Hamiltonian formalism for the Ablowitz--Ladik hierarchy we refer to \cite{GesztesyHoldenMichorTeschl:2007b}.

Next we define the $2\times 2$ zero-curvature matrices 
\begin{equation}
U(z) = \begin{pmatrix} z & \alpha  \\ z \beta & 1\\ \end{pmatrix}   \lb{AL2.03}
\end{equation}
and 
\begin{equation} \lb{AL_v}
V_{\ul p}(z) = i  \begin{pmatrix}
G_{\ul p}^-(z) & - F_{\ul p}^-(z)     \\[1.5mm]
H_{\ul p}^-(z) & - K_{\ul p}^-(z)  \\
\end{pmatrix},  \quad \ul p = (p_-,p_+) \in \bbN_0^2,
\end{equation}
for appropriate Laurent polynomials $F_{\ul p}(z)$, $G_{\ul p}(z)$, $H_{\ul p}(z)$, and $K_{\ul p}(z)$ in the  
spectral parameter $z\in \bbC\setminus\{0\}$ to be determined shortly. By postulating the stationary zero-curvature relation,  
\begin{equation}   \lb{ALstatzc}
0=U V_{\ul p} - V_{\ul p}^+ U,
\end{equation}
one concludes that \eqref{ALstatzc} is equivalent with
the following relations 
 \begin{align} \label{AL1,1}
z (G_{\ul p}^- - G_{\ul p}) + z \beta F_{\ul p} + \alpha H_{\ul p}^- &= 0,\\ \label{AL2,2}
z \beta F_{\ul p}^- + \alpha H_{\ul p} - K_{\ul p} + K_{\ul p}^- &= 0,\\
\label{AL1,2}
 - F_{\ul p} + z F_{\ul p}^- + \alpha (G_{\ul p} + K_{\ul p}^-) &= 0,\\ \label{AL2,1}
 z \beta (G_{\ul p}^- + K_{\ul p}) - z H_{\ul p} + H_{\ul p}^- &= 0.
\end{align}
In order to make the connection between the zero-curvature formalism and the recursion relations 
\eqref{AL0+}--\eqref{ALh_l-}, we now define Laurent polynomials $F_{\ul p}$, $G_{\ul p}$, $H_{\ul p}$, and $K_{\ul p}$ by\footnote{In this paper, a sum is interpreted as zero whenever the upper limit in the sum is strictly less than its lower limit.}
\begin{align}
F_{\ul p}(z) &= \sum_{\ell=1}^{p_-} f_{p_- -\ell,-} z^{-\ell} 
+ \sum_{\ell=0}^{p_+-1} f_{p_+ -1-\ell,+}z^\ell,  
\label{ALF_p} \\ 
G_{\ul p}(z) &= \sum_{\ell=1}^{p_-} g_{p_- -\ell,-}z^{-\ell}  
+ \sum_{\ell=0}^{p_+} g_{p_+ -\ell,+}z^\ell,  
 \label{ALG_p}  \\ 
H_{\ul p}(z) &= \sum_{\ell=0}^{p_- -1} h_{p_- -1-\ell,-}z^{-\ell}  
+ \sum_{\ell=1}^{p_+} h_{p_+ -\ell,+}z^\ell, 
 \label{ALH_p}  \\
K_{\ul p}(z) &= \sum_{\ell=0}^{p_-} g_{p_- -\ell,-}z^{-\ell}  
+  \sum_{\ell=1}^{p_+} g_{p_+ -\ell,+}z^\ell 
= G_{\ul p}(z)+g_{p_-,-}-g_{p_+,+}.   \label{ALK_p}
\end{align}

The corresponding homogeneous quantities are defined by ($\ell\in\bbN_0$)
\begin{align} 
\begin{split} 
\hatt F_{0,\mp} (z) & = 0, \quad 
\hatt F_{\ell,-}(z) = \sum_{k=1}^\ell \hat f_{\ell-k,-} z^{-k}, 
\quad  \hatt F_{\ell,+}(z) = \sum_{k=0}^{\ell-1} \hat f_{\ell-1-k,+}z^k,   \\ 
\hatt G_{0,-} (z) &  = 0,   \quad 
\hatt G_{\ell,-}(z) = \sum_{k=1}^\ell \hat g_{\ell-k,-}z^{-k}, \\ 
\hatt G_{0,+} (z) &= \f{1}{2},   \quad 
\hatt G_{\ell,+}(z) = \sum_{k=0}^\ell \hat g_{\ell-k,+}z^k,   \\ 
\hatt H_{0,\mp} (z) &  = 0, \quad 
\hatt H_{\ell,-}(z) = \sum_{k=0}^{\ell-1} \hat h_{\ell-1-k,-} z^{-k}, \quad  
\hatt H_{\ell,+}(z) = \sum_{k=1}^\ell \hat h_{\ell-k,+}z^k,   \label{ALhat_F_p} \\
\hatt K_{0,-} (z) & = \f{1}{2},  \quad    
\hatt K_{\ell,-}(z) = \sum_{k=0}^\ell \hat g_{\ell-k,-}z^{-k}  
= \hatt G_{\ell,-} (z)+\hat g_{\ell,-},  \\
\hatt K_{0,+} (z) &= 0, \quad \hatt K_{\ell,+}(z) = \sum_{k=1}^\ell \hat g_{\ell-k,+}z^k 
= \hatt G_{\ell,+} (z) - \hat g_{\ell,+}. 
\end{split} 
\end{align}

The stationary zero-curvature relation \eqref{ALstatzc}, $0=U V_{\ul p} - V_{\ul p}^+ U$, is then equivalent to 
\begin{align}
 -\alpha(g_{p_+,+} + g_{p_-,-}^-) + f_{p_+ -1,+} - f_{p_- -1,-}^-&=0,  \lb{AL2.50}\\ 
 \beta(g_{p_+,+}^- + g_{p_-,-}) + h_{p_+ -1,+}^- - h_{p_- -1,-} &=0.  \lb{AL2.51}
\end{align}
Thus, varying $p_\pm \in \bbN_0$,  equations \eqref{AL2.50} and \eqref{AL2.51} give rise to the stationary Ablowitz--Ladik (AL) hierarchy which we introduce as follows
\begin{align}\lb{ALstat}
\begin{split}
&\sAL_{\ul p}(\alpha, \beta) = 
\begin{pmatrix}- \alpha(g_{p_+,+} + g_{p_-,-}^-) + f_{p_+ -1,+} - f_{p_- -1,-}^-\\  
\beta(g_{p_+,+}^- + g_{p_-,-}) + h_{p_+ -1,+}^- - h_{p_- -1,-}  \end{pmatrix}=0,  \\ 
& \hspace*{6.9cm}  \ul p=(p_-,p_+)\in \bbN_0^2. 
\end{split}
\end{align}
Explicitly (recalling $\gamma=1-\alpha\beta$ and taking $p_-=p_+$ for simplicity), 
\begin{align} \no
\sAL_{(0,0)} (\alpha, \beta) &=  \begin{pmatrix}  -c_{(0,0)} \alpha\\[2mm]
c_{(0,0)}\beta\end{pmatrix} 
=0,\\ \no
\sAL_{(1,1)} (\alpha, \beta) &=  \begin{pmatrix} -\gamma (c_{0,-}\alpha^- + c_{0,+}\alpha^+) 
- c_{(1,1)} \alpha \\
 \gamma (c_{0,+}\beta^- + c_{0,-}\beta^+) +
c_{(1,1)} \beta\end{pmatrix}=0,\\ \no
\sAL_{(2,2)} (\alpha, \beta) &=  \begin{pmatrix}\begin{matrix}
-\gamma \big(c_{0,+}\alpha^{++} \gamma^+ + c_{0,-}\alpha^{--} \gamma^-
- \alpha (c_{0,+}\alpha^+\beta^- + c_{0,-}\alpha^-\beta^+)\\
- \beta (c_{0,-}(\alpha^-)^2 + c_{0,+}(\alpha^+)^2)\big)\end{matrix}\\[3mm] 
\begin{matrix}
 \gamma \big(c_{0,-}\beta^{++} \gamma^+ + c_{0,+}\beta^{--} \gamma^-
- \beta (c_{0,+}\alpha^+\beta^- + c_{0,-}\alpha^-\beta^+)\\
- \alpha (c_{0,+}(\beta^-)^2 + c_{0,-}(\beta^+)^2)\big)\end{matrix}\end{pmatrix}
 \\ & \quad+ \begin{pmatrix}
-\gamma (c_{1,-} \alpha^- + c_{1,+} \alpha^+) - c_{(2,2)} \alpha\\
 \gamma (c_{1,+} \beta^- + c_{1,-} \beta^+) + c_{(2,2)} \beta\end{pmatrix}
=0,  \, \text{ etc.,}
\end{align}
represent the first few equations of the stationary Ablowitz--Ladik hierarchy. 
Here we introduced  
\begin{equation}
c_{\ul p} = (c_{p_-,-} + c_{p_+,+})/2, \quad p_{\pm}\in\bbN_0.   \lb{ALdefcp}
\end{equation}
By definition, the set of solutions of \eqref{ALstat}, with $p_{\pm}$ ranging in 
$\bbN_0$ and $c_{\ell,\pm}\in\bbC$, $\ell\in\bbN_0$, represents the class of algebro-geometric Ablowitz--Ladik solutions. 

In the following we will frequently assume that $\alpha, \beta$ satisfy the $\ul p$th stationary AL system  
$\sAL_{\ul p}(\alpha, \beta) = 0$, supposing a particular choice of summation
constants $c_{\ell,\pm}\in\bbC$, $\ell=0,\dots,p_{\pm}$, $p_{\pm}\in\bbN_0$, has been made.

In accordance with our notation introduced in 
\eqref{AL2.04a}--\eqref{AL2.04c} and \eqref{ALhat_F_p}, the corresponding homogeneous stationary Ablowitz--Ladik equations are defined by  
\begin{equation}
\shAL_{\ul p} (\alpha, \beta) 
= \sAL_{\ul p} (\alpha, \beta)\big|_{c_{0,\pm}=1, \, c_{\ell,\pm}=0, \, \ell=1,\dots,p_{\pm}}, \quad \ul p=(p_-,p_+) \in \bbN_0^2.   \lb{ALstathom}
\end{equation}

In addition, one can show (cf.\ \cite[Lemma\ 2.2]{GesztesyHoldenMichorTeschl:2007}) that 
$g_{p_+,+} = g_{p_-,-}$ up to a lattice constant which can be set equal to zero without loss of generality. Thus, we will henceforth assume that
\begin{equation}
g_{p_+,+} = g_{p_-,-},   \lb{g+=g-}
\end{equation} 
which in turn implies that
\begin{equation}
K_{\ul p}= G_{\ul p}   \lb{ALK=G}
\end{equation}
and hence renders $V_{\ul p}$ in \eqref{AL_v} traceless in the stationary context. (We note that equations \eqref{g+=g-} and \eqref{ALK=G} cease to be valid in the time-dependent context, though.)

\begin{remark} \lb{rAL2.9}
$(i)$ The particular choice $c_{0,+}=c_{0,-}=1$ yields the stationary Ablowitz--Ladik equation. Scaling $c_{0,\pm}$ with the same constant then amounts to scaling $V_{\ul p}$ with this constant which drops out in the stationary zero-curvature equation 
\eqref{ALstatzc}. \\
$(ii)$ Different ratios between $c_{0,+}$  and $c_{0,-}$ will lead to different stationary hierarchies. In particular, the choice 
$c_{0,+}=2$, $c_{0,-}=\cdots=c_{p_- -1,-}=0$, $c_{p_-,-}\neq 0$, yields the stationary Baxter--Szeg\H o hierarchy considered in detail in 
\cite{GeronimoGesztesyHolden:2004}. However, in this case some parts from the recursion relation for the negative
coefficients still remain. In fact, \eqref{ALh_l-} reduces
to $g_{p_-,-}-g_{p_-,-}^- = \alpha h_{p_- -1,-}$, $h_{p_- -1,-}=0$ and thus
requires $g_{p_-,-}$ to be a constant in \eqref{ALstat} and \eqref{AL_p}. Moreover, 
$f_{p_- -1,-}=0$ in \eqref{ALstat} in this case.
\end{remark}

Next, taking into account \eqref{ALK=G}, one infers that 
the expression $R_{\ul p}$, defined as 
\begin{equation} \label{ALR}
R_{\ul p} = G_{\ul p}^2 - F_{\ul p} H_{\ul p},  
\end{equation}
is a lattice constant, that is, $R_{\ul p} - R_{\ul p}^- = 0$, by taking determinants 
in the stationary zero-curvature equation \eqref{ALstatzc}. Hence, $R_{\ul p}(z)$  only depends on $z$, and assuming in addition to \eqref{AL2.01} that 
\begin{equation}
c_{0,\pm} \in \bbC\setminus \{0\},  \quad  
\ul p = (p_-,p_+) \in\bbN_0^2 \setminus \{(0,0)\},    \lb{ALc0}
\end{equation}
one may write $R_{\ul p}$ as\footnote{We use the convention that a product is to be interpreted equal to $1$ whenever the upper limit of the product is strictly less than its lower limit.}  
\begin{align}  \label{ALE_m}
\begin{split}
& R_{\ul p}(z) = (c_{0,+}^2/4) z^{-2p_-} \prod_{m=0}^{2p+1}(z-E_m), \quad
\{E_m\}_{m=0}^{2p+1} \subset \bbC \setminus\{ 0\}, \\  
& \hspace*{5.9cm}   p=p_- + p_+ -1 \in\bbN_0.
\end{split}
\end{align}
Moreover, multiplying \eqref{ALR} by $z^{2p_-}$ and taking $z\to 0$ yields
\begin{equation}
\prod_{m=0}^{2p+1}E_m = \frac{c_{0,-}^2}{c_{0,+}^2}.   \label{ALprod E_m}
\end{equation}

Relation \eqref{ALR} allows one to introduce a hyperelliptic curve 
$\calK_p$ of (arithmetic) genus $p=p_- + p_+ -1$ (possibly with a singular affine part), where
\begin{equation} \label{ALKp}
\calK_p \colon\calF_p(z,y) = y^2 - 4c_{0,+}^{-2}z^{2p_-}R_{\ul p}(z) 
= y^2 - \prod_{m=0}^{2p+1}(z-E_m) = 0,  \quad p=p_- + p_+ -1. 
\end{equation}
 
Next we turn to the time-dependent Ablowitz--Ladik hierarchy. For that purpose the coefficients $\alpha$ and $\beta$ are now considered as functions of both the lattice point and time. For each system in the hierarchy, that is, for each 
$p_{\pm}$, we introduce a deformation (time) parameter $t_{\ul p}\in\bbR$ in 
$\alpha, \beta$, replacing $\alpha(n), \beta(n)$ by 
$\alpha(n,t_{\ul p}), \beta(n,t_{\ul p})$. Moreover, the definitions 
\eqref{AL2.03},  \eqref{AL_v}, and \eqref{ALF_p}--\eqref{ALK_p}  of $U, V_{\ul p}$, and $F_{\ul p}, G_{\ul p}, H_{\ul p}, K_{\ul p}$, respectively, still apply. Imposing the zero-curvature relation
\begin{equation}
U_{t_{\ul p}} + U V_{\ul p} - V_{\ul p}^+ U =0, \quad \ul p \in\bbN_0^2,   
\lb{ALzc p}
\end{equation}
then results in the equations
\begin{align}  \label{ALalphat}
\alpha_{t_{\ul p}} &= i \big(z F_{\ul p}^- + \alpha (G_{\ul p} + K_{\ul p}^-) - F_{\ul p}\big),    \\ \label{ALbetat}
\beta_{t_{\ul p}} &= - i \big(\beta (G_{\ul p}^- + K_{\ul p}) - H_{\ul p} 
+ z^{-1} H_{\ul p}^-\big), \\ \label{AL1,1r}
0 &= z (G_{\ul p}^- - G_{\ul p}) + z\beta F_{\ul p} + \alpha H_{\ul p}^-,   \\ \label{AL2,2r}
0 &= z \beta F_{\ul p}^- + \alpha H_{\ul p} + K_{\ul p}^- - K_{\ul p}.
\end{align}
Varying $\ul p \in \bbN_0^2$, the collection of evolution equations   
\begin{align}   \label{AL_p}
\begin{split}
& \AL_{\ul p} (\alpha, \beta) =
\begin{pmatrix}-i\alpha_{t_{\ul p}} - \alpha(g_{p_+,+} + g_{p_-,-}^-) + f_{p_+-1,+} 
- f_{p_- -1,-}^-\\  -i\beta_{t_{\ul p}}+ \beta(g_{p_+,+}^- + g_{p_-,-}) - h_{p_- -1,-} 
+ h_{p_+-1,+}^- \end{pmatrix}=0,  \\
& \hspace*{6.44cm} t_{\ul p}\in\bbR, \; \ul p=(p_-,p_+) \in\bbN_0^2,   
\end{split}
\end{align}
then defines the time-dependent Ablowitz--Ladik hierarchy. Explicitly, taking 
$p_-=p_+$ for simplicity,  
\begin{align} \no
&\AL_{(0,0)} (\alpha, \beta) =  \begin{pmatrix} -i \alpha_{t_{(0,0)}}- c_{(0,0)}\alpha \\[2mm]
-i\beta_{t_{(0,0)}}+c_{(0,0)}\beta \end{pmatrix} 
=0,\\ \no
&\AL_{(1,1)} (\alpha, \beta) =  \begin{pmatrix}  
-i \alpha_{t_{(1,1)}}- \gamma (c_{0,-}\alpha^- + c_{0,+}\alpha^+) 
- c_{(1,1)} \alpha \\
-i\beta_{t_{(1,1)}}+ \gamma (c_{0,+}\beta^- + c_{0,-}\beta^+) +
c_{(1,1)} \beta\end{pmatrix}=0,\\ 
&\AL_{(2,2)} (\alpha, \beta)    \\
& \quad =  \begin{pmatrix}\begin{matrix}-i \alpha_{t_{(2,2)}}-
\gamma \big(c_{0,+}\alpha^{++} \gamma^+ + c_{0,-}\alpha^{--} \gamma^-
- \alpha (c_{0,+}\alpha^+\beta^- + c_{0,-}\alpha^-\beta^+)\\
- \beta (c_{0,-}(\alpha^-)^2 + c_{0,+}(\alpha^+)^2)\big)\end{matrix}\\[3mm] 
\begin{matrix}-i\beta_{t_{(2,2)}}+
 \gamma \big(c_{0,-}\beta^{++} \gamma^+ + c_{0,+}\beta^{--} \gamma^-
- \beta (c_{0,+}\alpha^+\beta^- + c_{0,-}\alpha^-\beta^+)\\
- \alpha (c_{0,+}(\beta^-)^2 + c_{0,-}(\beta^+)^2)\big)\end{matrix}\end{pmatrix}
 \no \\ 
 & \qquad+ \begin{pmatrix}
-\gamma (c_{1,-} \alpha^- + c_{1,+} \alpha^+) - c_{(2,2)} \alpha\\
 \gamma (c_{1,+} \beta^- + c_{1,-} \beta^+) + c_{(2,2)} \beta\end{pmatrix}
=0, \, \text{ etc.,}   \no
\end{align}
represent the first few equations of the time-dependent Ablowitz--Ladik hierarchy. 
Here we recall the definition of $c_{\ul p}$ in \eqref{ALdefcp}.

By \eqref{AL_p}, \eqref{ALg_l+}, and \eqref{ALg_l-},
the time derivative of $\gamma=1-\alpha \beta$ is given by
\begin{equation} \lb{AL2.14}
\gamma_{t_{\ul p}} = i \gamma \big((g_{p_+,+} - g_{p_+,+}^-) 
- (g_{p_-,-} - g_{p_-,-}^-) \big).
\end{equation}

\begin{remark}  \lb{rAL2.13}
From \eqref{AL1,1}--\eqref{AL2,1} and the explicit computations of the coefficients 
$f_{\ell,\pm}$, $g_{\ell,\pm}$, and $h_{\ell,\pm}$, one concludes that the zero-curvature equation \eqref{ALzc p} 
and hence the Ablowitz--Ladik hierarchy is invariant under the scaling transformation 
\begin{equation}
\alpha \rightarrow \alpha_c = \{c\, \alpha(n)\}_{n\in\bbZ}, \quad 
\beta \rightarrow \beta_c = \{ \beta(n)/c\}_{n\in\bbZ}, \quad c \in \bbC\setminus \{0\}.
\end{equation}
Moreover, $R_{\ul p}=G_{\ul p}^2 - H_{\ul p}F_{\ul p}$ and hence $\{E_m\}_{m=0}^{2p+1}$ are 
invariant under this transformation. Furthermore, choosing $c=e^{i c_{\ul p} t}$,
one verifies that it is no restriction to assume $c_{\ul p}=0$. This also indicates
that stationary solutions $\alpha, \beta$ can only be constructed up to a
multiplicative constant $($compare Theorem~\ref{tAL3.7}, in particular, \eqref{AL3.66}, 
\eqref{AL3.68}$)$. 
\end{remark}

Finally, we briefly hint at explicit expressions of the Lax pair for the Ablowitz--Ladik hierarchy. More details will be presented in \cite{GesztesyHoldenMichorTeschl:2007b}.  First we need some notation. Let $T$ be a bounded operator in $\ell^2(\bbZ)$. Given the standard basis \eqref{ALbasis} in $\ell^2(\bbZ)$, we represent $T$ by
\begin{equation}
T=\big(T(m,n)\big)_{(m,n)\in\bbZ^2}, \quad 
T(m,n)=(\delta_m,T \, \delta_n), \quad (m, n) \in\bbZ^2. \lb{ALTop}
\end{equation}
Moreover, we introduce the upper and lower triangular parts $T_\pm$ of $T$ by
\begin{equation}
T_\pm=\big(T_\pm (m,n)\big)_{(m,n)\in\bbZ^2}, \quad
T_\pm (m,n)=\begin{cases} T(m,n), &\pm(n-m)>0, \\ 0, & \text{otherwise.}
\end{cases}
\lb{ALTpm}
\end{equation}

Next, consider the finite difference expression $P_{\ul p}$ defined by 
\begin{align}
P_{\ul p}& = \f{i}{2} \sum_{\ell=1}^{p_+} c_{p_+ -\ell,+} \big((L^\ell)_+ - (L^\ell)_- \big)
-\f{i}{2} \sum_{\ell=1}^{p_-}  c_{p_- -\ell,-} \big( (L^{-\ell})_+ - (L^{-\ell})_- \big) 
\no \\ 
& \quad - \f{i}{2} c_{\ul p} \, Q_d,  \quad  \ul p=(p_-,p_+) \in\bbN_0^2,    \lb{ALP_p} 
\end{align} 
with $L$ given by \eqref{ALLop}, $c_{\ul p}=(c_{p_-,-} + c_{p_+,+})/2$, and $Q_d$ denoting the doubly infinite diagonal matrix
\begin{equation} 
Q_d=\big((-1)^k \delta_{k,\ell} \big)_{k,\ell \in\bbZ}.  
\end{equation}
Then one can show that $(L,P_{\ul p})$ represents the Lax pair for the Ablowitz--Ladik equations \eqref{AL_p} for $\ul p \in\bbN_0^2\setminus\{(0,0\}$. In particular, the hierarchy of nonlinear Ablowitz--Ladik evolution equations 
\eqref{AL_p} then can alternatively be derived by imposing the Lax commutator equations  
\begin{equation}
L_{t_{\ul p}}(t_{\ul p}) - [P_{\ul p}(t_{\ul p}), L(t_{\ul p})] =0, \quad t_{\ul p}\in\bbR, \; 
\ul p \in\bbN_0^2.  \lb{ALLaxEq} 
\end{equation}
For additional representations of $P_{\ul p}$ in terms of $L$ and a particular factorization of $L$ we refer to \cite{GesztesyHoldenMichorTeschl:2007b}. The Ablowitz--Ladik Lax pair in the special defocusing case, where $\beta=\ol{\alpha}$, in the finite-dimensional context,  was recently derived by Nenciu \cite{Nenciu:2006}.  

In the special stationary case, where $P_{\ul p}$ and $L$ commute, $[P_{\ul p},L]=0$, they satisfy an algebraic relationship of the type  
\begin{align}
\begin{split}
& P_{\ul p}^2 + R_{\ul p}(L) = P_{\ul p}^2 
+ (c_{0,+}^2/4) L^{-2p_-} \prod_{m=0}^{2p+1}(L - E_m) =0,    \lb{ALBC} \\
& R_{\ul p}(z)= (c_{0,+}^2/4) z^{-2p_-} \prod_{m=0}^{2p+1}(z-E_m), 
\quad p=p_- + p_+ -1.  
\end{split} 
\end{align}
Thus, the expression $P_{\ul p}^2 + R_{\ul p}(L)$ in \eqref{ALBC} represents the Burchnall--Chaundy Laurent polynomial of the Lax pair $(L,P_{\ul p})$.

\section{The Stationary Ablowitz--Ladik Formalism}  \lb{sALstat}

This section is devoted to a detailed study of the stationary
Ablowitz--Ladik hierarchy and its algebro-geometric solutions. Our principal
tools are derived from combining the Laurent polynomial recursion formalism
introduced in Section \ref{sAL2} and a fundamental meromorphic function
$\phi$ on a hyperelliptic curve $\calK_p$. With the help of $\phi$ we
study the Baker--Akhiezer vector $\Psi$, trace formulas, and theta
function representations of $\phi$, $\Psi$, $\alpha$, and $\beta$. For proofs of the elementary results of the stationary formalism we refer to \cite{GesztesyHolden:2005}, \cite{GesztesyHoldenMichorTeschl:2007}.

Unless explicitly stated otherwise, we suppose throughout this section that 
\begin{equation} \label{ALneq 0,1}
\alpha, \beta \in \bbC^\bbZ, \quad 
\alpha(n)\beta(n) \notin \{0,1\}, \; n \in \bbZ, 
\end{equation}
and assume  \eqref{AL0+}--\eqref{ALminus}, 
\eqref{AL2.03}--\eqref{ALstatzc},
\eqref{ALF_p}--\eqref{ALK_p}, \eqref{ALstat}, \eqref{ALK=G}, \eqref{ALR}, 
\eqref{ALE_m}, keeping $\ul p \in\bbN_0^2\setminus\{(0,0\}$ fixed.

We recall the hyperelliptic curve 
\begin{align} \label{ALcalK_p}
& \calK_p \colon \calF_p(z,y) 
= y^2 - 4c_{0,+}^{-2}z^{2p_-}R_{\ul p}(z) = y^2 - \prod_{m=0}^{2p+1}(z-E_m) = 0,  \\
& R_{\ul p}(z) = \bigg(\frac{c_{0,+}}{2z^{p_-}}\bigg)^2\prod_{m=0}^{2p+1}(z-E_m), 
\quad  \{E_m\}_{m=0}^{2p+1} \subset \bbC \setminus\{ 0\}, \; p=p_- + p_+ -1,   \no 
\end{align}
as introduced in \eqref{ALKp}. Throughout this section we assume the affine part of 
$\calK_p$ to be nonsingular, that is, we suppose that 
\begin{equation}
E_m\neq E_{m'} \text{  for $m\neq m'$, \; $m,m'=0,1,\dots,2p+1$.}   \lb{ALEneqE}
\end{equation} 
$\calK_p$ is compactified by
joining two points $P_{\infty_\pm}$,
$P_{\infty_+}\neq P_{\infty_-}$, but for notational simplicity  the
compactification is also denoted by $\calK_p$. Points $P$ on
$\calK_p\setminus\{\Pinfp, \Pinfm\}$ are  represented as pairs $P=(z,y)$, where
$y(\dott)$ is the meromorphic function on $\calK_p$ satisfying
$\calF_p(z,y)=0$. The complex structure on $\calK_p$ is then defined in the usual way, see Appendix \ref{sAL.A}. Hence, $\calK_p$ becomes a two-sheeted hyperelliptic Riemann surface of genus $p$ in a standard manner.

We also emphasize that by fixing the curve $\calK_p$ (i.e., by fixing
$E_0,\dots,E_{2p+1}$), the summation constants $c_{1,\pm},\dots,c_{p_{\pm},\pm}$ in $f_{p_{\pm},\pm}$, $g_{p_{\pm},\pm}$, and $h_{p_{\pm},\pm}$ (and hence in the corresponding stationary $\sAL_{\ul p}$ equations) are uniquely determined as is clear from \eqref{ALBc} which establishes the summation constants $c_{\ell,\pm}$ as symmetric functions of 
$E_0^{\pm 1},\dots,E_{2p+1}^{\pm 1}$. 

For notational simplicity we will usually tacitly assume that $p\in\bbN$ and hence  
$\ul p\in\bbN_0^2 \setminus\{(0,0),(0,1),(1,0)\}$. 

We denote by $\{\mu_j(n)\}_{j=1,\dots,p}$ and $\{\nu_j(n)\}_{j=1,\dots,p}$ the zeros of $(\dott)^{p_-}F_{\ul p}(\dott,n)$ and 
$(\dott)^{p_- -1} H_{\ul p}(\dott,n)$, respectively.  Thus, we may write 
\begin{align}   
F_{\ul p}(z)&= - c_{0,+}\alpha^+ z^{-p_-}\prod_{j=1}^p(z-\mu_j),   \label{ALmu(n)}  \\
H_{\ul p}(z)&= c_{0,+}\beta z^{-p_- +1}\prod_{j=1}^p(z-\nu_j), \label{ALnu(n)}
\end{align}
and we recall that (cf.\ \eqref{ALR})
\begin{equation}  
R_{\ul p} - G_{\ul p}^2 = - F_{\ul p} H_{\ul p}.   \lb{ALquad}
\end{equation}
The next step is crucial; it permits us to ``lift'' the zeros $\mu_j$ and $\nu_j$ of 
$(\dott)^{p_-}F_{\ul p}$ and $(\dott)^{p_- -1}H_{\ul p}$ from the complex plane $\bbC$ to the curve $\calK_p$.
From \eqref{ALquad} one infers that
\begin{equation}
R_{\ul p}(z) -G_{\ul p}(z)^2=0, \quad
z\in\{\mu_j,\nu_k\}_{j,k=1,\dots,p}. \lb{3.3.7A}
\end{equation}
We now introduce $\{ \hat \mu_j \}_{j=1,\dots,p}\subset \calK_p$ and
$\{ \hat \nu_j \}_{j=1,\dots,p}\subset \calK_p$ by
\begin{equation} \label{ALhmu}
\hat \mu_j(n)=(\mu_j(n), (2/c_{0,+})\mu_j(n)^{p_-} G_{\ul p}(\mu_j(n),n)), \quad j=1, \dots, p, 
\; n\in\bbZ,   
\end{equation}
and 
\begin{equation}  \label{ALhnu}
\hat \nu_j(n)=(\nu_j(n), - (2/c_{0,+})\nu_j(n)^{p_-} G_{\ul p}(\nu_j(n),n)), \quad j=1, \dots, p, 
\; n\in\bbZ.
\end{equation}

We also introduce the points $P_{0,\pm}$ by 
\begin{equation}
    \Pzpm=(0,\pm (c_{0,-}/c_{0,+}))\in\calK_p, \quad 
    \f{c_{0,-}^2}{c_{0,+}^2} = \prod_{m=0}^{2p+1} E_m.   \lb{AL3.10}
\end{equation}
We emphasize that $\Pzpm$ and $\Pinfpm$ are not necessarily on the same
sheet of $\calK_p$.  

Next we introduce the fundamental meromorphic function
 on $\calK_p$ by
\begin{align} 
\phi(P,n) &= \frac{(c_{0,+}/2)z^{-p_-} y + G_{\ul p}(z,n)}{F_{\ul p}(z,n)}  
\label{ALphi} \\
&= \frac{-H_{\ul p}(z,n)}{(c_{0,+}/2)z^{-p_-} y - G_{\ul p}(z,n)},   
\label{ALphi1}  \\
& \hspace*{.9cm}  P=(z,y)\in \calK_p, \; n\in \bbZ,   \no 
\end{align}
with divisor  $(\phi(\dott,n))$ of $\phi(\dott,n)$ given by
\begin{equation} \label{AL(phi)}
(\phi(\dott,n)) = \calD_{P_{0,-} \hunu(n)} - \calD_{\Pinfm \humu(n)},   
\end{equation}
using \eqref{ALmu(n)} and \eqref{ALnu(n)}. Here we abbreviated 
\begin{equation}
\humu = \{\hat \mu_1, \dots, \hat \mu_p\}, \, 
\hunu = \{\hat \nu_1, \dots, \hat \nu_p\} \in\symq.   \lb{ALdiv} 
\end{equation}
Given $\phi(\dott,n)$, the meromorphic stationary Baker--Akhiezer vector 
$\Psi(\dott,n,n_0)$ on $\calK_p$ is then defined by
\begin{align} \no
\Psi(P,n,n_0) &= \binom{\psi_1(P,n,n_0)}{\psi_2(P,n,n_0)}, \\  \label{ALpsi1}
\psi_1(P,n,n_0) &= \begin{cases}      
\prod_{n'=n_0 + 1}^n \big(z + \alpha(n') \phi^-(P,n')\big), & n \geq n_0 +1, \\
1,                      &  n=n_0, \\
\prod_{n'=n + 1}^{n_0} \big(z + \alpha(n') \phi^-(P,n')\big)^{-1}, & n \leq n_0 -1,
\end{cases}   \\
\psi_2(P,n,n_0) &= \phi(P,n_0)
\begin{cases}      
\prod_{n'=n_0 + 1}^n \big(z \beta(n') \phi^-(P,n')^{-1} + 1\big), & n \geq n_0 +1, \\
1,                      &  n=n_0, \\
\prod_{n'=n + 1}^{n_0} \big(z \beta(n') \phi^-(P,n')^{-1} + 1\big)^{-1}, & n \leq n_0 -1.
\end{cases}         \label{ALpsi2}
\end{align}
Basic properties of $\phi$ and $\Psi$ are summarized in the following result.

\begin{lemma} [\cite{GesztesyHoldenMichorTeschl:2007}]  \lb{lAL3.1}
Suppose that $\alpha, \beta$ satisfy \eqref{ALneq 0,1} and the $\ul p$th stationary 
Ablowitz--Ladik system \eqref{ALstat}. Moreover, assume \eqref{ALcalK_p} and 
\eqref{ALEneqE} and let
$P=(z,y) \in \calK_p\setminus \{\Pinfp, \Pinfm,\Pzp,\Pzm\}$, $(n, n_0) \in \bbZ^2$.
Then $\phi$ satisfies the Riccati-type equation
\begin{align} \label{ALriccati} 
& \alpha \phi(P)\phi^-(P) - \phi^-(P) + z \phi(P) = z \beta,   \\
\intertext{as well as}
  \label{ALphi 1}
& \phi(P) \phi(P^*) = \frac{H_{\ul p}(z)}{F_{\ul p}(z)},\\ \label{ALphi 2}
& \phi(P) + \phi(P^*) = 2\frac{G_{\ul p}(z)}{F_{\ul p}(z)},\\ \label{ALphi 3}
& \phi(P) - \phi(P^*) = c_{0,+}z^{-p_-} \frac{y(P)}{F_{\ul p}(z)}.
\end{align}
The vector $\Psi$ satisfies
\begin{align} 
& U(z) \Psi^-(P)=\Psi(P),  \label{ALpsi 2} \\ 
\label{ALpsi 3}
& V_{\ul p}(z)\Psi^-(P)= - (i/2)c_{0,+} z^{-p_-} y \Psi^-(P), \\ 
& \psi_2(P,n,n_0) = \phi(P,n) \psi_1(P,n,n_0),   \label{ALpsi 1} \\ \label{ALpsi 4}
& \psi_1(P,n,n_0) \psi_1(P^*,n,n_0) = z^{n-n_0} \frac{F_{\ul p}(z,n)}{F_{\ul p}(z,n_0)} 
\pgam(n,n_0),
\\ \label{ALpsi 5}
& \psi_2(P,n,n_0) \psi_2(P^*,n,n_0) = z^{n-n_0} \frac{H_{\ul p}(z,n)}{F_{\ul p}(z,n_0)} 
\pgam(n,n_0),\\
& \psi_1(P,n,n_0) \psi_2(P^*,n,n_0) +\psi_1(P^*,n,n_0) \psi_2(P,n,n_0) \label{ALpsi 6} \\
& \quad =2 z^{n-n_0} \frac{G_{\ul p}(z,n)}{F_{\ul p}(z,n_0)} 
\pgam(n,n_0),\no \\
& \psi_1(P,n,n_0) \psi_2(P^*,n,n_0) -\psi_1(P^*,n,n_0) \psi_2(P,n,n_0) \label{ALpsi 7} \\
& \quad =-c_{0,+} z^{n-n_0-p_-} \frac{y}{F_{\ul p}(z,n_0)}  \pgam(n,n_0), \no
\end{align}
where we used the abbreviation 
\begin{equation} \lb{ALpgam}
\pgam(n,n_0) = \begin{cases}      
\prod_{n'=n_0 + 1}^n \gamma(n'), & n \geq n_0 +1, \\
1,                      &  n=n_0, \\
\prod_{n'=n + 1}^{n_0} \gamma(n')^{-1},  & n \leq n_0 -1.
\end{cases}
\end{equation}\end{lemma}

Combining the Laurent polynomial recursion approach of Section \ref{sAL2} with 
\eqref{ALmu(n)} and \eqref{ALnu(n)} readily yields trace formulas for $f_{\ell,\pm}$ and $h_{\ell,\pm}$ in terms of symmetric functions of the zeros $\mu_j$ and 
$\nu_k$ of $(\dott)^{p_-}F_{\ul p}$ and $(\dott)^{p_- -1}H_{\ul p}$, respectively. For simplicity we just record the simplest cases.

\begin{lemma}  [\cite{GesztesyHoldenMichorTeschl:2007}]  \lb{lAL3.2}
Suppose that $\alpha, \beta$ satisfy \eqref{ALneq 0,1} and the $\ul p$th stationary  
Ablowitz--Ladik system \eqref{ALstat}. Then, 
\begin{align} 
 \frac{\alpha}{\alpha^+}&= 
\prod_{j=1}^p\mu_j \bigg(\prod_{m=0}^{2p+1}E_m\bigg)^{-1/2},   \label{ALtr1} \\ 
\frac{\beta^+}{\beta}&= 
\prod_{j=1}^p\nu_j \bigg(\prod_{m=0}^{2p+1}E_m\bigg)^{-1/2},   \label{ALtr2} \\  
\sum_{j=1}^p\mu_j &= \alpha^+ \beta
- \gamma^+ \frac{\alpha^{++}}{\alpha^+} 
- \frac{c_{1,+}}{c_{0,+}},   \label{ALtr3} \\
\sum_{j=1}^p\nu_j &= \alpha^+ \beta
- \gamma \frac{\beta^-}{\beta} 
- \frac{c_{1,+}}{c_{0,+}}.  \label{ALtr4}
\end{align}
\end{lemma}

Next we turn to asymptotic properties of $\phi$ and $\Psi$ in a neighborhood of 
$\Pinfpm$ and $\Pzpm$.

\begin{lemma} [\cite{GesztesyHoldenMichorTeschl:2007}]  \label{lAL3.3}
Suppose that $\alpha, \beta$ satisfy \eqref{ALneq 0,1} and the $\ul p$th stationary 
Ablowitz--Ladik system \eqref{ALstat}.  Moreover, let
$P=(z,y)\in\calK_p\setminus\{\Pinfp,\Pinfm,\Pzp,\Pzm\}$, $(n,n_0)\in\bbZ^2$. 
Then $\phi$ has the asymptotic behavior 
\begin{align}  
\phi(P) \underset{\zeta\to 0}{=}&  \begin{cases} 
\beta + \beta^-\gamma \zeta + \Oh(\zeta^2), & \quad  P \rightarrow P_{\infty_+}, \\
- (\alpha^+)^{-1} \zeta^{-1} + (\alpha^+)^{-2}\alpha^{++}\gamma^+
+ \Oh(\zeta), & \quad  P \rightarrow P_{\infty_-}, 
\end{cases}
\quad \zeta=1/z,  \label{ALphi infty} \\ 
\phi(P) \underset{\zeta\to 0}{=}&  \begin{cases} 
\alpha^{-1} - \alpha^{-2} \alpha^-\gamma \zeta  + \Oh(\zeta^2), 
& \quad P \rightarrow P_{0,+}, \\  
- \beta^+ \zeta - \beta^{++}\gamma^+ \zeta^2 + \Oh(\zeta^3), & \quad P \rightarrow P_{0,-},
\end{cases}
\quad \zeta=z.   \label{ALphi zero}
\end{align}
The components of the Baker--Akhiezer vector $\Psi$ have the asymptotic behavior 
\begin{align}
\psi_1(P,n,n_0) \underset{\zeta\to 0}{=}& \begin{cases} 
\zeta^{n_0-n}(1 + \Oh(\zeta)), & P \rightarrow P_{\infty_+}, \\ 
\frac{\alpha^+(n)}{\alpha^+(n_0)}
\pgam(n,n_0) + \Oh(\zeta),
&P \rightarrow P_{\infty_-},  
\end{cases}  \quad  \zeta=1/z,   \lb{ALpsi_1 infty}  \\ 
\psi_1(P,n,n_0) \underset{\zeta\to 0}{=}& \begin{cases} 
\frac{\alpha(n)}{\alpha(n_0)} + \Oh(\zeta), & P \rightarrow P_{0,+}, \\  
 \zeta^{n-n_0} \pgam(n,n_0)(1 + \Oh(\zeta)),
&P \rightarrow P_{0,-}, 
\end{cases} \quad   \zeta=z,  \label{ALpsi_1 zero}  \\ 
\psi_2(P,n,n_0) \underset{\zeta\to 0}{=}& \begin{cases} 
\beta(n) \zeta^{n_0-n}(1 + \Oh(\zeta)),
&P \rightarrow P_{\infty_+}, \\
- \frac{1}{\alpha^+(n_0)}
\pgam(n,n_0) \zeta^{-1} (1 + \Oh(\zeta)), 
& P \rightarrow P_{\infty_-}, 
\end{cases} \quad   \zeta=1/z,   \label{ALpsi_2 infty} \\ 
\psi_2(P,n,n_0) \underset{\zeta\to 0}{=}& \begin{cases} 
\frac{1}{\alpha(n_0)} + \Oh(\zeta),
& P \rightarrow P_{0,+},   \\
- \beta^+(n)
\pgam(n,n_0) \zeta^{n+1-n_0}(1 + \Oh(\zeta)), 
& P \rightarrow P_{0,-}, 
\end{cases} \quad   \zeta=z.   \lb{ALpsi_2 zero} 
\end{align}
The divisors $(\psi_j)$ of $\psi_j$, $j=1,2$, are given by
\begin{align} \label{ALpsi1aa}
(\psi_1(\dott,n,n_0)) &= \calD_{\humu(n)} - \calD_{\humu(n_0)} 
+ (n-n_0)(\calD_{P_{0,-}} - \calD_{\Pinfp}), \\ 
(\psi_2(\dott,n,n_0)) &= \calD_{\hunu(n)} - \calD_{\humu(n_0)} 
+ (n-n_0)(\calD_{P_{0,-}} - \calD_{\Pinfp}) 
+ \calD_{P_{0,-}} - \calD_{\Pinfm}.   \label{ALpsi2aa}
\end{align}
\end{lemma}

Since nonspecial divisors play a fundamental role in this section and the
next, we now take a closer look at them.

\begin{lemma} [\cite{GesztesyHoldenMichorTeschl:2007}] \label{lAL3.4} 
Suppose that $\alpha, \beta$ satisfy \eqref{ALneq 0,1} and the
$\ul p$th stationary Ablowitz--Ladik system \eqref{ALstat}. Moreover, assume 
\eqref{ALcalK_p} and \eqref{ALEneqE} and let $n\in\bbZ$. Let $\calD_{\humu}$,
$\humu=\{\hmu_1,\dots,\hmu_p\}$, and $\calD_{\hunu}$,
$\hunu=\{\hunu_1,\dots,\hunu_p\}$, be the pole and zero divisors of degree
$p$, respectively, associated with $\alpha$, $\beta$, and $\phi$ defined
according to \eqref{ALhmu} and \eqref{ALhnu}, that is,
\begin{align}
\begin{split}
\hat\mu_j (n) &= (\mu_j (n), (2/c_{0,+}) \mu_j(n)^{p_-} G_{\ul p}(\mu_j(n),n)), 
\quad j=1,\dots,p,   \\
\hat\nu_j (n) &= (\nu_j (n),- (2/c_{0,+}) \nu_j(n)^{p_-} G_{\ul p}(\nu_j(n),n)), 
\quad j=1,\dots,p.
\end{split}
\end{align}
Then $\calD_{\humu(n)}$ and $\calD_{\hunu(n)}$ are nonspecial for all
$n\in\bbZ$.
\end{lemma}

Next, we shall provide an explicit representation of $\phi$,
$\Psi$, $\alpha$, and $\beta$ in terms of the Riemann
theta function associated with $\calK_p$. We freely employ
the notation established in 
Appendix \ref{sAL.A}. (We recall our tacit assumption $p\in\bbN$ to avoid the trivial case $p=0$.)

Let $\theta$ denote the Riemann theta function associated with $\calK_p$ 
and introduce a fixed homology basis $\{a_j,b_j\}_{j=1,\dots,p}$ on 
$\calK_p$. Choosing as a convenient fixed base point one of the branch points, $Q_0=(E_{m_0},0)$, the Abel
maps $\ua_{Q_0}$ and $\ual_{Q_0}$ are  defined by \eqref{aa46} 
and \eqref{aa47} and the Riemann vector $\underline{\Xi}_{Q_0}$ is given by
\eqref{aa55}.  Let $\omega_{P_+, P_-}^{(3)}$ be the normal differential of the third kind holomorphic on $\calK_p\setminus\{P_+,P_-\}$ 
with simple poles at $P_{\pm}$ and residues $\pm 1$, respectively. 
In particular, one obtains for $\omega_{P_{0,-}, \Pinfpm}^{(3)}$,  
\begin{equation} \label{AL3.57a}
\omega_{P_{0,-}, \Pinfpm}^{(3)} = \bigg(\frac{y+y_{0,-}}{z} \mp \prod_{j=1}^p
(z - \lambda_{\pm, j})\bigg) \frac{dz}{2y}, \quad P_{0,-}=(0, y_{0,-}),
\end{equation}
where the constants $\{\lambda_{\pm,j}\}_{j=1}^p\subset\bbC$ are uniquely determined by employing the normalization
\begin{equation} 
\int_{a_{j}}\omega_{P_{0,-}, \Pinfpm}^{(3)}=0, \quad j=1,\dots,p.  
\end{equation}
The explicit formula \eqref{AL3.57a} then implies the following
asymptotic expansions (using the local coordinate $\zeta=z$ near $\Pzpm$
and $\zeta=1/z$ near $\Pinfpm$),
\begin{align}
\int_{Q_0}^P \omega_{\Pzm, \Pinfm}^{(3)}&\underset{\zeta\to 0}{=}
\left\{\begin{matrix} 0 \\\ln(\zeta) \end{matrix} \right\}
+\omega_0^{0,\pm}(\Pzm,\Pinfm) + \Oh(\zeta)
\text{  as $P\to \Pzpm$}, \label{AL3.57c} \\
\int_{Q_0}^P \omega_{\Pzm, \Pinfm}^{(3)}&\underset{\zeta\to 0}{=}
\left\{\begin{matrix} 0 \\-\ln(\zeta) \end{matrix} \right\}
+\omega^{\infty_\pm}_0(\Pzm,\Pinfm) + \Oh(\zeta)
\text{  as $P\to \Pinfpm$}, \label{AL3.57d} \\
\int_{Q_0}^P \omega_{\Pzm, \Pinfp}^{(3)}&\underset{\zeta\to 0}{=}
\left\{\begin{matrix} 0 \\\ln(\zeta) \end{matrix} \right\}
+\omega_0^{0,\pm}(\Pzm,\Pinfp) + \Oh(\zeta)
\text{  as $P\to \Pzpm$}, \label{AL3.57e} \\
\int_{Q_0}^P \omega_{\Pzm, \Pinfp}^{(3)}&\underset{\zeta\to 0}{=}
\left\{\begin{matrix} -\ln(\zeta) \\ 0\end{matrix} \right\}
+\omega^{\infty_\pm}_0(\Pzm,\Pinfp) + \Oh(\zeta)
\text{  as $P\to \Pinfpm$}. \label{AL3.57f}
\end{align}

\begin{lemma} \label{lAL3.5}
With $\omega_0^{\infty_{\sigma}}(\Pzm,\Pinfpm)$ and
$\omega_0^{0,\sigma'}(\Pzm,\Pinfpm)$, $\sigma, \sigma' \in\{+,-\}$, defined as
in \eqref{AL3.57c}--\eqref{AL3.57f} one has   
\begin{align} 
\begin{split} 
& \exp\big(\omega_0^{0,-}(\Pzm,\Pinfpm)
-\omega_0^{\infty_+}(\Pzm,\Pinfpm) \\
&\qquad -\omega_0^{\infty_-}(\Pzm,\Pinfpm)
+\omega_0^{0,+}(\Pzm,\Pinfpm)\big)=1. \label{AL3.57g}
\end{split}
\end{align}
\end{lemma}
\begin{proof}
Pick $Q_{1,\pm}=(z_1,\pm y_1)\in\calK_p\setminus\{\Pinfpm\}$ in a
neighborhood of $\Pinfpm$ and $Q_{2,\pm}=(z_2,\pm
y_2)\in\calK_p\setminus\{\Pzpm\}$ in a neighborhood of $\Pzpm$.
Without loss of generality one may assume that $\Pinfp$ and $\Pzp$ lie
on the same sheet. Then by \eqref{AL3.57a},
\begin{align} \no
&\int_{Q_0}^{Q_{2,-}} \omega^{(3)}_{\Pzm,\Pinfm}
-\int_{Q_0}^{Q_{1,+}} \omega^{(3)}_{\Pzm,\Pinfm}
-\int_{Q_0}^{Q_{1,-}} \omega^{(3)}_{\Pzm,\Pinfm}
+\int_{Q_0}^{Q_{2,+}} \omega^{(3)}_{\Pzm,\Pinfm} \\
& \quad = \int_{Q_0}^{Q_{2,+}} \frac{dz}{z}
     -\int_{Q_0}^{Q_{1,+}} \frac{dz}{z} =\ln(z_2)-\ln(z_1)+2\pi ik, \label{AL3.57h}
\end{align}
for some $k\in\bbZ$. On the other hand, by
\eqref{AL3.57c}--\eqref{AL3.57f} one obtains
\begin{align} \no
&\int_{Q_0}^{Q_{2,-}} \omega^{(3)}_{\Pzm,\Pinfm}
-\int_{Q_0}^{Q_{1,+}} \omega^{(3)}_{\Pzm,\Pinfm}
-\int_{Q_0}^{Q_{1,-}} \omega^{(3)}_{\Pzm,\Pinfm}
+\int_{Q_0}^{Q_{2,+}} \omega^{(3)}_{\Pzm,\Pinfm} \\ \no
& \quad = \ln(z_2)+\ln(1/z_1)+\omega_0^{0,-}(\Pzm,\Pinfm)
-\omega_0^{\infty_+}(\Pzm,\Pinfm)\\
& \qquad
-\omega_0^{\infty_-}(\Pzm,\Pinfm)  +\omega_0^{0,+}(\Pzm,\Pinfm) + \Oh(z_2)+ \Oh(1/z_1), \label{AL3.57i}
\end{align}
and hence the part of \eqref{AL3.57g} concerning
$\omega^{(3)}_{\Pzm,\Pinfm}$ follows. The corresponding result for
$\omega^{(3)}_{\Pzm,\Pinfp}$ is proved analogously.
\end{proof}

In the following it will be convenient to use the abbreviation
\begin{align}
\begin{split}
   &\uz(P,\ul Q)=\uxi_{Q_0}-\ul A_{Q_0}(P)
+{\ual}_{Q_0}(\calD_{\ul Q}), \lb{3.3.35} \\
&P\in\calK_p, \; \ul Q=\{Q_1,\dots,Q_p\}\in\symq. 
\end{split}
\end{align}
We note that 
$\ul{z}(\dott,\ul{Q})$ is independent of the choice of base point $Q_0$.

For later use we state the following result.
\begin{lemma} \label{lAL3.6}
The following relations hold:
\begin{align}
\uz(\Pinfp,\humu^+)&=\uz(\Pinfm,\hunu)=\uz(\Pzm,\humu)=\uz(\Pzp,\hunu^+),
\label{AL3.72} \\
\uz(\Pinfp,\hunu^+)&=\uz(\Pzm,\hunu), \quad
\uz(\Pzp,\humu^+)=\uz(\Pinfm,\humu). \label{AL3.74}
\end{align}
\end{lemma}
\begin{proof}
We indicate the proof of some of the relations to be used in
\eqref{AL3.69a} and \eqref{AL3.69d}. Let 
$\hat{\underline\lambda}$ denote either $\humu$ or $\hunu$. Then,
\begin{align}
\uz(\Pzp,\hat{\underline\lambda}^+)&=\uxi_{Q_{0}}-\Amap(\Pzp)
+\amap(\calD_{\hat{\underline\lambda}^+}) \no\\
&=\uxi_{Q_{0}}-\Amap(\Pzp)
+\amap(\calD_{\hat{\underline\lambda}})+\underline{A}_{\Pzm}(\Pinfp) \no \\
&=\uxi_{Q_{0}}
-\Amap(\Pinfm)+\amap(\calD_{\hat{\underline\lambda}}) \no \\
&= \uz(\Pinfm,\hat{\underline\lambda}),   \label{AL3.74AA} \\
\uz(\Pinfp,\hat{\underline\lambda}^+)&=\uxi_{Q_{0}}-\Amap(\Pinfp)
+\amap(\calD_{\hat{\underline\lambda}^+}) \no\\
&=\uxi_{Q_{0}}-\Amap(\Pinfp)
+\amap(\calD_{\hat{\underline\lambda}})+\underline{A}_{\Pzm}(\Pinfp) \no \\
&=\uxi_{Q_{0}}
-\Amap(\Pzm)+\amap(\calD_{\hat{\underline\lambda}}) \no \\
&= \uz(\Pzm,\hat{\underline\lambda}), \, \text{ etc.}   \label{AL3.74BB}
\end{align}
Here we used ${\ul A}_{Q_0}(P^*)=-{\ul A}_{Q_0}(P)$, $P\in\calK_p$, since $Q_0$ is a branch point of $\calK_p$, and 
$\amap(\calD_{\hat{\underline\lambda}^+})=
\amap(\calD_{\hat{\underline\lambda}})+\underline{A}_{\Pzm}(\Pinfp)$. The latter equality immediately follows from \eqref{ALpsi1aa} in the case 
$\hat{\underline\lambda}=\humu$ and from combining \eqref{AL(phi)} and 
\eqref{ALpsi2aa} in the case $\hat{\underline\lambda}=\hunu$. 
\end{proof}

Given these preparations, the theta function representations of $\phi$, $\psi_1$, 
$\psi_2$, $\alpha$, and $\beta$ then read as follows.

\begin{theorem}\label{tAL3.7}
Suppose that $\alpha, \beta$ satisfy \eqref{ALneq 0,1} and the 
$\ul p$th stationary Ablowitz--Ladik system \eqref{ALstat}. Moreover, assume 
hypothesis \eqref{ALcalK_p} and \eqref{ALEneqE}, and let 
$P\in\calK_p\setminus\{\Pinfp,\Pinfm,\Pzp,\Pzm\}$ and
$(n,n_{0})\in \bbZ^2$. Then for each $n\in \bbZ$, $\calD_{\humu(n)}$
and $\calD_{\hunu(n)}$ are nonspecial. Moreover,
\begin{align}
\phi(P,n)&= C(n) \frac{\theta(\uz(P,\hunu(n)))}{\theta(\uz(P,\humu(n)))}
\exp\bigg(\int_{Q_0}^P \omega^{(3)}_{\Pzm, \Pinfm} \bigg), \label{AL3.59} \\
\psi_{1}(P,n,n_{0})&=C(n,n_{0})\frac{\theta(\uz(P,\humu(n)))}
{\theta(\uz(P,\humu(n_{0})))} \exp\bigg((n-n_{0})
\int_{Q_{0}}^{P} \omega^{(3)}_{\Pzm,\Pinfp} \bigg), \label{AL3.61} \\
\psi_{2}(P,n,n_{0})& =C(n) C(n,n_{0}) \frac{\theta(\uz(P,\hunu(n)))}{\theta(\uz(P,\humu(n_0)))} \no \\
& \quad\times
\exp\bigg(\int_{Q_{0}}^{P}
\omega^{(3)}_{\Pzm,\Pinfm}+(n-n_{0}) \int_{Q_{0}}^{P}
\omega^{(3)}_{\Pzm,\Pinfp} \bigg),  \label{AL3.61b}
\end{align}
where
\begin{align}
& C(n)=(-1)^{n-n_0}\exp\big((n-n_0)(\omega_0^{0,-}(\Pzm,\Pinfm)
-\omega_0^{\infty_+}(\Pzm,\Pinfm))\big) \no \\
& \qquad \quad \, \times
\frac{1}{\alpha(n_0)}\exp\big(-\omega_0^{0,+}(\Pzm,\Pinfm)\big)
\frac{\theta(\uz(\Pzp,\humu(n_0)))}{\theta(\uz(\Pzp,\hunu(n_0)))},
\label{AL3.61B} \\
&C(n,n_{0})= \exp\big(-(n-n_{0})\omega_0^{\infty_+}(\Pzm,\Pinfp)\big)
\frac{\theta(\uz(\Pinfp,\humu(n_{0})))}
{\theta(\uz(\Pinfp,\humu(n)))}. \label{AL3.62}
\end{align}
The Abel map linearizes the auxiliary divisors $\calD_{\humu(n)}$
and $\calD_{\hunu(n)}$ in the sense that
\begin{align}
\amap(\calD_{\humu(n)}) &=\amap(\calD_{\humu(n_{0})})
+ (n-n_{0}) \underline{A}_{\Pzm}(\Pinfp), \label{AL3.63} \\
\amap(\calD_{\hunu(n)})&=\amap(\calD_{\hunu(n_{0})})
+ (n-n_{0}) \underline{A}_{\Pzm}(\Pinfp), \label{AL3.64}
\end{align}
in addition,
\begin{align}  
\amap(\calD_{\hunu(n)}) &= \amap(\calD_{\humu(n)}) - 
\Amap(P_{0,-}) + \Amap(\Pinfm)  \no \\
&= \amap(\calD_{\humu(n)}) + \ul{A}_{P_{0,-}}(\Pinfm).   \lb{AL3.64A}
\end{align}
Finally, $\alpha,\beta$ are of the form
\begin{align}  
&\alpha(n)=
\alpha(n_0)(-1)^{n-n_0}\exp\big(-(n-n_0)(\omega_0^{0,-}(\Pzm,\Pinfm)
-\omega_0^{\infty_+}(\Pzm,\Pinfm))\big) \no \\ & \hspace*{1.1cm} \times
\frac{\theta(\uz(\Pzp,\hunu(n_0)))\theta(\uz(\Pzp,\humu(n)))}
{\theta(\uz(\Pzp,\humu(n_0)))\theta(\uz(\Pzp,\hunu(n)))}, \label{AL3.66} \\
&\beta(n)= \beta(n_0) (-1)^{n-n_0}
\exp\big((n-n_0)(\omega_0^{0,-}(\Pzm,\Pinfm)
-\omega_0^{\infty_+}(\Pzm,\Pinfm))\big) \no \\
& \hspace*{1.1cm} \times
\frac{\theta(\uz(\Pinfp,\humu(n_0)))\theta(\uz(\Pinfp,\hunu(n)))}
{\theta(\uz(\Pinfp,\hunu(n_0)))\theta(\uz(\Pinfp,\humu(n)))}, \label{AL3.68}
\\
& \alpha(n)\beta(n)=\exp\big(\omega_0^{\infty_+}(\Pzm,\Pinfm)
-\omega_0^{0,+}(\Pzm,\Pinfm)\big) \no \\
& \hspace*{1.8cm} \times
\frac{\theta(\uz(\Pzp,\humu(n)))\theta(\uz(\Pinfp,\hunu(n)))}
{\theta(\uz(\Pzp,\hunu(n)))\theta(\uz(\Pinfp,\humu(n)))}, \label{AL3.67}
\\  
&\pgam(n,n_0)= \exp\big((n-n_0)(\omega_0^{0,-}(\Pzm,\Pinfp)
-\omega_0^{\infty_+}(\Pzm,\Pinfp))\big)  \no \\
& \hspace*{1.8cm} \times
\frac{\theta(\uz(\Pzm,\humu(n)))\theta(\uz(\Pinfp,\humu(n_{0})))}
{\theta(\uz(\Pzm,\humu(n_0)))\theta(\uz(\Pinfp,\humu(n)))}. \label{AL3.74D}
\end{align}
\end{theorem}
\begin{proof}
Applying Abel's theorem (cf.\ Theorem \ref{thm1}, \eqref{a43}) to \eqref{AL(phi)} 
proves \eqref{AL3.64A}, and applying it to \eqref{ALpsi1aa}, \eqref{ALpsi2aa} results in 
\eqref{AL3.63} and \eqref{AL3.64}. 
By Lemma \ref{lAL3.4}, $\calD_{\humu}$ and
$\calD_{\hunu}$ are nonspecial. By equation \eqref{AL(phi)} and Theorem \ref{taa17a},
$\phi(P,n)\exp\big(-\int_{Q_0}^P \omega^{(3)}_{\Pzm, \Pinfm} \big)$ must
be of the type
\begin{equation}
                 \phi(P,n)\exp\bigg(-\int_{Q_{0}}^{P}
                 \omega^{(3)}_{\Pzm,\Pinfm} \bigg)= C(n)
\frac{\theta(\uz(P,\hunu(n)))}{\theta(\uz(P,\humu(n)))} \label{AL3.69}
\end{equation}
for some constant $C(n)$. A comparison of \eqref{AL3.69} and the
asymptotic relations \eqref{ALphi infty} then yields, with the
help of \eqref{AL3.57c}, \eqref{AL3.57d} and \eqref{AL3.72}, \eqref{AL3.74}, 
the following expressions for $\alpha$ and $\beta$:
\begin{align}
(\alpha^+)^{-1}&= C^+ e^{\omega_0^{0,+}(\Pzm,\Pinfm)}
\frac{\theta(\uz(\Pzp,\hunu^+))}{\theta(\uz(\Pzp,\humu^+))} \no \\
&=C^+ e^{\omega_0^{0,+}(\Pzm,\Pinfm)}
\frac{\theta(\uz(\Pinfm,\hunu))}{\theta(\uz(\Pinfm,\humu))} \no \\
&=-C e^{\omega_0^{\infty_-}(\Pzm,\Pinfm)}
\frac{\theta(\uz(\Pinfm,\hunu))}{\theta(\uz(\Pinfm,\humu))}. \label{AL3.69a}
\end{align}
Similarly one obtains
\begin{align}
\beta^+&= C^+ e^{\omega_0^{\infty_+}(\Pzm,\Pinfm)}
\frac{\theta(\uz(\Pinfp,\hunu^+))}{\theta(\uz(\Pinfp,\humu^+))} \no \\
&=C^+ e^{\omega_0^{\infty_+}(\Pzm,\Pinfm)}
\frac{\theta(\uz(\Pzm,\hunu))}{\theta(\uz(\Pzm,\humu))} \no \\
&=-C e^{\omega_0^{0,-}(\Pzm,\Pinfm)}
\frac{\theta(\uz(\Pzm,\hunu))}{\theta(\uz(\Pzm,\humu))}.  \label{AL3.69d}
\end{align}
Here we used \eqref{AL3.63} and \eqref{AL3.64}, more precisely,
\begin{equation}
\amap(\calD_{\humu^+})=\amap(\calD_{\humu})
+\underline{A}_{\Pzm}(\Pinfp), \quad
\amap(\calD_{\hunu^+})=\amap(\calD_{\hunu})
+\underline{A}_{\Pzm}(\Pinfp).   
\end{equation}
Thus, one concludes
\begin{equation}
C(n+1)=-\exp\big[\omega_0^{0,-}(\Pzm,\Pinfm)
-\omega_0^{\infty_+}(\Pzm,\Pinfm)\big]C(n), \quad n\in\bbZ,  \label{AL3.69e}
\end{equation}
and
\begin{equation}
C(n+1)=-\exp\big[\omega_0^{\infty_-}(\Pzm,\Pinfm)
-\omega_0^{0,+}(\Pzm,\Pinfm)\big]C(n), \quad n\in\bbZ,  \label{AL3.69f}
\end{equation}
which is consistent with \eqref{AL3.57g}.
The first-order difference equation \eqref{AL3.69e} then implies
\begin{align}
C(n)&=(-1)^{(n-n_0)}\exp\big[(n-n_0)
(\omega_0^{0,-}(\Pzm,\Pinfm)
-\omega_0^{\infty_+}(\Pzm,\Pinfm))\big]C(n_0), \no \\
& \hspace*{8.5cm}  n,n_0\in\bbZ. \label{AL3.69g}
\end{align}
Thus one infers \eqref{AL3.66} and \eqref{AL3.68}.
Moreover, \eqref{AL3.69g} and taking $n=n_0$ in the first line in
\eqref{AL3.69a} yield \eqref{AL3.61B}. Dividing the first line in
\eqref{AL3.69d} by the first line in \eqref{AL3.69a} then proves
\eqref{AL3.67}.

By \eqref{ALpsi1aa} and Theorem \ref{taa17a}, $\psi_1(P,n,n_{0})$ must
be of the type \eqref{AL3.61}. A comparison of \eqref{ALpsi1},
\eqref{ALphi infty}, and \eqref{AL3.61} as $P\to\Pinfp$ (with local coordinate 
$\zeta=1/z$) then yields
\begin{equation} \label{AL3.74A}
\psi_1(P,n,n_0)\underset{\zeta\to 0}{=}\zeta^{n_0-n}(1+\Oh(\zeta))
\end{equation}
and
\begin{align}
\psi_1(P,n,n_0) & \underset{\zeta\to 0}{=}  C(n,n_0)
\frac{\theta(\uz(\Pinfp,\humu(n)))}{\theta(\uz(\Pinfp,\humu(n_0)))} \no \\
& \quad \;\;\; \times
\exp\big[(n-n_0)\omega_0^{\infty_+}(\Pzm,\Pinfp)\big]
\zeta^{n_0-n}(1+\Oh(\zeta)) \label{AL3.71}
\end{align}
proving \eqref{AL3.62}. Equation \eqref{AL3.61b} is clear from
\eqref{ALpsi 1}, \eqref{AL3.59}, and \eqref{AL3.61}.

Finally, a comparison of \eqref{ALpsi_1 zero} and \eqref{AL3.61} as $P\to\Pzm$
(with local coordinate $\zeta=z$) yields
\begin{align}
\psi_1(P,n,n_0)&\underset{\zeta\to 0}{=}
\pgam(n,n_0)\zeta^{n-n_0}(1+\Oh(\zeta)) \lb{AL3.74B} \\ \no
&\underset{\zeta\to 0}{=} C(n,n_0)
\frac{\theta(\uz(\Pzm,\humu(n)))}{\theta(\uz(\Pzm,\humu(n_0)))}
\exp\big((n-n_0)\omega_0^{0,-}(\Pzm,\Pinfp)\big) \no \\
& \qquad \times \zeta^{n-n_0}(1+\Oh(\zeta)) \lb{AL3.74C}
\end{align}
and hence
\begin{align}
\pgam(n,n_0)&=C(n,n_0)
\frac{\theta(\uz(\Pzm,\humu(n)))}{\theta(\uz(\Pzm,\humu(n_0)))}
\exp\big((n-n_0)\omega_0^{0,-}(\Pzm,\Pinfp)\big) \no \\
&= \exp\big((n-n_0)(\omega_0^{0,-}(\Pzm,\Pinfp)
-\omega_0^{\infty_+}(\Pzm,\Pinfp))\big)  \no \\
& \quad \times
\frac{\theta(\uz(\Pzm,\humu(n)))\theta(\uz(\Pinfp,\humu(n_{0})))}
{\theta(\uz(\Pzm,\humu(n_0)))\theta(\uz(\Pinfp,\humu(n)))}
\end{align}
using \eqref{AL3.62}.
\end{proof}

We note that the apparent $n_0$-dependence of $C(n)$ in the right-hand side of 
\eqref{AL3.61B} actually drops out to ensure the $n_0$-independence of $\phi$ in 
\eqref{AL3.59}.  

The theta function representations \eqref{AL3.66}, \eqref{AL3.68} for $\alpha, \beta$ and that for $\Gamma$ in \eqref{AL3.74D} also show that 
$\gamma(n)\notin \{0,1\}$ for all $n\in\bbZ$, and hence condition \eqref{ALneq 0,1} is 
satisfied for the stationary algebro-geometric AL solutions discussed in 
this section, provided the associated divisors $\calD_{\humu}(n)$ and $\calD_{\hunu}(n)$ stay away from $\Pinfpm, \Pzpm$ for all $n\in\bbZ$. 

The stationary algebro-geometric initial value problem for the Ablowitz--Ladik hierarchy
with complex-valued initial data, that is, the construction of $\alpha$ and 
$\beta$ by starting from a set of initial data (nonspecial divisors) of full measure, will be presented in \cite{GesztesyHoldenMichorTeschl:2007a}.

\section{The Time-Dependent Ablowitz--Ladik Formalism} \label{sALtime}
 
In this section we extend the algebro-geometric analysis of Section 
\ref{sALstat} to the time-dependent Ablowitz--Ladik hierarchy. For proofs of the elementary results of the time-dependent formalism we refer to 
\cite{GesztesyHolden:2005}, \cite{GesztesyHoldenMichorTeschl:2007}.

For most of this section we assume the following hypothesis.

\begin{hypothesis} \label{hAL4.1}
$(i)$ Suppose that $\alpha, \beta$ satisfy
\begin{align}
\begin{split}
& \alpha(\dott,t), \beta(\dott,t)\in \bbC^\bbZ,\; t\in\bbR,
\quad 
\alpha(n,\dott), \, \beta(n,\dott) \in C^1(\bbR), \; n\in\bbZ,  \\
& \alpha(n,t)\beta(n,t)\notin\{0,1\}, \; (n,t)\in\bbZ\times\bbR.   \lb{AL4.1A}
\end{split}
\end{align}
$(ii)$  Assume that the hyperelliptic curve $\calK_p$ satisfies 
\eqref{ALcalK_p} and \eqref{ALEneqE}.
\end{hypothesis}

The basic problem in the analysis of algebro-geometric solutions of the 
Ablowitz--Ladik hierarchy consists of solving the time-dependent $\ul r$th 
Ablowitz--Ladik flow with initial data a stationary solution of the $\ul p$th system in the hierarchy. More precisely, given $\ul p \in \bbN_0^2\setminus\{(0,0\}$ we consider a solution $\alpha^{(0)}$, $\beta^{(0)}$ of the $\ul p$th stationary Ablowitz--Ladik system $\sAL_{\ul p} (\alpha^{(0)}, \beta^{(0)}) = 0$, 
$\ul p =(p_-,p_+)\in\bbN_0^2\setminus\{(0,0\}$, associated with the hyperelliptic 
curve $\calK_p$ and a corresponding set of summation constants 
$\{c_{\ell,\pm}\}_{\ell=1,\dots,p_{\pm}}\subset\bbC$. Next, let 
$\ul r=(r_-,r_+)\in\bbN_0^2$; we intend to construct a solution $\alpha, \beta$ of the Ablowitz--Ladik flow 
$\AL_{\ul r}(\alpha,\beta)=0$ with $\alpha(t_{0,\ul r})=\alpha^{(0)}$, 
$\beta(t_{0,\ul r})=\beta^{(0)}$ for some $t_{0,\ul r}\in\bbR$.  To emphasize that the summation constants in the definitions of the stationary and the time-dependent Ablowitz--Ladik equations are independent of each other, we indicate
this by adding a tilde on all the time-dependent quantities. Hence we shall employ the notation $\ti V_{\ul r}$, $\ti F_{\ul r}$, $\ti G_{\ul r}$, $\ti H_{\ul r}$, 
$\ti K_{\ul r}$, $\tilde f_{s,\pm}$,
$\tilde g_{s,\pm}$, $\tilde h_{s,\pm}$, $\tilde c_{s,\pm}$, in order to distinguish them from $V_{\ul p}$, $F_{\ul p}$, $G_{\ul p}$, $H_{\ul p}$, $K_{\ul p}$, $f_{\ell,\pm}$, $g_{\ell,\pm}$, 
$h_{\ell,\pm}$, $c_{\ell,\pm}$, in the following. In addition, we will follow a more elaborate notation inspired by Hirota's $\tau$-function approach and indicate the individual $\ul r$th 
Ablowitz--Ladik flow by  a separate time variable $t_{\ul r} \in \bbR$. 

Summing up, we are looking for solutions $\alpha, \beta$ of the time-dependent
algebro-geometric initial value problem
\begin{align}
\begin{split}
& \ti \AL_{\ul r} (\alpha, \beta) =
\begin{pmatrix}-i\alpha_{t_{\ul r}} - \alpha(\tilde g_{r_+,+} + \tilde g_{r_-,-}^-) + \tilde f_{r_+ -1,+} - \tilde f_{r_- -1,-}^-\\
-i\beta_{t_{\ul r}}+ \beta(\tilde g_{r_+,+}^- + \tilde g_{r_-,-}) 
- \tilde h_{r_- -1,-} + \tilde h_{r_+ -1,+}^-     \end{pmatrix} = 0,  \lb{ALal_ivp}  \\
& (\alpha, \beta)\big|_{t=t_{0,\ul r}} = \big(\alpha^{(0)}, \beta^{(0)}\big) ,  
\end{split} \\
& \sAL_{\ul p} \big(\alpha^{(0)}, \beta^{(0)}\big) =  \begin{pmatrix} 
-\alpha^{(0)}(g_{p_+,+} + g_{p_-,-}^-) + f_{p_+ -1,+} - f_{p_- -1,-}^-\\
  \beta^{(0)}(g_{p_+,+}^- + g_{p_-,-}) - h_{p_- -1,-} + h_{p_+ -1,+}^-       
\end{pmatrix}=0   \lb{AL4.3A}
\end{align}
for some $t_{0,\ul r}\in\bbR$, where $\alpha=\alpha(n,t_{\ul r})$, 
$\beta=\beta(n,t_{\ul r})$ satisfy \eqref{AL4.1A} and a fixed curve $\calK_p$ is associated with the stationary solutions $\alpha^{(0)}, \beta^{(0)}$ in \eqref{AL4.3A}. 
Here,
\begin{equation}
\ul p =(p_-,p_+) \in \bbN_0^2\setminus\{(0,0)\}, \quad 
\ul r =(r_-,r_+) \in \bbN_0^2, \quad p=p_- + p_+ -1. 
\end{equation}
In terms of the zero-curvature formulation this amounts to solving 
\begin{align} 
U_{t_{\ul r}}(z,t_{\ul r}) + U(z,t_{\ul r}) \ti V_{\ul r}(z,t_{\ul r}) 
- \ti V_{\ul r}^+(z,t_{\ul r}) U(z,t_{\ul r}) &= 0, 
\label{ALzc tilde}  \\
U(z,t_{0,\ul r}) V_{\ul p}(z,t_{0,\ul r}) - V_{\ul p}^+(z,t_{0,\ul r}) 
U(z,t_{0,\ul r}) &= 0. 
\lb{ALzcstat}
\end{align}
One can show (cf.\ \cite{GesztesyHoldenMichorTeschl:2007a}) that the stationary Ablowitz--Ladik system \eqref{ALzcstat} is actually satisfied for all times 
$t_{\ul r}\in\bbR$. Thus, we impose
\begin{align} 
U_{t_{\ul r}}(z,t_{\ul r}) + U(z,t_{\ul r}) \ti V_{\ul r}(z,t_{\ul r}) 
- \ti V_{\ul r}^+(z,t_{\ul r}) U(z,t_{\ul r}) & = 0, 
\label{ALzctilde}  \\ 
U(z,t_{\ul r}) V_{\ul p}(z,t_{\ul r}) - V_{\ul p}^+(z,t_{\ul r}) U(z,t_{\ul r}) & = 0,  \lb{ALzctstat}
\end{align}
instead of \eqref{ALzc tilde} and \eqref{ALzcstat}. 
For further reference, we recall the relevant quantities here 
(cf.\ \eqref{AL2.03}, \eqref{AL_v}, \eqref{ALF_p}--\eqref{ALH_p}, \eqref{ALK=G}):
\begin{align}
\begin{split}
U(z) &= \begin{pmatrix}
z & \alpha     \\
z \beta & 1\\
\end{pmatrix},    \\
V_{\ul p}(z) &= i  \begin{pmatrix}
G_{\ul p}^-(z) & - F_{\ul p}^-(z)     \\[1.5mm]
H_{\ul p}^-(z)  & - G_{\ul p}^-(z)  \\
\end{pmatrix}, \quad 
\ti V_{\ul r}(z) = i  \begin{pmatrix}
\ti G_{\ul r}^-(z)  & - \ti F_{\ul r}^-(z)     \\[1.5mm]
\ti H_{\ul r}^-(z) & - \ti K_{\ul r}^-(z)  \\
\end{pmatrix},   \lb{ALv_osv} 
\end{split}
\end{align}
and 
\begin{align}
F_{\ul p}(z) &= \sum_{\ell=1}^{p_-} f_{p_- -\ell,-} z^{-\ell}  
+ \sum_{\ell=0}^{p_+ -1} f_{p_+ -1-\ell,+} z^\ell 
=- c_{0,+}\alpha^+ z^{-p_-}\prod_{j=1}^p(z-\mu_j),  \no \\ \no
G_{\ul p}(z) &= \sum_{\ell=1}^{p_-} g_{p_- -\ell,-} z^{-\ell} 
+ \sum_{\ell=0}^{p_+} g_{p_+ -\ell,+} z^\ell,  \\ \no
H_{\ul p}(z) &= \sum_{\ell=0}^{p_- -1} h_{p_- -1-\ell,-} z^{-\ell} 
+ \sum_{\ell=1}^{p_+} h_{p_+ -\ell,+} z^\ell 
= c_{0,+}\beta z^{-p_- +1}\prod_{j=1}^p(z-\nu_j), \\
\ti F_{\ul r}(z) &= \sum_{s=1}^{r_-} \tilde f_{r_- -s,-} z^{-s}  
+ \sum_{s=0}^{r_+ -1} \tilde f_{r_+ -1-s,+} z^s,   \lb{AL4.9}  \\ \no
\ti G_{\ul r}(z) &= \sum_{s=1}^{r_-} \tilde g_{r_- -s,-} z^{-s}  
+ \sum_{s=0}^{r_+} \tilde g_{r_+ -s,+} z^s,\\ \no
\ti H_{\ul r}(z) &= \sum_{s=0}^{r_- -1} \tilde h_{r_- -1-s,-} z^{-s}  
+ \sum_{s=1}^{r_+} \tilde h_{r_+ -s,+} z^s, \\
\ti K_{\ul r}(z) &= \sum_{s=0}^{r_-} \tilde g_{r_- -s,-} z^{-s}  
+  \sum_{s=1}^{r_+} \tilde g_{r_+ -s,+} z^s 
= \ti G_{\ul r}(z)+\tilde g_{r_-,-}-\tilde g_{r_+,+}   \no 
\end{align}
for fixed $\ul p\in\bbN_0^2\setminus\{(0,0)\}$, $\ul r\in\bbN_0^2$. Here 
$f_{\ell,\pm}$, $\tilde f_{s,\pm}$, 
$g_{\ell,\pm}$, $\tilde g_{s,\pm}$, $h_{\ell,\pm}$, and $\tilde h_{s,\pm}$ are defined as in \eqref{AL0+}--\eqref{ALh_l-} with appropriate sets of summation constants 
$c_{\ell,\pm}$, $\ell\in\bbN_0$, and $\tilde c_{k,\pm}$, $k\in\bbN_0$. Explicitly, 
\eqref{ALzctilde} and \eqref{ALzctstat} are equivalent to 
(cf.\ \eqref{AL1,1}--\eqref{AL2,1},  \eqref{ALalphat}--\eqref{AL2,2r}),
\begin{align}  \label{ALalpha_t}
\alpha_{t_{\ul r}} &= i \big(z \ti F_{\ul r}^- + \alpha (\ti G_{\ul r} + \ti{K}_{\ul r}^-)
- \ti F_{\ul r}\big),\\ \label{ALbeta_t}
\beta_{t_{\ul r}} &= - i \big(\beta (\ti G_{\ul r}^- + \ti{K}_{\ul r}) - \ti H_{\ul r} 
+ z^{-1} \ti H_{\ul r}^-\big), \\ \label{AL1,1 r}
0 &= z (\ti G_{\ul r}^- - \ti G_{\ul r}) + z\beta \ti F_{\ul r} + \alpha \ti H_{\ul r}^-,\\ \label{AL2,2 r}
0 &= z \beta \ti F_{\ul r}^- + \alpha \ti H_{\ul r} + \ti{K}_{\ul r}^- - \ti{K}_{\ul r}, \\
0 &= z (G_{\ul p}^- - G_{\ul p}) + z \beta F_{\ul p} + \alpha H_{\ul p}^-,  \lb{AL11} \\  
0 &=z \beta F_{\ul p}^- + \alpha H_{\ul p} - G_{\ul p} + G_{\ul p}^-,  \lb{AL12} \\
0 &= - F_{\ul p} + z F_{\ul p}^- + \alpha (G_{\ul p} + G_{\ul p}^-), \lb{AL21}  \\  
0 &= z \beta (G_{\ul p} + G_{\ul p}^-) - z H_{\ul p} + H_{\ul p}^-,  \lb{AL22}
\end{align}
respectively. In particular, \eqref{ALR} holds in the present $t_{\ul r}$-dependent setting, that is,  
\begin{equation} \label{ALR_t}
G_{\ul p}^2 - F_{\ul p} H_{\ul p} = R_{\ul p}.
\end{equation}

As in the stationary context \eqref{ALhmu}, \eqref{ALhnu} we introduce
\begin{align}
\begin{split}
\hat \mu_j(n,t_{\ul r})&=(\mu_j(n,t_{\ul r}), (2/c_{0,+}) 
\mu_j(n,t_{\ul r})^{p_-} G_{\ul p}(\mu_j(n,t_{\ul r}),n,t_{\ul r}))
\in\calK_p, \\ 
& \hspace*{4.45cm} j=1, \dots, p, \; (n,t_{\ul r})\in\bbZ\times\bbR,   
\lb{AL4.20}
\end{split}
\end{align}
and
\begin{align}
\begin{split}
\hat \nu_j(n,t_{\ul r})&=(\nu_j(n,t_{\ul r}), - (2/c_{0,+}) 
\nu_j(n,t_{\ul r})^{p_-} G_{\ul p}(\nu_j(n,t_{\ul r}),n,t_{\ul r}))
\in\calK_p, \\
& \hspace*{4.6cm}  j=1, \dots, p, \; (n,t_{\ul r})\in\bbZ\times\bbR,   \lb{AL4.21}
\end{split}
\end{align}
and note that the regularity assumptions \eqref{AL4.1A} on $\alpha, \beta$ imply
continuity of $\mu_j$ and $\nu_k$ with respect to $t_{\ul r}\in\bbR$ (away from collisions of these zeros, $\mu_j$ and $\nu_k$ are of course $C^\infty$).  

In analogy to \eqref{ALphi}, \eqref{ALphi1}, one defines the following
meromorphic function $\phi (\dott,n,t_{\ul r})$ on $\calK_p$,
\begin{align}
\phi(P,n,t_{\ul r}) & = \frac{(c_{0,+}/2) z^{-p_-} y + G_{\ul p}(z,n,t_{\ul r})}{F_{\ul p}(z,n,t_{\ul r})}    \lb{AL4.22}  \\
& = \frac{-H_{\ul p}(z,n,t_{\ul r})}{(c_{0,+}/2) z^{-p_-} y - 
G_{\ul p}(z,n,t_{\ul r})},    \lb{AL4.23}  \\ 
&  P=(z,y)\in\calK_p, \; (n,t_{\ul r})\in\bbZ\times\bbR,  \no
\end{align}
with divisor $(\phi(\dott,n,t_{\ul r}))$ of $\phi(\dott,n,t_{\ul r})$ given by 
\begin{equation} 
(\phi(\dott,n,t_{\ul r})) = \calD_{P_{0,-} \hunu(n,t_{\ul r})} - \calD_{\Pinfm \humu(n,t_{\ul r})}.  
\end{equation}
The time-dependent Baker--Akhiezer vector is then defined in terms of $\phi$ by 
\begin{align}
& \Psi(P,n,n_0,t_{\ul r},t_{0,\ul r}) = \begin{pmatrix}\psi_1(P,n,n_0,t_{\ul r},t_{0,\ul r})\\
\psi_2(P,n,n_0,t_{\ul r},t_{0,\ul r})\end{pmatrix}, \\ 
& \psi_1(P,n,n_0,t_{\ul r},t_{0,\ul r}) = \exp \bigg(i \int_{t_{0,\ul r}}^{t_{\ul r}} ds 
\big(\ti G_{\ul r} (z,n_0,s) - \ti F_{\ul r}(z,n_0,s) \phi(P,n_0,s)\big)\bigg) \no  \\  
&\quad \times 
\begin{cases}
\prod_{n'=n_0+1}^{n} \big(z + \alpha(n',t_{\ul r}) \phi^-(P,n',t_{\ul r})\big), & n \geq n_0 +1,\\
1, & n=n_0, \\
\prod_{n'=n+1}^{n_0} \big(z + \alpha(n',t_{\ul r}) \phi^-(P,n',t_{\ul r})\big)^{-1}, & n \leq n_0 -1, 
\end{cases} 
\\
& \psi_2(P,n,n_0,t_{\ul r},t_{0,\ul r}) = \exp \bigg(i \int_{t_{0,\ul r}}^{t_{\ul r}} ds 
\big(\ti G_{\ul r} (z,n_0,s) - \ti F_{\ul r}(z,n_0,s)  \phi(P,n_0,s)\big)\bigg) \no \\  
& \;\;  \times \phi(P,n_0,t_{\ul r})
\begin{cases}
\prod_{n'=n_0+1}^{n} \big(z \beta(n',t_{\ul r})\phi^-(P,n',t_{\ul r})^{-1} + 1\big), 
& n \geq n_0 +1,\\
1, & n=n_0, \\
\prod_{n'=n+1}^{n_0}\big(z \beta(n',t_{\ul r})\phi^-(P,n',t_{\ul r})^{-1} + 1\big)^{-1}, 
& n \leq n_0 -1, 
\end{cases}  \\  
&\hspace*{2.3cm} P=(z,y)\in\calK_p\setminus\{\Pinfp,\Pinfm,\Pzp,\Pzm\}, 
\; (n,t_{\ul r})\in\bbZ\times\bbR.   \no
\end{align}
One observes that
\begin{align} 
\begin{split}
& \psi_1(P,n,n_0,t_{\ul r},\tilde t_{\ul r}) = \psi_1(P,n_0,n_0,t_{\ul r},\tilde t_{\ul r})\psi_1(P,n,n_0,t_{\ul r},t_{\ul r}), \label{ALpsi t n_0}  \\
& P=(z,y)\in\calK_p\setminus
\{\Pinfp,\Pinfm,\Pzp,\Pzm\}, \; (n,n_0,t_{\ul r},\tilde t_{\ul r})\in\bbZ^2\times\bbR^2.
\end{split}
\end{align}

The following lemma records basic properties of $\phi$ and $\Psi$ in analogy to the stationary case discussed in Lemma \ref{lAL3.1}. 

\begin{lemma} [\cite{GesztesyHoldenMichorTeschl:2007}]  \lb{lAL4.2}
Assume Hypothesis \ref{hAL4.1} and suppose that 
\eqref{ALzctilde}, \eqref{ALzctstat} hold. In addition, let
$P=(z,y)\in\calK_p\setminus\{\Pinfp, \Pinfm\}$, 
$(n,n_0,t_{\ul r},t_{0,\ul r})\in\bbZ^2\times\bbR^2$.\ Then $\phi$ satisfies 
\begin{align} \label{ALriccati_time} 
& \alpha \phi(P)\phi^-(P) - \phi^-(P) + z \phi(P) = z \beta, \\
& \phi_{t_{\ul r}}(P) =i \ti F_{\ul r} \phi^2(P) - 
i \big(\ti G_{\ul r}(z) + \ti K_{\ul r}(z)\big)\phi(P) +  i \ti H_{\ul r}(z),    \label{ALphi_t}  \\
  \label{ALphi 1t}
& \phi(P) \phi(P^*) = \frac{H_{\ul p}(z)}{F_{\ul p}(z)},\\ \label{ALphi 2t}
& \phi(P) + \phi(P^*) = 2\frac{G_{\ul p}(z)}{F_{\ul p}(z)},\\ \label{ALphi 3t}
& \phi(P) - \phi(P^*) = c_{0,+} z^{-p_-} \frac{y(P)}{F_{\ul p}(z)}.
\end{align}
Moreover, assuming  
$P=(z,y)\in\calK_p\setminus\{\Pinfp, \Pinfm,\Pzp,\Pzm\}$, then 
$\Psi$ satisfies
\begin{align}  \label{ALpsi t 1}
& \psi_2(P,n,n_0,t_{\ul r},t_{0,\ul r})= \phi(P,n,t_{\ul r}) 
\psi_1(P,n,n_0,t_{\ul r},t_{0,\ul r}),\\ 
\label{ALpsi 2t}
& U(z) \Psi^-(P)=\Psi(P),\\ \label{ALpsi 3t}
& V_{\ul p}(z)\Psi^-(P) = - (i/2) c_{0,+} z^{-p_-} y \Psi^-(P), \\
& \Psi_{t_{\ul r}}(P) = \ti V_{\ul r}^+(z) \Psi(P),   
\lb{ALtime_AL} \\
& \psi_1(P,n,n_0,t_{\ul r},t_{0,\ul r}) \psi_1(P^*,n,n_0,t_{\ul r},t_{0,\ul r}) = z^{n-n_0} \frac{F_{\ul p}(z,n,t_{\ul r})}{F_{\ul p}(z,n_0,t_{0,\ul r})}  \pgam(n,n_0,t_{\ul r}), \label{ALpsi 4t}\\ 
& \psi_2(P,n,n_0,t_{\ul r},t_{0,\ul r}) \psi_2(P^*,n,n_0,t_{\ul r},t_{0,\ul r}) 
= z^{n-n_0} \frac{H_{\ul p}(z,n,t_{\ul r})}{F_{\ul p}(z,n_0,t_{0,\ul r})}  \pgam(n,n_0,t_{\ul r}),
\label{ALpsi 5t} \\
& \psi_1(P,n,n_0,t_{\ul r},t_{0,\ul r}) \psi_2(P^*,n,n_0,t_{\ul r},t_{0,\ul r}) 
+\psi_1(P^*,n,n_0,t_{\ul r},t_{0,\ul r}) \psi_2(P,n,n_0,t_{\ul r},t_{0,\ul r}) \no \\
&\quad =2 z^{n-n_0} \frac{G_{\ul p}(z,n,t_{\ul r})}{F_{\ul p}(z,n_0,t_{0,\ul r})}
\pgam(n,n_0,t_{\ul r}), \label{ALpsi 6t} \\
& \psi_1(P,n,n_0,t_{\ul r},t_{0,\ul r}) \psi_2(P^*,n,n_0,t_{\ul r},t_{0,\ul r}) 
-\psi_1(P^*,n,n_0,t_{\ul r},t_{0,\ul r}) \psi_2(P,n,n_0,t_{\ul r},t_{0,\ul r}) \no\\
& \quad =-c_{0,+} z^{n-n_0-p_-} \frac{y}{F_{\ul p}(z,n_0,t_{0,\ul r})} 
\pgam(n,n_0,t_{\ul r}), 
\label{ALpsi 7t}
\end{align}
where  
\begin{equation} \lb{ALpgamt}
\pgam(n,n_0,t_{\ul r}) = \begin{cases}      
\prod_{n'=n_0 + 1}^n \gamma(n',t_{\ul r}), & n \geq n_0 +1, \\
1,                      &  n=n_0, \\
\prod_{n'=n + 1}^{n_0} \gamma(n',t_{\ul r})^{-1},  & n \leq n_0 -1.
\end{cases}
\end{equation}
In addition, as long as the zeros $\mu_j(n_0,s)$ of $(\dott)^{p_-}F_{\ul p} (\dott,n_0,s)$ are all  simple and distinct from zero for $s \in\calI_{\mu}$, 
$\calI_{\mu}\subseteq\bbR$ an open interval, 
$\Psi(\dott,n,n_0,t_{\ul r},t_{0,\ul r})$ is meromorphic on $\calK_p\setminus \{\Pinfp,\Pinfm,$ $\Pzp,\Pzm\}$ for $(n,t_{\ul r},t_{0,\ul r})\in\bbZ\times\calI_{\mu}^2$.
\end{lemma}

Next we consider the $t_{\ul r}$-dependence of $F_{\ul p}$, $G_{\ul p}$, and $H_{\ul p}$.

\begin{lemma} [\cite{GesztesyHoldenMichorTeschl:2007}]  \lb{lAL4.3}
Assume Hypothesis \ref{hAL4.1} and suppose that 
\eqref{ALzctilde}, \eqref{ALzctstat} hold. In addition, let
$(z,n,t_{\ul r})\in\bbC\times\bbZ\times\bbR$. Then, 
\begin{align} \label{ALF_t}
F_{\ul p,t_{\ul r}} &= - 2 i G_{\ul p} \ti F_{\ul r} 
+ i \big(\ti G_{\ul r} + \ti K_{\ul r}\big)F_{\ul p}, \\ \label{ALG_t}
G_{\ul p,t_{\ul r}} &= i F_{\ul p} \ti H_{\ul r} - i H_{\ul p} \ti F_{\ul r}, \\  \label{ALH_t}
H_{\ul p,t_{\ul r}} &= 2 i G_{\ul p} \ti H_{\ul r} - i \big(\ti G_{\ul r} + \ti K_{\ul r}\big)H_{\ul p}.
\end{align}
In particular, \eqref{ALF_t}--\eqref{ALH_t} are equivalent to 
\begin{equation} \label{ALWp t}
V_{\ul p,t_{\ul r}} = \big[\ti V_{\ul r}, V_{\ul p}\big].
\end{equation}
\end{lemma}

Next we turn to the Dubrovin equations for the time variation of the zeros $\mu_j$ of 
$(\dott)^{p_-}F_{\ul p}$ and $\nu_j$ of $(\dott)^{p_- -1}H_{\ul p}$ governed by the 
$\widetilde{\AL}_{\ul r}$ flow. 

\begin{lemma} [\cite{GesztesyHoldenMichorTeschl:2007}]  \lb{lAL4.5}
Assume Hypothesis \ref{hAL4.1} and suppose that \eqref{ALzctilde}, 
\eqref{ALzctstat} hold on $\bbZ\times \calI_{\mu}$ with $\calI_{\mu} \subseteq\bbR$ an open interval. In addition, assume that the
zeros $\mu_j$, $j=1,\dots,p$, of $(\dott)^{p_-}F_{\ul p}(\dott)$ remain distinct and nonzero on 
$\bbZ\times \calI_{\mu}$. Then $\{\hmu_j\}_{j=1,\dots,p}$, defined in \eqref{AL4.20}, satisfies the following first-order system of differential equations on 
$\bbZ\times \calI_{\mu}$, 
\begin{equation}
\mu_{j,t_{\ul r}} = - i \ti F_{\ul r}(\mu_j)  y(\hmu_j) (\alpha^+)^{-1} 
\prod_{\substack{k=1\\k\neq j}}^p (\mu_j-\mu_k)^{-1},  \quad 
j=1,\dots,p,   \lb{AL4.69a}
\end{equation}
with 
\begin{equation}
\hmu_j(n,\cdot)\in C^\infty(\calI_\mu,\calK_p), \quad j=1,\dots,p, 
\; n\in\bbZ.   \lb{AL4.69b}
\end{equation} 
For the zeros $\nu_j$, $j=1,\dots,p$, of $(\dott)^{p_- -1}H_{\ul p}(\dott)$, identical statements hold with 
$\mu_j$ and $\calI_{\mu}$ replaced by $\nu_j$ and $\calI_{\nu}$, etc. $($with 
$\calI_{\nu} \subseteq\bbR$ an open interval\,$)$. In particular, 
$\{\hat \nu_j\}_{j=1,\dots,p}$, defined in \eqref{AL4.21}, satisfies the first-order 
system on $\bbZ\times \calI_{\nu}$, 
\begin{equation}
\nu_{j,t_{\ul r}} = i \ti H_{\ul r}(\nu_j) y(\hnu_j) ( \beta \nu_j)^{-1} 
\prod_{\substack{k=1\\k\neq j}}^p (\nu_j-\nu_k)^{-1},  \quad 
j=1,\dots,p,   \lb{AL4.71a}
\end{equation}
with 
\begin{equation}
\hnu_j(n,\cdot)\in C^\infty(\calI_\nu,\calK_p), \quad j=1,\dots,p, 
\; n\in\bbZ.   \lb{AL4.71b}
\end{equation} 
\end{lemma}

When attempting to solve the Dubrovin systems \eqref{AL4.69a} and \eqref{AL4.71a}, they must be augmented with appropriate divisors 
$\calD_{\humu(n_0,t_{0,\ul r})}\in\sym^p \calK_p$, $t_{0,\ul r}\in \calI_\mu$, and 
$\calD_{\hunu(n_0,t_{0,\ul r})}\in\sym^p \calK_p$, $t_{0,\ul r}\in \calI_\nu$,  
as initial conditions. The algebro-geometric initial value problem for the AL hierarchy with appropriate initial divisors will be discussed in detail in 
\cite{GesztesyHoldenMichorTeschl:2007a}. 

Next, we turn to the asymptotic expansions of $\phi$ and $\Psi$ in a neighborhood of $\Pinfpm$ and $\Pzpm$. Since this is a bit more involved we provide some details.

\begin{lemma}  \lb{lAL4.6}
Assume Hypothesis \ref{hAL4.1} and suppose that \eqref{ALzctilde}, \eqref{ALzctstat} hold. Moreover, let $P=(z,y)\in\calK_p\setminus\{\Pinfp,\Pinfm,\Pzp,\Pzm\}$, 
$(n,n_0,t_{\ul r},t_{0,\ul r})\in\bbZ^2\times\bbR^2$. Then $\phi$ has the asymptotic behavior 
\begin{align}  
\phi(P) \underset{\zeta\to 0}{=}&  \begin{cases} 
\beta + \beta^-\gamma \zeta + \Oh(\zeta^2), & \quad  P \rightarrow P_{\infty_+}, \\
- (\alpha^+)^{-1} \zeta^{-1} + (\alpha^+)^{-2}\alpha^{++}\gamma^+
+ \Oh(\zeta), & \quad  P \rightarrow P_{\infty_-}, 
\end{cases}
\quad \zeta=1/z,  \label{ALphi inftyt} \\ 
\phi(P) \underset{\zeta\to 0}{=}&  \begin{cases} 
\alpha^{-1} - \alpha^{-2} \alpha^-\gamma \zeta  + \Oh(\zeta^2), 
& \quad P \rightarrow P_{0,+}, \\  
- \beta^+ \zeta - \beta^{++}\gamma^+ \zeta^2 + \Oh(\zeta^3), & \quad P \rightarrow P_{0,-},
\end{cases}
\quad \zeta=z.   \label{ALphi zerot}
\end{align}
The component $\psi_1$ of the Baker--Akhiezer vector $\Psi$ has the asymptotic behavior 
\begin{align} 
& \psi_1(P,n,n_0,t_{\ul r},t_{0,\ul r}) \underset{\zeta\to 0}{=} 
\exp\bigg(\pm \frac{i}{2}(t_{\ul r}-t_{0,\ul r})\sum_{s=0}^{r_+} 
\tilde c_{r_+ -s,+}\zeta^{-s} \bigg)
(1+\Oh(\zeta))  \no \\
&\quad \times\begin{cases}
\zeta^{n_0-n}, &  P\to\Pinfp, \\[1mm]
\begin{matrix}\pgam(n,n_0,t_{\ul r})\frac{\alpha^+(n,t_{\ul r})}{\alpha^+(n_0,t_{0,\ul r})}\\
\times\exp\Big(i \int_{t_{0,\ul r}}^{t_{\ul r}} ds 
\big(\tilde g_{r_+,+}(n_0,s) - \tilde g_{r_-,-}(n_0,s)\big)\Big),
\end{matrix} &P\to\Pinfm, 
\end{cases} 
\quad \zeta=1/z,  \label{ALpsi infty t}\\
& \psi_1(P,n,n_0,t_{\ul r},t_{0,\ul r}) \underset{\zeta\to 0}{=}
\exp\bigg(\pm \frac{i}{2}(t_{\ul r}-t_{0,\ul r})\sum_{s=0}^{r_-} 
\tilde c_{r_- -s,-}\zeta^{-s} \bigg)
(1 + \Oh(\zeta)) \no \\
&\quad \times\begin{cases}
\frac{\alpha(n,t_{\ul r})}{\alpha(n_0,t_{0,\ul r})}, & P\to P_{0,+}, \\[1mm]
\begin{matrix} \pgam(n,n_0,t_{\ul r})\zeta^{n-n_0} \\
\times \exp\Big(i \int_{t_{0,\ul r}}^{t_{\ul r}} ds 
\big(\tilde g_{r_+,+}(n_0,s) - \tilde g_{r_-,-}(n_0,s)\big)\Big),
\end{matrix} &P\to P_{0,-}, 
\end{cases} 
\quad \zeta=z.  \label{ALpsi 0 t}
\end{align} 
\end{lemma}
\begin{proof} 
Since by the definition of $\phi$ in \eqref{AL4.22} the time parameter $t_{\ul r}$ can be viewed as an additional but fixed parameter, the asymptotic behavior of $\phi$ remains the same as in Lemma \ref{lAL3.3}. Similarly, also the asymptotic behavior of 
$\psi_1(P,n,n_0,t_{\ul r},t_{\ul r})$ is derived in an identical fashion to that in Lemma \ref{lAL3.3}.
This proves \eqref{ALpsi infty t} and \eqref{ALpsi 0 t} for $t_{0,\ul r}=t_{\ul r}$, that is, 
\begin{align}
\psi_1(P,n,n_0,t_{\ul r},t_{\ul r}) 
&\underset{\zeta\to 0}{=} \begin{cases}
\zeta^{n_0-n}(1+\Oh(\zeta)), & P\to\Pinfp, \\
\pgam(n,n_0,t_{\ul r})\frac{\alpha^+(n,t_{\ul r})}{\alpha^+(n_0,t_{\ul r})}+\Oh(\zeta), 
&  P\to\Pinfm, \end{cases}  
\qquad \zeta=1/z,    \\
\psi_1(P,n,n_0,t_{\ul r},t_{\ul r})
&\underset{\zeta\to 0}{=} \begin{cases}
\frac{\alpha(n,t_{\ul r})}{\alpha(n_0,t_{\ul r})} +\Oh(\zeta), &  P\to\Pzp, \\
\pgam(n,n_0,t_{\ul r})\zeta^{n-n_0}(1+\Oh(\zeta)), 
& P\to\Pzm, \end{cases}  
\qquad \zeta=z.   
\end{align}
Remaining to be investigated is  
\begin{equation}
\psi_1(P,n_0,n_0,t_{\ul r},t_{0,\ul r}) = \exp \bigg(i \int_{t_{0,\ul r}}^{t_{\ul r}} dt 
\big(\ti G_{\ul r} (z,n_0,t) - \ti F_{\ul r}(z,n_0,t) \phi(P,n_0,t)\big) \bigg).   \lb{AL4.83A}
\end{equation}
The asymptotic expansion of the integrand is derived using Theorem \ref{tALB.2}. 
Focusing on the homogeneous coefficients first, one computes as 
$P \to P_{\infty_{\pm}}$,
\begin{align}
&\hatt G_{s,+} - \hatt F_{s,+} \phi = \hatt G_{s,+} - \hatt F_{s,+} 
\frac{G_{\ul p} + (c_{0,+}/2) z^{-p_-}y}{F_{\ul p}}   \no \\
& \quad =
\hatt G_{s,+} - \hatt F_{s,+} \bigg(\f{2z^{p_-}}{c_{0,+}} \frac{G_{\ul p}}{y} + 1 \bigg)
\bigg(\f{2z^{p_-}}{c_{0,+}} \frac{F_{\ul p}}{y}\bigg)^{-1}   \no \\
& \;\; \underset{\zeta\to 0}{=}
\pm \frac{1}{2} \zeta^{-s} + \frac{\hat g_{0,+} \mp \tfrac{1}{2}}{\hat f_{0,+}}
\hat f_{s,+} + \Oh(\zeta), \quad P \to P_{\infty_{\pm}}, \; \zeta=1/z.
\end{align}
Since
\begin{equation}
\ti F_{\ul r} \underset{\zeta\to 0}{=} \sum_{s=0}^{r_+} \tilde c_{r_+ -s,+}\hatt F_{s,+} 
+ \Oh(\zeta),  \quad
\ti G_{\ul r} \underset{\zeta\to 0}{=} \sum_{s=0}^{r_+} \tilde c_{r_+ -s,+}\hatt G_{s,+} 
+ \Oh(\zeta),
\end{equation}
one infers from \eqref{ALphi inftyt}
\begin{equation}
\ti G_{\ul r} - \ti F_{\ul r} \phi \underset{\zeta\to 0}{=}
\f{1}{2}\sum_{s=0}^{r_+} \tilde c_{r_+ -s,+} \zeta^{-s} + \Oh(\zeta),
\quad  P \to P_{\infty_+}, \; \zeta=1/z.    \lb{AL4.85a}
\end{equation}
Insertion of \eqref{AL4.85a} into \eqref{AL4.83A} then proves \eqref{ALpsi infty t} as
$P\to\Pinfp$.

As $P \to P_{\infty_-}$, we need one additional term in the asymptotic expansion of
$\ti F_{\ul r}$, that is, we will use
\begin{equation}
\ti F_{\ul r} \underset{\zeta\to 0}{=} \sum_{s=0}^{r_+} \tilde c_{r_+ -s,+}\hatt F_{s,+}
 + \sum_{s=0}^{r_-} \tilde c_{r_- -s,-} \hat f_{s-1,-} \zeta + \Oh(\zeta^2).
\end{equation}
This then yields
\begin{equation}
\ti G_{\ul r} - \ti F_{\ul r} \phi \underset{\zeta\to 0}{=}
-\f{1}{2}\sum_{s=0}^{r_+} \tilde c_{r_+ -s,+} \zeta^{-s} 
- (\alpha^+)^{-1}(\tilde f_{r_+,+} - \tilde f_{r_- -1,-})
+ \Oh(\zeta).  \lb{AL4.85b}
\end{equation}
Invoking \eqref{ALf_l+} and  \eqref{ALal_ivp} one concludes that 
\begin{equation}
\tilde f_{r_- -1,-} - \tilde f_{r_+,+}  = -i \alpha_{t_{\ul r}}^+  
+ \alpha^+ (\tilde g_{r_+,+} - \tilde g_{r_-,-}) \lb{AL4.85c}
\end{equation}
and hence 
\begin{equation}
\ti G_{\ul r} - \ti F_{\ul r} \phi \underset{\zeta\to 0}{=}
-\f{1}{2}\sum_{s=0}^{r_+} \tilde c_{r_+ -s,+} \zeta^{-s} 
 - \f{i \alpha_{t_{\ul r}}^+}{\alpha^+}  + \tilde g_{r_+,+} - \tilde g_{r_-,-} + \Oh(\zeta).  
 \lb{AL4.85d}
\end{equation}
Insertion of \eqref{AL4.85d} into \eqref{AL4.83A} then proves 
\eqref{ALpsi infty t} as $P\to\Pinfm$.

Using Theorem \ref{tALB.2} again, one obtains in the same manner as
$P \to P_{0,\pm}$,
\begin{equation} \label{AL4.87a}
\hatt G_{s,-} - \hatt F_{s,-} \phi \underset{\zeta\to 0}{=}  \pm \frac{1}{2} \zeta^{-s} - \hat g_{s,-} +
\frac{\hat g_{0,-} \pm \tfrac{1}{2}}{\hat f_{0,-}}
\hat f_{s,-} + \Oh(\zeta).
\end{equation}
Since
\begin{align}
\ti F_{\ul r} & \underset{\zeta\to 0}{=} \sum_{s=0}^{r_-} \tilde c_{r_- -s,-}
\hatt F_{s,-}  + \tilde f_{r_+ -1,+} + \Oh(\zeta),
\quad P \to \Pzpm, \; \zeta=z,   \lb{AL4.88A } \\
\ti G_{\ul r} & \underset{\zeta\to 0}{=} \sum_{s=0}^{r_-} \tilde c_{r_- -s,-}
\hatt G_{s,-} + \tilde g_{r_+,+} + \Oh(\zeta),
\quad P \to \Pzpm, \; \zeta=z,   \lb{AL4.88B}
\end{align}
\eqref{AL4.87a}--\eqref{AL4.88B} yield
\begin{equation}
\ti G_{\ul r} - \ti F_{\ul r}  \phi \underset{\zeta\to 0}{=}
\pm \frac{1}{2} \sum_{s=0}^{r_-} \tilde c_{r_- -s,-} \zeta^{-s} + \tilde g_{r_+,+} 
- \tilde g_{r_-,-}
- \frac{\hat g_{0,-} \pm \tfrac{1}{2}}{\hat f_{0,-}}
(\tilde f_{r_+ -1,+} - \tilde f_{r_-,-}) + \Oh(\zeta),   
\end{equation}
where we again used \eqref{ALphi zerot}, \eqref{ALhat f}, and \eqref{ALal_ivp}. 
As $P\to\Pzm$, one thus obtains
\begin{align}
\ti G_{\ul r} - \ti F_{\ul r}  \phi \underset{\zeta\to 0}{=}
- \frac{1}{2} \sum_{s=0}^{r_-} \tilde c_{r_- -s,-} \zeta^{-s} + \tilde g_{r_+,+} 
- \tilde g_{r_-,-},  \quad P \to \Pzm, \; \zeta=z.  \lb{AL4.95}
\end{align}
Insertion of \eqref{AL4.95} into \eqref{AL4.83A} then proves \eqref{ALpsi 0 t} 
as $P\to\Pzm$.

As $P\to\Pzp$, one obtains  
\begin{align}
\ti G_{\ul r} - \ti F_{\ul r}  \phi  & \underset{\zeta\to 0}{=}
\frac{1}{2} \sum_{s=0}^{r_-} \tilde c_{r_- -s,-} \zeta^{-s} + \tilde g_{r_+,+} 
- \tilde g_{r_-,-}
- \frac{1}{\alpha} (\tilde f_{r_+ -1,+} - \tilde f_{r_-,-}) + \Oh(\zeta),   \no \\
& \underset{\zeta\to 0}{=}
\frac{1}{2} \sum_{s=0}^{r_-} \tilde c_{r_- -s,-} \zeta^{-s} 
- \frac{i \alpha_{t_{\ul r}}}{\alpha} + \Oh(\zeta),   \quad P \to \Pzp, \; \zeta=z,  \lb{AL4.96}
\end{align}
using $\tilde f_{r_-,-}=\tilde f_{r_- -1,-}^- + \alpha(\tilde g_{r_-,-} - \tilde g_{r_-,-}^-)$ 
(cf.\ \eqref{ALf_l-}) and \eqref{ALal_ivp}. 
Insertion of \eqref{AL4.96} into \eqref{AL4.83A} then proves \eqref{ALpsi 0 t} 
as $P\to\Pzp$. 
\end{proof}

Next, we note that Lemmas \ref{lAL3.2} and \ref{lAL3.4} on trace formulas and nonspecial divisors in the stationary context immediately extend to the present time-dependent situation since $t_{\ul r} \in \bbR$ just plays the role of a parameter. We thus omit the details.

\medskip

Finally, we turn to the principal result of this section, the
representation of $\phi$, $\Psi$, $\alpha$, and $\beta$ in terms
of the Riemann theta function associated with $\calK_p$, assuming $p\in\bbN$ for the remainder of this section.

In addition to \eqref{AL3.57a}--\eqref{AL3.57g}, let $\omega_{\Pinfpm,q}^{(2)}$ and 
 $\omega_{\Pzpm,q}^{(2)}$ 
be the normalized differentials of the second kind with a unique pole at $\Pinfpm$ and $\Pzpm$, respectively, and principal parts  
\begin{align}
\omega_{\Pinfpm,q}^{(2)}&\underset{\zeta\to 0}{=}
\big(\zeta^{-2-q}+\Oh(1) \big)d\zeta, \quad P\to\Pinfpm, \; \zeta=1/z, \; 
q\in\bbN_0, \label{AL4.69}  \\
\omega_{\Pzpm,q}^{(2)}&\underset{\zeta\to 0}{=}
\big(\zeta^{-2-q}+\Oh(1) \big)d\zeta, \quad P\to\Pzpm, \; \zeta=z, \; 
q\in\bbN_0, \label{AL4.69A}
\end{align}
with vanishing $a$-periods,
\begin{equation}
\int_{a_j}\omega_{\Pinfpm,q}^{(2)} = \int_{a_j}\omega_{\Pzpm,q}^{(2)} =0, 
\quad j=1,\dots,p.
\label{AL4.70}
\end{equation}
Moreover, we define
\begin{align} \no
\ti \Omega_{\ul r}^{(2)}&=\frac{i}{2}\bigg(\sum_{s=1}^{r_-} s \tilde c_{r_- -s, -}
\big(\omega_{P_{0,+},s-1}^{(2)}-\omega_{P_{0,-},s-1}^{(2)} \big) \\
&\hspace*{1cm} + \sum_{s=1}^{r_+} s \tilde c_{r_+ -s, +}
\big(\omega_{\Pinfp,s-1}^{(2)}-\omega_{\Pinfm,s-1}^{(2)} \big)\bigg),  \label{AL4.71}
\end{align}
where $\tilde c_{\ell,\pm}$ are the summation constants in $\ti F_{\ul r}$. 
The corresponding vector of $b$-periods of $\ti\Omega_{\ul r}^{(2)}/(2\pi i)$ is 
then denoted by 
\begin{equation}
\ti {\underline U}_{\ul r}^{(2)}
=\big(\ti U_{\ul r,1}^{(2)},\dots,\ti U_{\ul r,p}^{(2)}
\big), \quad \ti U_{\ul r,j}^{(2)}
=\frac{1}{2\pi i}\int_{b_j}\ti \Omega_{\ul r}^{(2)}, \quad j=1,\dots,p.
\label{AL4.74}
\end{equation}

Finally, we abbreviate
\begin{align}
\ti\Omega_{\ul r}^{\infty_{\pm}} & =
\lim_{P \rightarrow P_{\infty_\pm}}\bigg(\int_{Q_0}^P\ti \Omega_{\ul r}^{(2)} \pm \frac{i}{2}
\sum_{s=0}^{r_+} \tilde c_{r_+ -s,+} \zeta^{-s} \bigg),   \lb{AL4.75}  \\
\ti \Omega_{\ul r}^{0,\pm} & =
\lim_{P \rightarrow P_{0,\pm}}\bigg(\int_{Q_0}^P\ti \Omega_{\ul r}^{(2)} \pm \frac{i}{2}
\sum_{s=0}^{r_-} \tilde c_{r_- -s,-} \zeta^{-s} \bigg).    \lb{AL4.75a}
\end{align}

\begin{theorem}  \lb{tAL4.7}
Assume Hypothesis \ref{hAL4.1} and suppose that \eqref{ALzctilde}, 
\eqref{ALzctstat} hold. In addition, let  
$P\in\calK_p\setminus\{\Pinfp,\Pinfm,\Pzp,\Pzm\}$ and
$(n,n_{0},t_{\ul r},t_{0,\ul r})\in\bbZ^2\times\bbR^2$. Then for
each $(n,t_{\ul r})\in\bbZ\times\bbR$, $\calD_{\humu(n,t_{\ul r})}$ and
$\calD_{\hunu(n,t_{\ul r})}$ are nonspecial. Moreover, 
\begin{align}
\phi(P,n,t_{\ul r})&= C(n,t_{\ul r})
\frac{\theta(\uz(P,\hunu(n,t_{\ul r})))}{\theta(\uz(P,\humu(n,t_{\ul r})))}
\exp\bigg(\int_{Q_0}^P \omega^{(3)}_{\Pzm, \Pinfm} \bigg), \label{AL4.76} \\
\psi_{1}(P,n,n_{0},t_{\ul r},t_{0,\ul r})&=C(n,n_{0},t_{\ul r},t_{0,\ul r})
\frac{\theta(\uz(P,\humu(n,t_{\ul r})))}
           {\theta(\uz(P,\humu(n_{0},t_{0,\ul r})))} \label{AL4.78} \\
& \quad \times \exp\bigg((n-n_{0})
\int_{Q_{0}}^{P} \omega^{(3)}_{\Pzm,\Pinfp}
-(t_{\ul r}-t_{0,\ul r})\int_{Q_0}^P \ti \Omega_{\ul r}^{(2)}\bigg), \no \\
\psi_{2}(P,n,n_{0},t_{\ul r},t_{0,\ul r})& =C(n,t_{\ul r}) C(n,n_{0},t_{\ul r},t_{0,\ul r})
\frac{\theta(\uz(P,\hunu(n,t_{\ul r})))} {\theta(\uz(P,\humu(n_0,t_{0,\ul r})))}  
 \label{AL4.79}  \\ 
& \hspace*{-2cm} \times \exp\bigg(\int_{Q_{0}}^{P}
\omega^{(3)}_{\Pzm,\Pinfm}+(n-n_{0}) \int_{Q_{0}}^{P}
\omega^{(3)}_{\Pzm,\Pinfp} -(t_{\ul r}-t_{0,\ul r})\int_{Q_0}^P \ti \Omega_{\ul r}^{(2)}\bigg),  
\no 
\end{align}
where
\begin{align}  \label{AL4.80}
& C(n,t_{\ul r})=\frac{(-1)^{n-n_0}}{\alpha(n_0,t_{\ul r})}\exp\big((n-n_0)(\omega_0^{0,-}(\Pzm,\Pinfm)
-\omega_0^{\infty_+}(\Pzm,\Pinfm))\big) \\ \no 
& \hspace*{1.65cm}\times \exp\big(-\omega_0^{0,+}(\Pzm,\Pinfm)\big)
\frac{\theta(\uz(\Pzp,\humu(n_0,t_{\ul r})))}{\theta(\uz(\Pzp,\hunu(n_0,t_{\ul r})))},
\\ \label{AL4.81}
&C(n,n_{0},t_{\ul r},t_{0,\ul r})=
\frac{\theta(\uz(\Pinfp,\humu(n_0,t_{0,\ul r})))}
{\theta(\uz(\Pinfp,\humu(n,t_{\ul r})))} \\ \no
& \hspace*{2.8cm} \times
\exp\big((t_{\ul r}-t_{0,\ul r}) \ti\Omega^{\infty_+}_{\ul r} 
-(n-n_{0})\omega_0^{\infty_+}(\Pzm,\Pinfp)\big).
\end{align}
The Abel map linearizes the auxiliary divisors $\calD_{\humu(n,t_{\ul r})}$ and
$\calD_{\hunu(n,t_{\ul r})}$ in the sense that
\begin{align}
\amap(\calD_{\humu(n,t_{\ul r})})
&=\amap(\calD_{\humu(n_0,t_{0,\ul r})})+ (n-n_0) \underline A_{\Pzm}(\Pinfp)
+ (t_{\ul r}-t_{0,\ul r}) \ti{\underline U}^{(2)}_{\ul r}, \label{AL4.82} \\
\amap(\calD_{\hunu(n,t_{\ul r})})
&=\amap(\calD_{\hunu(n_0,t_{0,\ul r})})
+ (n-n_0) \underline A_{\Pzm}(\Pinfp)+ (t_{\ul r}-t_{0,\ul r}) 
\ti{\underline U}^{(2)}_{\ul r}. 
\label{AL4.83}
\end{align}
Finally, $\alpha,\beta$ are of the form 
\begin{align} 
&\alpha(n,t_{\ul r}) = 
\alpha(n_0,t_{0,\ul r})
\exp\big((n-n_0)(\omega_0^{0,+}(\Pzm,\Pinfp)
-\omega_0^{\infty_+}(\Pzm,\Pinfp))\big)   \label{AL4.84} \\
& \quad \times
\exp\big((t_{\ul r}-t_{0,\ul r}) (\ti\Omega_{\ul r}^{\infty_+} - \ti\Omega_{\ul r}^{0,+})\big)  
\frac{\theta(\uz(\Pzp,\humu(n,t_{\ul r})))
\theta(\uz(P_{\infty_+},\humu(n_0,t_{0,\ul r})))}
{\theta(\uz(\Pzp,\humu(n_0,t_{0,\ul r})))
\theta(\uz(P_{\infty_+},\humu(n,t_{\ul r})))},  \no  \\  
&\beta(n,t_{\ul r}) = \frac{1}{\alpha(n_0,t_{0,\ul r})}
\exp\big((n-n_0)(\omega_0^{0,+}(\Pzm,\Pinfp)
-\omega_0^{\infty_+}(\Pzm,\Pinfp))\big)  \no  \\  
& \quad \times 
\exp\big(\omega_0^{\infty_+}(\Pzm,\Pinfm)-\omega_0^{0,+}(\Pzm,\Pinfm)\big)
 \label{AL4.85}  \\ 
& \quad \times 
\exp\big(-(t_{\ul r}-t_{0,\ul r}) (\ti\Omega_{\ul r}^{\infty_+} - \ti\Omega_{\ul r}^{0,+})\big) 
\frac{\theta(\uz(\Pzp,\humu(n_0,t_{0,\ul r})))
\theta(\uz(P_{\infty_+},\hunu(n,t_{\ul r})))}
{\theta(\uz(\Pzp,\hunu(n_0,t_{0,\ul r})))
\theta(\uz(P_{\infty_+},\humu(n,t_{\ul r})))},  \no
\end{align}
and
\begin{align}
\begin{split}
\alpha(n,t_{\ul r})\beta(n,t_{\ul r})
&=\exp\big(\omega_0^{\infty_+}(\Pzm,\Pinfm)
-\omega_0^{0,+}(\Pzm,\Pinfm)\big)  \\
& \quad \times
\frac{\theta(\uz(\Pzp,\humu(n,t_{\ul r})))
\theta(\uz(\Pinfp,\hunu(n,t_{\ul r})))}
{\theta(\uz(\Pzp,\hunu(n,t_{\ul r})))
\theta(\uz(\Pinfp,\humu(n,t_{\ul r})))}. \label{AL4.87}
\end{split}
\end{align}
\end{theorem}
\begin{proof}
As in Theorem \ref{tAL3.7} one concludes that $\phi(P,n,t_{\ul r})$ is of the form
\eqref{AL4.76} and that for $t_{0,\ul r}=t_{\ul r}$, $\psi_1(P,n,n_0,t_{\ul r},t_{\ul r})$ is
of the form
\begin{equation}
\psi_{1}(P,n,n_{0},t_{\ul r},t_{\ul r})=C(n,n_{0},t_{\ul r},t_{\ul r})
\frac{\theta(\uz(P,\humu(n,t_{\ul r})))}
    {\theta(\uz(P,\humu(n_0,t_{\ul r})))}\exp\bigg((n-n_0)
\int_{Q_{0}}^{P} \omega^{(3)}_{\Pzm,\Pinfp}\bigg). \label{AL4.88}
\end{equation}
To discuss $\psi_1(P,n,n_0,t_{\ul r},t_{0,\ul r})$ we recall \eqref{ALpsi t n_0}, that is, 
\begin{equation}
\psi_1(P,n,n_0,t_{\ul r},t_{0,\ul r})=\psi_1(P,n,n_0,t_{\ul r},t_{\ul r})
\psi_1(P,n_0,n_0,t_{\ul r},t_{0,\ul r}),  \label{AL4.89}
\end{equation}
and hence remaining to be studied is 
\begin{equation}
\psi_{1}(P,n_0,n_{0},t_{\ul r},t_{0,\ul r})=\exp\bigg(i \int_{t_{0,\ul r}}^{t_{\ul r}} ds 
\big(\ti G_{\ul r}(z,n_0,s)-\ti F_{\ul r}(z,n_0,s)\phi(P,n_0,s)\big)\bigg).
\label{AL4.90}
\end{equation}
Introducing $\hat\psi_1(P)$ on
$\calK_p\setminus\{\Pinfp,\Pinfm\}$ by
\begin{align} 
\hat\psi_1(P,n_0,t_{\ul r},t_{0,\ul r})&= C(n_0,n_0,t_{\ul r},t_{0,\ul r})
\frac{\theta(\uz(P,\humu(n_0,t_{\ul r})))}{\theta(\uz(P,\humu(n_{0},t_{0,\ul r})))} \no \\
& \quad \times 
\exp\bigg(-(t_{\ul r}-t_{0,\ul r})\int_{Q_{0}}^{P} \ti \Omega^{(2)}_{\ul r}\bigg),
\label{AL4.91}
\end{align}
we intend to prove that
\begin{align}
\begin{split}
&\psi_1(P,n_0,n_0,t_{\ul r},t_{0,\ul r})=\hat\psi_1(P,n_0,t_{\ul r},t_{0,\ul r}),
\label{AL4.92} \\
&P\in\calK_p\setminus\{\Pinfp, \Pinfm\}, \; n_0\in\bbZ,
\;  t_{\ul r}, t_{0,\ul r} \in \bbR, 
\end{split}
\end{align}
for an appropriate choice of the normalization constant 
$C(n_0,n_0,t_{\ul r},t_{0,\ul r})$ in \eqref{AL4.91}. We start by noting that a comparison of 
\eqref{ALpsi infty t}, \eqref{ALpsi 0 t} and \eqref{AL4.69}, \eqref{AL4.69A},  
\eqref{AL4.71}, \eqref{AL4.78} shows that $\psi_1$
and $\hat\psi_1$ have the same essential singularities at $\Pinfpm$ and $\Pzpm$. 
Thus, we turn to the local behavior of $\psi_1$ and $\hat\psi_1$.  By \eqref{AL4.91},
$\hat\psi_1$ has zeros and poles at $\humu(n_{0},t_{\ul r})$ and
$\humu(n_{0},t_{0,\ul r})$, respectively.  Similarly, by \eqref{AL4.90},
$\psi_1$ can have zeros and poles only at poles of $\phi(P,n_0,s)$,
$s\in[t_{0,\ul r},t_{\ul r}]$ (resp., $s\in[t_{\ul r},t_{0,\ul r}]$). In the following we
temporarily restrict $t_{0,\ul r}$ and $t_{\ul r}$ to a sufficiently small nonempty
interval $I\subseteq \bbR$ and pick $n_0\in\bbZ$ such that for all $s\in
I$, $\mu_j(n_0,s)\neq\mu_k(n_0,s)$ for all $j\neq k$, $j,k=1,\dots,p$.
One computes
\begin{align}
& i \ti G_{\ul r}(z,n_0,s) - i \ti F_{\ul r}(z,n_0,s)\phi(P,n_0,s) \no \\
&\quad = i \ti G_{\ul r}(z,n_0,s)- i \ti F_{\ul r}(z,n_0,s)
\frac{(c_{0,+}/2) z^{-p_-} y+G_{\ul p}(z,n_0,s)}{F_{\ul p}(z,n_0,s)}\no \\
&\qquad \underset{P\to\hmu_j(n_0,s)}{=}\frac{i \ti F_{\ul r}(\mu_j(n_0,s),n_0,s)
y(\hmu_j(n_0,s))}{\alpha^+(n_0,s)(z-\mu_j(n_0,s))
\prod_{\substack{k=1\\k\neq j}}^p (\mu_j(n_0,s)-\mu_k(n_0,s))} +\Oh(1)
\no \\
&\qquad \underset{P\to\hmu_j(n_0,s)}{=} \frac{\partial}{\partial
s}\ln\big(\mu_j(n_0,s)-z \big) +\Oh(1). \label{AL4.98}
\end{align}
Restricting $P$ to a sufficiently small neighborhood $\mathcal U_j(n_0)$ of
$\{\hmu_j(n_0,s)\in\calK_p\,|\, s\in[t_{0,\ul r},t_{\ul r}]\subseteq I \}$ such
that $\hmu_k(n_0,s)\not\in\mathcal U_j(n_0)$ for all
$s\in[t_{0,\ul r},t_{\ul r}]\subseteq I$ and all
$k\in\{1,\dots,p\}\setminus\{j\}$,
\eqref{AL4.91} and \eqref{AL4.98} imply
\begin{align} \no
&\psi_1(P,n_0,n_0,t_{\ul r},t_{0,\ul r}) \\ 
&\quad =\begin{cases}
(\mu_j(n_0,t_{\ul r})-z)\Oh(1) &\text{as $P\to\hmu_j(n_0,t_{\ul r})
\neq\hmu_j(n_0,t_{0,\ul r})$}, \\
\Oh(1)  &\text{as $P\to\hmu_j(n_0,t_{\ul r})
=\hmu_j(n_0,t_{0,\ul r})$}, \\
(\mu_j(n_0,t_{0,\ul r})-z)^{-1}\Oh(1) &\text{as $P\to\hmu_j(n_0,t_{0,\ul r})
\neq\hmu_j(n_0,t_{\ul r})$,}
\end{cases} \label{AL4.99} \\
&\hspace*{7.6cm} P=(z,y)\in\calK_p, \no
\end{align}
with $\Oh(1)\neq 0$. Thus, $\psi_{1}$ and $\hat\psi_{1}$ have the same
local behavior and  identical essential singularities at $\Pinfpm$ and $\Pzpm$. By
Lemma \ref{la6}, $\psi_{1}$ and $\hat\psi_{1}$ coincide up to a multiple
constant (which may depend on $n_{0},t_{\ul r},t_{0,\ul r}$). This proves
\eqref{AL4.92} for $t_{0,\ul r}, t_{\ul r}\in I$ and for $n_0$ as restricted above.
By continuity with respect to divisors this extends to all $n_0\in\bbZ$
since by hypothesis $\calD_{\humu(n,s)}$ remain
nonspecial for all $(n,s)\in\bbZ\times\bbR$. Moreover, since by
\eqref{AL4.90}, for fixed $P$ and $n_0$, $\psi_1(P,n_0,n_0,.,t_{0,\ul r})$
is entire in $t_{\ul r}$ (and this argument is symmetric in $t_{\ul r}$ and
$t_{0,\ul r}$), \eqref{AL4.92} holds for all $t_{\ul r},t_{0,\ul r}\in\bbR$ (for an
appropriate choice of $C(n_0,n_0,t_{\ul r},t_{0,\ul r})$). Together with
\eqref{AL4.89}, this proves \eqref{AL4.78} for all
$(n,t_{\ul r}),(n_{0},t_{0,\ul r})\in\bbZ\times \bbR$. The expression
\eqref{AL4.79} for $\psi_2$ then immediately follows from
\eqref{AL4.76} and \eqref{AL4.78}.

To determine the constant $C(n,n_0,t_{\ul r},t_{0,\ul r})$ one compares the  
asymptotic expansions of $\psi_1(P,n,n_0,t_{\ul r},t_{0,\ul r})$ for $P \to P_{\infty_+}$
in \eqref{ALpsi infty t} and \eqref{AL4.78}
\begin{align} \label{AL4.124}  \no
C(n,n_0,t_{\ul r},t_{0,\ul r}) &=
\exp\big((t_{\ul r}-t_{0,\ul r}) \ti\Omega_{\ul r}^{\infty_+} 
- (n-n_0)\omega_0^{\infty_+}(\Pzm,\Pinfp)\big)\\
& \quad \times
\frac{\theta(\uz(P_{\infty_+},\humu(n_0,t_{0,\ul r})))}{\theta(\uz(P_{\infty_+},\humu(n,t_{\ul r})))}.
\end{align}
Remaining to be computed are the expressions for $\alpha$ and $\beta$. 
Comparing the asymptotic expansions of $\psi_1(P,n,n_0,t_{\ul r},t_{0,\ul r})$ for $P \to P_{0,+}$
in \eqref{ALpsi 0 t} and \eqref{AL4.78} shows
\begin{align} \no
\frac{\alpha(n,t_{\ul r})}{\alpha(n_0,t_{0,\ul r})} &= C(n,n_0,t_{\ul r},t_{0,\ul r}) 
\exp\big((n-n_0)(\omega_0^{0,+}(\Pzm,\Pinfp) -(t_{\ul r}-t_{0,\ul r}) \ti\Omega_{\ul r}^{0,+}\big)\\
& \quad \times
\frac{\theta(\uz(\Pzp,\humu(n,t_{\ul r})))}{\theta(\uz(\Pzp,\humu(n_0,t_{0,\ul r})))}
\end{align}
and inserting $C(n,n_0,t_{\ul r},t_{0,\ul r})$ proves \eqref{AL4.84}.
Equation \eqref{AL4.80} for $C(n,t_{\ul r})$ follows as in the stationary case since
$t_{\ul r}$ can be viewed as an additional but fixed parameter. 
By the first line of \eqref{AL3.69a},
\begin{equation}
\alpha(n,t_{\ul r})=\frac{1}{C(n,t_{\ul r})}\exp\big(-\omega_0^{0,+}(\Pzm,\Pinfm)\big)
\frac{\theta(\uz(\Pzp,\humu(n,t_{\ul r})))}{\theta(\uz(\Pzp,\hunu(n,t_{\ul r})))}. 
\label{AL4.131}
\end{equation}
Inserting the result \eqref{AL4.84} for $\alpha(n,t_{\ul r})$ into \eqref{AL4.131} 
then yields (using Lemma \ref{lAL3.6})
\begin{align} \no
& C(n,t_{\ul r})= \frac{1}{\alpha(n_0,t_{0,\ul r})} 
\frac{\theta(\uz(\Pzp,\humu(n_0,t_{0,\ul r})))}{\theta(\uz(\Pinfp,\humu(n_0,t_{0,\ul r})))} \exp\big((t_{\ul r}-t_{0,\ul r})(\ti\Omega_{\ul r}^{0,+} - \ti\Omega_{\ul r}^{\infty_+})\big) \\ 
& \quad \times 
\exp\big((n-n_0)(\omega_0^{\infty_+}(\Pzm,\Pinfp) - \omega_0^{0,+}(\Pzm,\Pinfp)) - \omega_0^{0,+}(\Pzm,\Pinfm)\big).   \label{ALC(n,t)}
\end{align}
Also, since the first line of \eqref{AL3.69d} holds,
\begin{equation}
\beta(n,t_{\ul r})=C(n,t_{\ul r})\exp\big(\omega_0^{\infty_+}(\Pzm,\Pinfm)\big)
\frac{\theta(\uz(\Pinfp,\hunu(n,t_{\ul r})))}{\theta(\uz(\Pinfp,\humu(n,t_{\ul r})))},
\label{AL4.132}
\end{equation}
an insertion of \eqref{ALC(n,t)} into \eqref{AL4.132}, observing
Lemma \ref{lAL3.6}, yields equation \eqref{AL4.85} for $\beta(n,t_{\ul r})$. 
Finally,
multiplying \eqref{AL4.131} and \eqref{AL4.132} proves \eqref{AL4.87}.

Single-valuedness of $\psi_1(\dott,n_0,n_0,t_{\ul r},t_{0,\ul r})$ on $\calK_p$ implies 
\begin{equation}
\amap(\calD_{\humu(n_0,t_{\ul r})}) = \amap(\calD_{\humu(n_0,t_{0,\ul r})}) 
+ i (t_{\ul r}-t_{0,\ul r}) \underline{\ti U}_{\ul r}^{(2)}.  \lb{AL4.126}
\end{equation}
Inserting \eqref{AL4.126} into \eqref{AL3.63},
\begin{equation}
\amap(\calD_{\humu(n,t_{\ul r})}) =\amap(\calD_{\humu(n_0,t_{\ul r})})
+\underline{A}_{\Pzm}(\Pinfp)(n-n_0),
\end{equation}
one obtains the result \eqref{AL4.82}.
\end{proof}

Again we note that the apparent $n_0$-dependence of $C(n,t_{\ul r})$ in the right-hand side of \eqref{AL4.80} actually drops out to ensure the $n_0$-independence of $\phi$ in 
\eqref{AL4.76}.  

The theta function representations \eqref{AL4.84} and \eqref{AL4.85} for $\alpha$ and 
$\beta$, and the one for $\Gamma(\dott,\dott,t_{\ul r})$ analogous to that in \eqref{AL4.87} also show that 
$\gamma(n,t_{\ul r})\notin \{0,1\}$ for all $(n,t_{\ul r})\in\bbZ\times\bbR$. Hence, condition 
\eqref{AL4.1A} is satisfied for the time-dependent algebro-geometric AL solutions discussed in this section, provided the associated divisors $\calD_{\humu}(n,t_{\ul r})$ and 
$\calD_{\hunu}(n,t_{\ul r})$ stay away from $\Pinfpm, \Pzpm$ for all 
$(n,t_{\ul r})\in\bbZ\times\bbR$. 

The time-dependent algebro-geometric initial value problem for the Ablowitz--Ladik hierarchy with complex-valued initial data, that is, the construction of $\alpha$ and 
$\beta$ by starting from a set of initial data (nonspecial divisors) of full measure, will be presented in \cite{GesztesyHoldenMichorTeschl:2007a}. 

\appendix
\section{Hyperelliptic Curves and Their Theta Functions}\lb{sAL.A}
\renewcommand{\theequation}{A.\arabic{equation}}
\renewcommand{\thetheorem}{A.\arabic{theorem}}
\setcounter{theorem}{0}
\setcounter{equation}{0}

We provide a very brief summary of some of the fundamental properties
and notations needed from the theory of hyperelliptic curves.  More details can be found in some of the standard textbooks \cite{FarkasKra:1992}, 
\cite{Fay:1973}, and \cite{Mumford:1984}, as well as monographs
dedicated to integrable systems such as
\cite[Ch.\ 2]{BelokolosBobenkoEnolskiiItsMatveev:1994}, 
\cite[App.\ A, B]{GesztesyHolden:2003}.

Fix $\N\in\bbN$. The hyperelliptic curve $\calK_\N$
of genus $\N$ used in Sections \ref{sAL2} and \ref{sALstat} is
defined by
\begin{align}
&\calK_\N\colon \calF_\N(z,y)=y^2-R_{2\N+2}(z)=0, \quad
R_{2\N+2}(z)=\prod_{m=0}^{2\N+1}(z-E_m), \lb{b0} \\
& \{E_m\}_{m=0,\dots,2\N+1}\subset\bbC, \quad
E_m \neq E_{m'} \text{ for } m \neq m', \, m,m'=0,\dots,2\N+1.
\label{b1}
\end{align}
The curve \eqref{b0} is compactified by adding the
points $\Pinfp$ and $\Pinfm$,
$\Pinfp \neq \Pinfm$, at infinity.
One then introduces an appropriate set of
$\N+1$ nonintersecting cuts $\calC_j$ joining
$E_{m(j)}$ and $E_{m^\prime(j)}$ and denotes
\begin{equation}
\calC=\bigcup_{j\in \{1,\dots,\N+1 \}}\calC_j,
\quad
\calC_j\cap\calC_k=\emptyset,
\quad j\neq k.\label{b2}
\end{equation}
Defining the cut plane
\begin{equation}
\Pi=\bbC\setminus\calC, \label{b3}
\end{equation}
and introducing the holomorphic function
\begin{equation}
R_{2\N+2}(\dott)^{1/2}\colon \Pi\to\bbC, \quad
z\mapsto \left(\prod_{m=0}^{2\N+1}(z-E_m) \right)^{1/2}\label{b4}
\end{equation}
on $\Pi$ with an appropriate choice of the square root
branch in \eqref{b4}, one considers 
\begin{equation}
\calM_{\N}=\{(z,\sigma R_{2\N+2}(z)^{1/2}) \,|\, 
z\in\bbC,\; \sigma\in\{\pm 1\}
\}\cup\{\Pinfp,\Pinfm\} \label{b5}
\end{equation}
by extending $R_{2\N+2}(\dott)^{1/2}$ to $\calC$. The
hyperelliptic curve $\calK_\N$ is then the set
$\calM_{\N}$ with its natural complex structure obtained
upon gluing the two sheets of $\calM_{\N}$
crosswise along the cuts. The set of branch points
$\calB(\calK_\N)$ of $\calK_\N$ is given by
\begin{equation}
\calB(\calK_\N)=\{(E_m,0)\}_{m=0,\dots,2\N+1} \lb{5a}
\end{equation}
and finite points $P$ on $\calK_\N$ are denoted by
$P=(z,y)$, where $y(P)$ denotes the meromorphic function
on $\calK_\N$ satisfying $\calF_\N(z,y)=y^2-R_{2\N+2}(z)=0$.
Local coordinates near $P_0=(z_0,y_0)\in\calK_\N\setminus
(\calB(\calK_\N)\cup\{\Pinfp,\Pinfm\})$ are
given by $\zeta_{P_0}=z-z_0$, near $\Pinfpm$ by
$\zeta_{\Pinfpm}=1/z$, and near branch points
$(E_{m_0},0)\in\calB(\calK_\N)$ by
$\zeta_{(E_{m_0},0)}=(z-E_{m_0})^{1/2}$. The Riemann surface
$\calK_\N$ defined in this manner has topological genus $\N$.
Moreover, we introduce the holomorphic sheet exchange map
(involution)
\begin{equation}
* \colon \calK_\N \to \calK_\N,\quad
P=(z,y)\mapsto P^*=(z,-y),\;
P_{\infty_\pm} \mapsto P_{\infty_\pm}^*=P_{\infty_\mp}.    \lb{a12}
\end{equation}

One verifies that $dz/y$ is a holomorphic differential
on $\calK_\N$ with zeros of order $\N-1$ at $\Pinfpm$
and hence
\begin{equation}
\eta_j=\frac{z^{j-1}dz}{y}, \quad j=1,\dots,\N,   \lb{b24}
\end{equation}
form a basis for the space of holomorphic differentials
on $\calK_\N$.  Introducing the
invertible matrix $C$ in $\bbC^\N$,
\begin{align}
\begin{split}
& C =(C_{j,k})_{j,k=1,\dots,\N}, \quad C_{j,k}
= \int_{a_k} \eta_j, \\
& \underline{c} (k) = (c_1(k), \dots,
c_p(k)), \quad c_j (k) =
C_{j,k}^{-1}, \;\, j,k=1,\dots,\N, \lb{A.7}
\end{split}
\end{align}
the corresponding basis of normalized holomorphic
differentials $\omega_j$, $j=1,\dots,\N$ on $\calK_\N$ is given by
\begin{equation}
\omega_j = \sum_{\ell=1}^\N c_j (\ell) \eta_\ell,
\quad \int_{a_k} \omega_j =
\delta_{j,k}, \quad j,k=1,\dots,\N. \lb{b26}
\end{equation}
Here $\{a_j,b_j\}_{j=1,\dots,\N}$ is a homology basis for
$\calK_\N$ with intersection matrix of the cycles satisfying
\begin{equation}
a_j \circ b_k=\delta_{j,k}, \; a_j \circ a_k=0,
\; b_j \circ b_k=0, \quad j,k=1,\dots,\N. \lb{c26}
\end{equation}

Associated with the homology basis
$\{a_j, b_j\}_{j=1,\dots,\N}$ we
also recall the canonical dissection of $\calK_\N$
along its cycles yielding
the simply connected interior $\hatt \calK_\N$ of the
fundamental polygon $\partial {\hatt \calK}_\N$ given by
\begin{equation}
\partial  {\hatt \calK}_\N =a_1 b_1 a_1^{-1} b_1^{-1}
a_2 b_2 a_2^{-1} b_2^{-1} \cdots
a_\N^{-1} b_\N^{-1}.
\lb{a25}
\end{equation}
Let $\calM (\calK_\N)$ and $\calM^1 (\calK_\N)$ denote the
set of meromorphic
functions (0-forms) and meromorphic
differentials (1-forms)
on $\calK_\N$. Holomorphic
differentials are also called Abelian differentials
of the first kind. Abelian differentials of the
second kind, $\omega^{(2)} \in \calM^1 (\calK_\N)$, are characterized
by the property that all their residues vanish.  They will usually be
normalized by demanding that all their $a$-periods vanish, that is,
$\int_{a_j} \omega^{(2)} =0$, $j=1,\dots,\N$. 
Any meromorphic differential $\omega^{(3)}$ on
$\calK_\N$ not of the first or
second kind is said to be of the third
kind. A differential of the third kind $\omega^{(3)} \in \calM^1 (\calK_\N)$
is usually normalized by the vanishing of its
$a$-periods, that is, $\int_{a_j} \omega^{(3)} =0$, $j=1,\dots, \N$. 
A normal differential of the third kind $\omega_{P_1, P_2}^{(3)}$ associated
with two points $P_1$, $P_2 \in \hatt \calK_\N$, $P_1 \neq P_2$, by definition, 
has simple poles at
$P_j$ with residues $(-1)^{j+1}$, $j=1,2$ and
vanishing $a$-periods.  

Next, define the matrix $\tau=(\tau_{j,\ell})_{j,\ell=1,\dots,\N}$ by
\begin{equation}
\tau_{j,\ell}=\int_{b_\ell}\omega_j, \quad j,\ell=1,
\dots,\N. \label{b8}
\end{equation}
Then
\begin{equation}
\Im(\tau)>0 \quad \text{and} \quad \tau_{j,\ell}=\tau_{\ell,j},
\quad j,\ell =1,\dots,\N.  \lb{a18a}
\end{equation}
Associated with $\tau$ one introduces the period lattice
\begin{equation}
L_\N = \{ \ul z \in\bbC^\N \,|\, \ul z = \ul m + \ul n\tau,
\; \ul m, \ul n \in\bbZ^\N\}
\lb{a28}
\end{equation}
and the Riemann theta function associated with $\calK_\N$ and
the given homology basis $\{a_j,b_j\}_{j=1,\dots,\N}$,
\begin{equation}
\theta(\ul z)=\sum_{\ul n\in\bbZ^\N}\exp\big(2\pi
i(\ul n,\ul z)+\pi i(\ul n, \ul n\tau)\big),
\quad \ul z\in\bbC^\N, \label{b9}
\end{equation}
where $(\ul u, \ul v)=\ol {\ul u}\, \ul v^\top=\sum_{j=1}^\N \ol{u_j}\, v_j$
denotes the scalar product in $\bbC^\N$. It has the fundamental properties
\begin{align}
& \theta(z_1, \ldots, z_{j-1}, -z_j, z_{j+1},
\ldots, z_\N) =\theta
(\ul z), \lb{a27}\\
& \theta (\ul z +\ul m + \ul n\tau)
=\exp \big(-2 \pi i (\ul n,\ul z) -\pi i (\ul n, \ul n\tau) \big) \theta
(\ul z), \quad \ul m, \ul n \in\bbZ^\N.
\lb{aa51}
\end{align}

Next, fix a base point $Q_0\in\calK_\N\setminus
\{\Pzpm,\Pinfpm\}$, denote by
$J(\calK_\N) = \bbC^\N/L_\N$ the Jacobi variety of $\calK_\N$,
and define the
Abel map $\underline{A}_{Q_0}$ by
\begin{equation}
\underline{A}_{Q_0} \colon \calK_\N \to J(\calK_\N), \quad
\underline{A}_{Q_0}(P)=
\bigg(\int_{Q_0}^P \omega_1,\dots,\int_{Q_0}^P \omega_\N \bigg)
\pmod{L_\N}, \quad P\in\calK_\N. \label{aa46}
\end{equation}
Similarly, we introduce
\begin{equation}
\ul \alpha_{Q_0}  \colon
\Div(\calK_\N) \to J(\calK_\N),\quad
\calD \mapsto \ul \alpha_{Q_0} (\calD)
=\sum_{P \in \calK_\N} \calD (P) \ul A_{Q_0} (P),
\label{aa47}
\end{equation}
where $\Div(\calK_\N)$ denotes the set of
divisors on $\calK_\N$. Here $\calD \colon \calK_\N \to \bbZ$
is called a divisor on $\calK_\N$ if $\calD(P)\neq0$ for only
finitely many $P\in\calK_\N$. (In the main body of this paper
we will choose $Q_0$ to be one of the branch points, i.e.,
$Q_0\in\calB(\calK_\N)$, and for simplicity we will always choose
the same path of integration from $Q_0$ to $P$ in all Abelian
integrals.) 

In connection with divisors on $\calK_\N$ we shall employ the
following
(additive) notation,
\begin{align} \lb{A.17}
&\calD_{Q_0\ul Q}=\calD_{Q_0}+\calD_{\ul Q}, \quad \calD_{\ul
Q}=\calD_{Q_1}+\cdots +\calD_{Q_m}, \\
& {\ul Q}=\{Q_1, \dots ,Q_m\} \in \sym^m \calK_\N,
\quad Q_0\in\calK_\N, \; m\in\bbN, \no
\end{align}
where for any $Q\in\calK_\N$,
\begin{equation} \lb{A.18}
\calD_Q \colon  \calK_\N \to\bbN_0, \quad
P \mapsto  \calD_Q (P)=
\begin{cases} 1 & \text{for $P=Q$},\\
0 & \text{for $P\in \calK_\N\setminus \{Q\}$}, \end{cases}
\end{equation}
and $\sym^n \calK_\N$ denotes the $n$th symmetric product of
$\calK_\N$. In particular, $\sym^m \calK_\N$ can be
identified with
the set of nonnegative
divisors $0 \leq \calD \in \Div(\calK_\N)$ of degree $m$.

For $f\in \calM (\calK_\N) \setminus \{0\}$,
$\omega \in \calM^1 (\calK_\N) \setminus \{0\}$ the
divisors of $f$ and $\omega$ are denoted
by $(f)$ and
$(\omega)$, respectively.  Two
divisors $\calD$, $\calE\in \Div(\calK_\N)$ are
called equivalent, denoted by
$\calD \sim \calE$, if and only if $\calD -\calE
=(f)$ for some
$f\in\calM (\calK_\N) \setminus \{0\}$.  The divisor class
$[\calD]$ of $\calD$ is
then given by $[\calD]
=\{\calE \in \Div(\calK_\N) \,|\, \calE \sim \calD\}$.  We
recall that
\begin{equation}
\deg ((f))=0,\, \deg ((\omega)) =2(\N-1),\,
f\in\calM (\calK_\N) \setminus
\{0\},\,  \omega\in \calM^1 (\calK_\N) \setminus \{0\},
\lb{a38}
\end{equation}
where the degree $\deg (\calD)$ of $\calD$ is given
by $\deg (\calD)
=\sum_{P\in \calK_\N} \calD (P)$.  It is customary to call
$(f)$ (respectively,
$(\omega)$) a principal (respectively, canonical)
divisor.

Introducing the complex linear spaces
\begin{align}
\calL (\calD) & =\{f\in \calM (\calK_\N) \,|\, f=0
        \text{ or } (f) \geq \calD\}, \quad
r(\calD) =\dim \calL (\calD),
\lb{a39}\\
\calL^1 (\calD) & =
        \{ \omega\in \calM^1 (\calK_\N) \,|\, \omega=0
        \text{ or } (\omega) \geq
\calD\}, \quad i(\calD) =\dim \calL^1 (\calD),  \lb{a40}
\end{align}
with $i(\calD)$ the index of speciality of $\calD$, one infers
that $\deg(\calD)$, $r(\calD)$, and $i(\calD)$ only depend on
the divisor class $[\calD]$ of $\calD$.  Moreover, we recall the
following fundamental facts.

\begin{theorem} \lb{thm1}
Let $\calD \in \Div(\calK_\N)$,
$\omega \in \calM^1 (\calK_\N) \setminus \{0\}$. Then, 
\begin{equation}
        i(\calD) =r(\calD-(\omega)), \quad \N\in\bbN_0.
\lb{a41}
\end{equation}
The Riemann--Roch theorem reads
\begin{equation}
r(-\calD) =\deg (\calD) + i (\calD) -\N+1,
\quad \N\in\bbN_0.
\lb{a42}
\end{equation}
By Abel's theorem, $\calD\in \Div(\calK_\N)$,
$\N\in\bbN$ is principal
if and only if
\begin{equation}
\deg (\calD) =0 \text{ and } \ul \alpha_{Q_0} (\calD)
=\ul{0}.
\lb{a43}
\end{equation}
Finally, assume
$\N\in\bbN$. Then $\ul \alpha_{Q_0}
: \Div(\calK_\N) \to J(\calK_\N)$ is surjective
$($Jacobi's inversion theorem$)$.
\end{theorem}

\begin{theorem} \lb{thm3}
Let $\calD_{\ul Q} \in \sym^\N \calK_\N$,
$\ul Q=\{Q_1, \ldots, Q_\N\}$.  Then,
\begin{equation}
1 \leq i (\calD_{\ul Q} ) =s   \lb{a46}
\end{equation}
if and only if $\{Q_1,\ldots, Q_\N\}$ contains $s$ pairings of the type 
$\{P, P^*\}$. $($This includes, of course, branch
points for which $P=P^*$.$)$ One has $s\leq \N/2$.
\end{theorem}

Denote by $\ul \Xi_{Q_0}=(\Xi_{Q_{0,1}}, \dots,
\Xi_{Q_{0,\N}})$ the vector of Riemann constants,
\begin{equation}
\Xi_{Q_{0,j}}=\frac12(1+\tau_{j,j})-
\sum_{\substack{\ell=1 \\ \ell\neq j}}^\N\int_{a_\ell}
\omega_\ell(P)\int_{Q_0}^P\omega_j,
\quad j=1,\dots,\N. \lb{aa55}
\end{equation}

\begin{theorem} \lb{taa17a}
Let $\ul Q =\{Q_1,\dots,Q_\N\}\in \sym^\N \calK_\N$ and
assume $\calD_{\ul Q}$ to be nonspecial, that is,
$i(\calD_{\ul Q})=0$. Then
\begin{equation}
\theta(\ul {\Xi}_{Q_0} -\ul {A}_{Q_0}(P) + \alpha_{Q_0}
(\calD_{\ul Q}))=0 \text{ if and only if }
P\in\{Q_1,\dots,Q_\N\}. \lb{aa55a}
\end{equation}
\end{theorem}

\begin{lemma} \lb{la6}
\cite[Lemmas 5.4 and 6.1]{BullaGesztesyHoldenTeschl:1997} Let $(n,t_{\ul r}),
(n_0,t_{0,\ul r})\in\Omega$ for some
$\Omega\subseteq\bbZ\times\bbR$.  Assume
$\psi(\dott,n,t_{\ul r})$  to be meromorphic on ${\calK}_\N\setminus
\{\Pinfp, \Pinfm, \Pzp, \Pzm\}$ with possible essential singularities at $\Pinfpm$,
$\Pzpm$ such that $\tilde \psi(\dott,n,t_{\ul r})$ defined by
\begin{equation}
\tilde \psi (P,n,t_{\ul r}) =\psi (P,n,t_{\ul r})
\exp \bigg((t_{\ul r}-t_{0,\ul r})\int_{Q_0}^P
\ti\Omega_{\ul r}^{(2)}\bigg) \lb{342a}
\end{equation}
is meromorphic on ${\calK}_\N$ and its divisor satisfies
\begin{equation}
(\tilde \psi (\dott,n,t_{\ul r}))\geq -{\calD}_{\humu (n_{0},t_{0,\ul r})}
+(n-n_0)\big(\calD_{\Pzm}-\calD_{\Pinfp}\big)
\lb{342b}
\end{equation}
for some positive divisor ${\calD}_{\humu (n_{0},t_{0,\ul r})}$ of degree
$\N$. Here $\ti\Omega_{\ul r}^{(2)}$ is defined in \eqref{AL4.71} and the
path of integration is chosen  identical to that in the Abel 
maps\footnote{This is to avoid multi-valued
expressions and hence the use of the multiplicative
Riemann--Roch theorem in the proof of Lemma \ref{la6}.} 
\eqref{aa46} and \eqref{aa47}. Define a divisor ${\calD}_0 (n,t_{\ul r})$ by
\begin{equation}
(\tilde \psi (\dott,n,t_{\ul r}))={\calD}_0 (n,t_{\ul r})
-{\calD}_{\humu (n_{0},t_{0,\ul r})}+(n-n_0)\big(\calD_{\Pzm}
-\calD_{\Pinfp}\big). \lb{342c}
\end{equation}
Then
\begin{equation}
{\calD}_0 (n,t_{\ul r}) \in\sym^{\N} {\calK}_{\N}, \quad {\calD}_0
(n,t_{\ul r}) > 0, \quad \deg ({\calD}_0 (n,t_{\ul r}))=p. \lb{342d}
\end{equation}
Moreover, if ${\calD}_0 (n,t_{\ul r})$ is nonspecial for all
$(n,t_{\ul r})\in\Omega$, that is, if
\begin{equation}
i ({\calD}_0 (n,t_{\ul r}) ) =0, \quad (n,t_{\ul r})\in\Omega, \lb{342e}
\end{equation}
then $\psi (\dott,n,t_{\ul r})$ is unique up to a constant
multiple $($which may depend on the parameters
$(n,t_{\ul r}), (n_{0},t_{0,\ul r})\in\Omega$$)$.
\end{lemma}

\section{Asymptotic Spectral Parameter Expansions} \lb{ALApp.high}
\renewcommand{\theequation}{B.\arabic{equation}}
\renewcommand{\thetheorem}{B.\arabic{theorem}}
\setcounter{theorem}{0}
\setcounter{equation}{0}

In this appendix we consider asymptotic spectral parameter expansions of
$F_{\ul p}/y$, $G_{\ul p}/y$, and $H_{\ul p}/y$, the resulting recursion relations for
the homogeneous coefficients  $\hat f_\ell$, $\hat g_\ell$, and $\hat
h_\ell$, their connection with the nonhomogeneous coefficients $f_\ell$,
$g_\ell$, and $h_\ell$, and the connection between $c_{\ell,\pm}$ and
$c_{\ell}(\ul E^{\pm 1})$ (cf.\ \eqref{ALB2.26h}). For detailed proofs of the material in this section we refer to \cite{GesztesyHolden:2005},  
\cite{GesztesyHoldenMichorTeschl:2007}. We will employ the notation
\begin{equation}
 \ul E^{\pm 1}=\big(E_0^{\pm 1},\dots,E_{2p+1}^{\pm 1}\big).   \lb{ALEpm}
\end{equation}

We start with the following elementary results (consequences of the  binomial
expansion) assuming $\eta\in\bbC$ such that
$|\eta|<\min\{|E_0|^{-1},\dots, |E_{2p+1}|^{-1}\}$:
\begin{equation}
\left(\prod_{m=0}^{2p+1} \big(1-{E_m}{\eta}\big)
\right)^{1/2}=\sum_{k=0}^{\infty}c_k(\ul
E)\eta^{k}, \lb{ALB2.26g}
\end{equation}
where
\begin{align}
c_0(\ul E)&=1,\no \\
c_k(\ul E)&=\!\!\!\!\!\!\!\!\sum_{\substack{j_0,\dots,j_{2p+1}=0\\
       j_0+\cdots+j_{2p+1}=k}}^{k}\!\!
\f{(2j_0)!\cdots(2j_{2p+1})!\, E_0^{j_0}\cdots E_{2p+1}^{j_{2p+1}}}
{2^{2k} (j_0!)^2\cdots (j_{2p+1}!)^2 (2j_0-1)\cdots(2j_{2p+1}-1)},
\quad k\in\bbN.  \label{ALB2.26h}
\end{align}
The first few coefficients explicitly are given by
\begin{align}
c_0(\ul E)=1, \;
c_1(\ul E)=-\f12\sum_{m=0}^{2p+1} E_m, \;
c_2(\ul E)=\f14\sum_{\substack{m_1,m_2=0\\ m_1< m_2}}^{2p+1}
E_{m_1} E_{m_2}-\f18 \sum_{m=0}^{2p+1} E_m^2,
\quad \text{etc.} \lb{ALB2.26i}
\end{align}

Next we turn to asymptotic expansions. We recall the convention
$y(P)= \mp \zeta^{-p-1}+\Oh(\zeta^{-p})$ near $\Pinfpm$
(where $\zeta=1/z$) and $y(P) = \pm (c_{0,-}/c_{0,+})+\Oh(\zeta)$ near $\Pzpm$ 
(where $\zeta=z$). 

\begin{theorem} [\cite{GesztesyHoldenMichorTeschl:2007}] \lb{tALB.2}
Assume \eqref{ALneq 0,1}, $\sAL_{\ul p}(\alpha,\beta)=0$, and suppose
$P=(z,y)\in\calK_p\setminus\{\Pinfp,\Pinfm\}$. Then $z^{p_-} F_{\ul p}/y$,
$z^{p_-} G_{\ul p}/y$, and $z^{p_-} H_{\ul p}/y$ have the following convergent expansions 
as $P\to \Pinfpm$, respectively, $P\to\Pzpm$,  
\begin{align} 
\frac{z^{p_-}}{c_{0,+}} \frac{F_{\ul p}(z)}{y} &= \begin{cases} 
\mp \sum_{\ell=0}^\infty \hat f_{\ell,+} \zeta^{\ell+1},  &
P\to \Pinfpm, \qquad \zeta=1/z, \\
\pm \sum_{\ell=0}^\infty \hat f_{\ell,-} \zeta^\ell,  &
P\to \Pzpm, \qquad \zeta=z,
\end{cases}  \label{ALF/y 0} \\
\frac{z^{p_-}}{c_{0,+}} \frac{G_{\ul p}(z)}{y} &= \begin{cases}
\mp \sum_{\ell=0}^\infty \hat g_{\ell,+} \zeta^\ell,  &
P\to \Pinfpm, \qquad \zeta=1/z, \\
\pm \sum_{\ell=0}^\infty \hat g_{\ell,-} \zeta^\ell,  &
P\to \Pzpm, \qquad \zeta=z, 
\end{cases}    \label{ALG/y 0} \\
\frac{z^{p_-}}{c_{0,+}} \frac{H_{\ul p}(z)}{y} &= \begin{cases}
\mp \sum_{\ell=0}^\infty \hat h_{\ell,+} \zeta^\ell,  &
P\to \Pinfpm, \qquad \zeta=1/z, \\
\pm \sum_{\ell=0}^\infty \hat h_{\ell,-} \zeta^{\ell+1},  &
P\to \Pzpm, \qquad \zeta=z,
\end{cases}    \label{ALH/y 0} 
\end{align}
where $\zeta=1/z$ $($resp., $\zeta=z$$)$ is the local coordinate near $\Pinfpm$ 
$($resp., $\Pzpm$$)$ and $\hat f_{\ell,\pm}$, $\hat g_{\ell,\pm}$, and $\hat h_{\ell,\pm}$ are the homogeneous versions of the coefficients $f_{\ell,\pm}$, $g_{\ell,\pm}$, 
and $h_{\ell,\pm}$ introduced in \eqref{AL2.04a}--\eqref{AL2.04c}. 

Moreover, the $E_m$-dependent summation constants
$c_{\ell,\pm}$, $\ell=0,\dots, p_{\pm}$, in $F_{\ul p}$, $G_{\ul p}$, and 
$H_{\ul p}$ are given by 
\begin{equation}
c_{\ell,\pm}= c_{0,\pm} c_\ell(\ul E^{\pm 1}), \quad \ell=0,\dots,p_{\pm}.   \lb{ALBc}
\end{equation}
\end{theorem}

{\bf Acknowledgments.}
We thank P.\ D.\ Miller for helpful discussions in connection with the Ablowitz--Ladik hierarchy. F.G., J.M., and G.T. gratefully acknowledge the extraordinary hospitality of the Department of Mathematical Sciences of the Norwegian University of Science and Technology, Trondheim, during extended stays in the summers of 2004--2006, where parts of this paper were written.


\end{document}